\documentclass[twoside]{IEEEtran}
\IEEEoverridecommandlockouts
\usepackage{cite}
\usepackage{amsmath,amssymb,amsfonts}
\usepackage{algorithmic}
\usepackage{graphicx}
\usepackage{subfig}
\usepackage{epstopdf}
\usepackage{textcomp}
\usepackage{xcolor}
\usepackage{bbold}
\usepackage{bm}
\usepackage{dsfont}
\usepackage{algorithm}
\usepackage{algorithmic}
\usepackage{multirow}
\usepackage{multicol}
\usepackage{array}
\usepackage{booktabs}
\usepackage{threeparttable}
\usepackage{gensymb}
\usepackage{amsthm}
\usepackage{fancyhdr}
\def\BibTeX{{\rm B\kern-.05em{\sc i\kern-.025em b}\kern-.08em
    T\kern-.1667em\lower.7ex\hbox{E}\kern-.125emX}}

\newtheorem{proposition}{Proposition}

\fancypagestyle{firstpage}
{
    \pagestyle{fancy}
    \fancyhead[LO]{\textbf{This work has been submitted to the IEEE for possible publication. Copyright may be transferred without notice, after which this version may no longer be accessible.}}
}

\begin{document}

\title{\huge{Robust Multitarget Tracking in Interference Environments: A Message-Passing Approach}}

%\author{\IEEEauthorblockN{Xianglong Bai, Quan Pan}\\
%\IEEEauthorblockA{\textit{School of Automation} \\
%\textit{Northwestern Polytechnical University}\\
%Xi'an, P.~R.~China \\
%baixianglong@mail.nwpu.edu.cn}
%}
\author{{Xianglong Bai}\thanks{The authors are with the School of Automation, Northwestern Polytechnical University, Xi’an 710129, China.
This work was supported in part
by the National Natural Science Foundation of China under Grant 61873211, Grant 61790552, Grant U21B2008, and in part by the Natural Science Basic Research Plan in Shaanxi Province of China under Grant 2021JM-06.}, Hua Lan, Zengfu Wang*\thanks{*~Corresponding author:~Zengfu Wang.}, Quan Pan, Yuhang Hao, Can Li}

\maketitle
\thispagestyle{firstpage}

\begin{abstract}
  Multitarget tracking in the interference environments suffers from the nonuniform, unknown and time-varying clutter, resulting in dramatic performance deterioration.
  We address this challenge by proposing a robust multitarget tracking algorithm, which estimates the states of clutter and targets simultaneously by the message-passing (MP) approach.
  We define the non-homogeneous clutter with a finite mixture model containing a uniform component and multiple nonuniform components.
  The measured signal strength is utilized to estimate the mean signal-to-noise ratio (SNR) of targets and the mean clutter-to-noise ratio (CNR) of clutter, which are then used as additional feature information of targets and clutter to improve the performance of discrimination of targets from clutter.
  We also present a hybrid data association which can reason over correspondence between targets, clutter, and measurements.
  Then, a unified MP algorithm is used to infer the marginal posterior probability distributions of targets, clutter, and data association by splitting the joint probability distribution into a mean-field approximate part and a belief propagation part.
  As a result, a closed-loop iterative optimization of the posterior probability distribution can be obtained, which can effectively deal with the coupling between target tracking, clutter estimation and data association.
  Simulation results demonstrate the performance superiority and robustness of the proposed multitarget tracking algorithm compared with the probability hypothesis density (PHD) filter and the cardinalized PHD (CPHD) filter.
\end{abstract}


\section{INTRODUCTION}\label{sec:INTRODUCTION}
% \subsection{Joint Clutter Detection and Target Tracking}\label{subsec:Clutter Estimation and Target Tracking}
In radar target tracking, interference is often present in the received signals.
Interference may arise in different forms, such as objects that are not of interest~(e.g., precipitation, vegetation, soil), and electronic countermeasures~(e.g., suppression jamming, chaff jamming), etc~\cite{Richards2010}.
In many scenarios, interference suppression and clutter elimination techniques may not be as effective as expected and result in some nonuniform, unknown and time-varying clutter.
The state-of-the-art algorithms assume that the distributions of both the spatial position and the number of clutter are known and fixed.
Specifically, clutter is uniformly distributed in the
space of radar measurements,
the number of which follows a Possion distribution with a fixed mean~\cite{Bar1995Multitarget}.
In specific applications, if the tracking algorithm uses a clutter distribution that does not match the real clutter distribution, it may lead to missed target tracking, increased false tracks and computational complexity.
To this end, robust multitarget tracking (RMTT) can be achieved by joint clutter estimation and target tracking (JCETT), which estimates the time-varying states of moving targets and unknown clutter simultaneously from measurements, leading to a significant performance improvement of multitarget tracking~(MTT).
Furthermore, to obtain a better performance in dense clutter and low SNR environments, it is desired to incorporate signal strength information into MTT.

However, RMTT in interference environments is complicated by the following factors.
(a) \emph{Clutter modelling and estimation}: Modern radars suffer from strong unwanted interference from natural environments and countermeasures.
For example, atmosphere, e.g., water, fog, snow, smoke, is usually volumetric scattering and  can result in illuminating clutter.
The received signals of atmosphere may vary by several reasons, such as frequency, suspended particle sizes, and concentrations of atmospheric particles.
In addition, the chaff interference is ejected from an aircraft or ship and blooms into a large reflector hovering in the cloud, which is affected by a couple of factors, such as radar cross section~(RCS), shape of chaff clouds and atmospheric phenomena, etc~\cite{Richards2010}.
Consequently, there are different types of clutter distribution in the background of target tracking: one uniform distribution generated by noise; several nonuniform distributions generated by interference, each of which we refer to as a clutter component.
The non-homogeneous distributions make clutter modelling and estimation very challenging.
(b) \emph{Data association}: MTT in interference environments is difficult due to the unknown association between measurements and targets as well as the unknown association between measurements and clutter components.
It is often assumed that at each time, a target can generate at most one measurement, a clutter component can potentially generates multiple measurements, and a measurement can be generated by either one target or one clutter component.
Unfortunately, the exponential complexity of data association makes it incredibly challenging.
In addition, the data association and the state estimation of target and clutter are highly coupled, i.e., erroneous data association deteriorates the state estimation of target and clutter, and the inaccurate state estimation of target and clutter leads to the data association risk.

The majority of the RMTT literature considering clutter estimation focused on finite set statistics.
Mahler \emph{et al.}~\cite{Mahler2011} learned the clutter intensity while target tracking by proposing an adaptive PHD/CPHD filter.
Assuming unknown and time-varying clutter, Beard \emph{et al.}~\cite{Beard2013} proposed a bootstrap CPHD filter, which performed comparatively as well as the matched CPHD filter.
Kim and Song~\cite{Kim2017} proposed a PHD filter with clutter intensity estimation in dynamic cluttered environments.
In environments with unknown and non-homogeneous clutter, Vo \emph{et al.}~\cite{Vo2013} proposed a robust multi-Bernoulli~(MB) filter to adaptively learn clutter parameters by modelling clutter based on multiple independent generators.
In the vein of \cite{Vo2013}, Gostar \emph{et al.}~\cite{Gostar2015} proposed an MB filter with clutter estimation for sensor selection.
In~\cite{Lian2010, Chen2012, Liu2018}, the clutter intensity was modelled by a finite mixture model, and estimated in the random finite set scheme by the expectation maximization method and the Markov chain Monte Carlo method.

There is a large volume of published studies devoted to MTT using signal strength-related information to achieve RMTT.
To improve the discrimination ability for closely spaced targets, assuming that the target SNR was known and fixed during tracking, the problem of data association between measurements and targets based on measured spatial and amplitude information was addressed in~\cite{Lerro1993, Ehrman2009}.
Then, the amplitude likelihood was marginalized over a range of possible target SNR values and the corresponding amplitude information was integrated into the general PHD and CPHD filters~\cite{Clark2010}, and cardinality balanced multitarget multi-Bernoulli (CBMeMBer) filter~\cite{Feng2016}.
The amplitude information was used in multi-object filtering to estimate the target kinematic state and RCS in \cite{Mertens2016} for robust ground target tracking.
Later, the amplitude information was used in MTT for a cluttered environment, and SNR estimation algorithms were proposed based on the maximum a posteriori method~\cite{Bae2017} and the sequential Monte Carlo method~\cite{Bae2019}.
To improve the discrimination between targets and clutter, the joint amplitude likelihood of the sea clutter neighbouring cells is calculated, which is then integrated into a labelled MB filter~\cite{Sun2019a}.
Yang \emph{et al.}~\cite{Yang2018} incorporated the amplitude information into the simultaneously clutter estimation and target tracking.
Ristic \emph{et al.}~\cite{Ristic2021} proposed an MB filter for maritime target tracking using amplitude information.

However, most of the existing RMTT methods for JCETT only estimate the spatial distribution or the amplitude distribution of clutter, but not both.
As far as we know, the previously proposed JCETT method estimating both the spatial state and return power of clutter is limited to the method proposed in~\cite{Yang2018}.
The method in~\cite{Yang2018} adopted the random finite set filters for clutter estimation and target tracking, where the spatial and SNR of target and clutter were modelled as an inverse Gamma Gaussian (IGG) distribution.
Unfortunately, the method~\cite{Yang2018} did not provide details on the spatial state estimation for nonuniform clutter and also resulted in an unappealing separate and sequential estimation framework for clutter and target.
As we mentioned before, the target and clutter state estimation and data association are highly coupled and affect each other.
It is highly demanded to develop a closed-loop iterative optimization framework for JCETT.

The Bayesian inference algorithms for probabilistic graphical models, such as (loopy) belief propagation (BP)~\cite{Yedidia2005} and variational inference~\cite{Zhang2019}, have been used for MTT.
Chen \emph{et al.}~\cite{Chen2006Data} considered the data association problem as a maximum a posteriori configuration problem, and solved it by the max-product BP.
In contrast to~\cite{Chen2006Data}, Williams \emph{et al.}~\cite{Williams2014} considered the data association problem as a posteriori probability estimation problem, and solved it by the sum-product BP.
Later, the data association method proposed in~\cite{Williams2014} was extended to the multi-scan version~\cite{Williams2018}, the multi-path version~\cite{Sun2016IF}, and labeled MB filtering~\cite{Kropfreiter2019}.
Meyer \emph{et al.}~\cite{meyer2017scalable,Meyer2018} proposed a scalable multi-sensor MTT algorithm by using BP, which was then extended to the scheme of self-tuning the unknown model parameters~\cite{Soldi2019}.
In addition, the BP method was also used to extended target tracking~\cite{Meyer2020,meyer2021scalable}, cooperative self-localization and MTT~\cite{Sharma2019}, joint registration and fusion of heterogeneous sensors~\cite{Cormack2019}, fusion of sensor measurements and target-provided information in MTT~\cite{Gaglione2022}.

In our previous work, the unified MP method~\cite{Riegler2013}, which combines the virtues of (loopy) belief propagation (BP)~\cite{Yedidia2005} and Mean-field (MF) approximation~\cite{Zhang2019} while circumventing their drawbacks, has been used for multi-path environment~\cite{Lan2019}, maneuvering target tracking~\cite{Lan-2020-107621} and over-the-horizon radar network fusion~\cite{Lan2020}.
These algorithms were derived by representing the joint probability density functions (PDFs) associated with the system models by a factor graph.
We decomposed the factor graph into an MF part and a BP part, and the posterior PDFs of the corresponding hidden variables were approximated by BP and MF.
As a result, a closed-loop iterative optimization of the posterior probability distribution were obtained, which can effectively deal with the coupling between latent variables.

In this paper, we propose an MP-based framework and algorithm for RMTT in interference environments using strength information of measurements.
We formulate RMTT as an JCETT problem including all target states, clutter states and data association.
In particular, we use the Swerling-I and Swerling-III models to represent the RCS fluctuations of target and clutter, and adopt the Rayleigh likelihood for the strength information.
Furthermore, the clutter intensity is modelled as a finite mixture model.
An enabling technique for our methods is the combined formulation of data association which can reason over correspondence between targets, clutter, and measurements.
By this new formulation of the RMTT problem, the statistical structure of RMTT is represented by a factor graph.
Finally, we use the unified MP algorithm to solve the problem, in which the MF approximation and BP are used in the MF part and the BP part of the factor graph, respectively.
Different from our previous work in~\cite{Lan-2020-107621,Lan2020}, the factor graph constructed in this paper has several new subgraphs, including the target SNR subgraph, the clutter spatial state and mean CNR subgraph, the clutter mixing weight subgraph, and the combined data association subgraph.
The modelling and massage passing of these subgraphs makes the problem more challenging.
The summary contributions of this paper are as follows:
\begin{itemize}
  \item We formulate a Bayesian statistical framework for the RMTT problem involving all the hidden variables of targets, clutter and data association.
  In particular, the target and clutter state are modelled in both the spatial and the power domains, which can improve the performance of RMTT.
  We also develop a combined data association which can reason over correspondence between targets, clutter, and measurements.
  \item We use the MP approach combined with the MF approximation and BP to solve the RMTT problem, referred to MP-RMTT.
  In MP-RMTT, the estimations of all hidden variables are optimized by a closed-loop iterative architecture, i.e., the target and clutter state estimations in the previous iteration are utilized to improve the data association, and then the new data association is used to optimize the target and clutter state estimations, which is capable for handling the coupling issue between the hidden variables.
\end{itemize}

% \subsection{Paper Organization}\label{subsec:Contributions, Paper Organization and Notations}
The rest of the paper is organized as follows. The problem formulation of RMTT is  described in Section \ref{sec:system}.
In Section \ref{sec:MP-Based Approach}, the proposed MP-RMTT algorithm is derived.
Section \ref{sec:EXPERIMENTAL} evaluates the performance of MP-RMTT via simulations.
At last, Section \ref{sec:CONCLUSIONS} concludes this paper.

\section{PROBLEM FORMULATION}\label{sec:system}
In this section, we present the system models firstly.
After that, we state the RMTT problem to be solved.

\subsection{Target State Modelling}\label{subsec:Modeling of target and clutter}
At time $k$, let $\bm{X}^{\rm t}_{k}=\{ \bm{X}^{\rm t}_{i,k} \}_{i=1} ^{N_T}$ be the target \emph{joint augmented state}, where $N_T$ is the number of targets and is determined in initialization stage, and $\bm{X}^{\rm t}_{i,k}=[({{\bm{x}}} _{i,k}^{\rm {t}})^ {\rm{T}} \ \sigma_{i,k}^{\rm t}]^{\rm{T}}$ is the \emph{augmented state} of target $i$, consisting of the target kinematic state ${\bm{x}}_{i,k}^{\rm t}$ and mean SNR $\sigma_{i,k}^{\rm t}$.
The kinematic state ${\bm{x}}_{i,k}^{\rm t}$ contains the position and velocity of target $i$.
In addition, let target mean SNR $\sigma_{i,k}^{\rm t} = S_{i,k}^{\rm t}/N_0$, where $S_{i,k}^{\rm t}$ is the expected target signal power and $N_0$ is the expected noise power.
The SNR represented in log scale is SNR(dB) = $10 \log_{10} (\sigma_{i,k}^{\rm t})$~\cite{Skolnik2002}.
For each target $i$, we define the target kinematic state sequence and the target mean SNR sequence over time $1$ to $K$ as ${\bm{x}} _{i,1:K}^{\rm t}=\{{\bm{x}} _{i,k}^{\rm t}\}_{k=1} ^{K}$ and ${\bm{\sigma}} _{i,1:K}^{\rm t}=\{{\sigma} _{i,k}^{\rm t}\}_{k=1} ^{K}$, respectively.
We define the target joint augmented state sequence over time $1$ to time $K$ as $\bm{X} _{1:K} ^{\rm t} = \{\bm{X} _{k} ^{\rm t}\}_{k=1} ^{K}$.

For a same target, assuming that the PDF for the kinematic state, $p(\bm{x} _{i,k}^{\rm t})$, and the PDF for the mean SNR, $p(\sigma _{i,k} ^{\rm t})$, are independent, the PDF of target augmented state is factorized as $p(\bm{X}^{\rm t}_{i,k})= p(\bm{x} _{i,k}^{\rm t})p(\sigma _{i,k} ^{\rm t})$.
$p(\bm{x} _{i,k}^{\rm t})$ is chosen to be a Gaussian distribution which is conjugate prior of the spatial measurement likelihood (which will be detailed in Section~\ref{subsec: Measurement model}).
$p(\sigma _{i,k} ^{\rm t})$ is chosen to be the IG distribution, which is conjugate prior of the strength measurement likelihood (which will be detailed in Section~\ref{subsec: Measurement model})~\cite{Mertens2016}.
We refer the reader to~\cite{Mertens2016, Yang2018} for details of the IG distribution.

By assuming that each target augmented state evolves independently with a first-order Markov dynamic model, the PDF of $\bm{X} _{1:K} ^{\rm t}$ is
\begin{equation}\label{equ:target transition}
\begin{split}
p(\bm{X} _{1:K} ^{\rm t}) =\prod_{i = 1}^{N_T} p(\bm{X}_{i,1}^{\rm {t}}) \prod_{k = 2}^{K} p(\bm{X}_{i,k}^{\rm {t}}|\bm{X}_{i,k-1}^{\rm {t}}), \\
\end{split}
\end{equation}
where $p(\bm{X}_{i,1}^{\rm {t}})$ is a prior PDF at time $1$ and $p(\bm{X}_{i,k}^{\rm {t}}|\bm{X}_{i,k-1}^{\rm {t}})$ is the transition PDF of target augmented state.
Assuming that the target kinematic state transition PDFs $p(\bm{x}_{i,k}^{\rm t}|\bm{x}_{i,k-1}^{\rm t})$ and the target mean SNR transition PDFs $p(\sigma ^{\rm t} _{i,k}| \sigma ^{\rm t} _{i,k-1})$ are independent, one has $p(\bm{X} _{i,k} ^{\rm t} |\bm{X} _{i,k-1} ^{\rm t}) =p(\bm{x} _{i,k} ^{\rm t} |\bm{x} _{i,k-1} ^{\rm t}) p(\sigma ^{\rm t} _{i,k}| \sigma ^{\rm t} _{i,k-1})$.
The state transition PDF of target kinematic state $p(\bm{x}_{i,k}^{\rm t}|\bm{x}_{i,k-1}^{\rm t})$ can be deteiminted by the dynamic model of each target~\cite{Lan-2020-107621}.
Assume that the target mean SNR varies slowly.
Given the PDF of target mean SNR at time $k-1$, we define $p(\sigma ^{\rm t} _{i,k}| \sigma ^{\rm t} _{i,k-1}) = \mathcal{I} (\sigma _{i,k} ^{\rm t}; \alpha _{i,k|k-1} ^{\rm t}, \beta _{i,k|k-1} ^{\rm t})$, where $\alpha ^{\rm t} _{i,k|k-1} = ({{\alpha ^{\rm t} _{i,k-1} +u ^{\rm t}-1}})/ {u^{\rm t}}$, $\beta ^{\rm t} _{i,k|k-1} ={\beta ^{\rm t} _{i,k-1}}/ {u^{\rm t}}$ with $\alpha ^{\rm t} _{i,k-1}$ and $\beta ^{\rm t} _{i,k-1}$ being parameters of the PDF $p(\sigma ^{\rm t} _{i,k-1})$, and $u^{\rm t}$ being a forgetting factor~\cite{Yang2018}.
We assume that the target mean SNR can be predicted in reverse by the same transition PDFs as the forward prediction.

At time $k$, let $\bm{S}_k=\{s_{i,k} \}_{i=1} ^{N_T}$ be the target joint visibility state, where $s_{i,k}  \in \{0,1\}$ is a binary random variable and indicates the presence of target $i$ if $s_{i,k}=1$ or the absence of target $i$ if $s_{i,k}=0$.
We define the target visibility state sequence over time $1$ to time $K$ of target $i$ as ${\bm{s}} _{i,1:K}=\{{{s}} _{i,k}\}_{k=1} ^{K}$.
We also define the sequence of target joint visibility state over  time $1$ to time $K$ as $\bm{S} _{1:K} = \{\bm{S} _{k} \} _{k=1} ^{K}$ .
By assuming that the appearance or the disappearance of each targets are independent and the visibility state of each target transits based on the Markov process, the PDF of $\bm{S} _{1:K}$ can be written as
\begin{equation}\label{equ:target visibility}
p(\bm{S} _{1:K})= \prod_{i = 1}^{N_T} p({s}_{i,1}) \prod_{k = 2}^{K} p({s}_{i,k}|s_{i,k-1}), \\
\end{equation}
where $p({s}_{i,1})$ is the prior PDF as a Bernoulli distribution, and the transition PDF $p({s}_{i,k}|s_{i,k-1})$ is represented by a matrix
\begin{equation}\label{equ:presence transition}
\begin{split}
  \bm{T}_k = & \left[ \begin{array}{cc} p(s_{i,k}=1|s_{i,k}=1) & p(s_{i,k}=1|s_{i,k}=0) \\ p(s_{i,k}=0|s_{i,k}=1) & p(s_{i,k}=0|s_{i,k}=0) \\ \end{array} \right] \\
  = & \left[ \begin{array}{cc} p_s & p_b \\ 1-p_s & 1-p_b \\ \end{array} \right],
\end{split}
\end{equation}
where $p_{\rm s}$ is the target survival probability and $p_{\rm b}$ is the target birth probability.

\subsection{Clutter Modelling}\label{subsec: Clutter Modelling}
The clutter is modelled in the space of measurements.
We assume that there are one uniform clutter component and multiple nonuniform clutter components distributed over the entire surveillance region.
Let $\tau=0$ and $\tau=1,\ldots,N_C$ be the indices of the uniform and the nonuniform clutter component respectively, and $N_C$ is the maximum possible number of nonuniform clutter components and is determined in the initialization stage.

We define the clutter joint augmented state as $\bm{X}_k^{\rm c}= \{\bm{X} _{\tau,k} ^{\rm c}\} _{\tau=0} ^{N_C}$, where $\bm{X}_{0,k}^{\rm c} =\sigma _{0,k} ^{\rm c}$ is the state of the uniform clutter component and $\bm{X}_{\tau,k} ^{\rm c} = \{ \tilde{\bm{x}} _{\tau,k} ^{\rm c} \ \sigma _{\tau,k} ^{\rm c} \}$ is the augmented state of nonuniform clutter component $\tau$.
Here, $\sigma _{\tau,k} ^{\rm c} = S_{\tau,k}^{\rm c}/N_0$ denotes the mean clutter-to-noise ration (CNR) of the clutter component $\tau$, $\tau=0,\ldots,N_C$, where $S_{\tau,k}^{\rm c}$ is the expected clutter power.
The CNR represented in log scale is CNR(dB) = $10 \log_{10} (\sigma_{\tau,k}^{\rm c})$~\cite{Skolnik2002}.
Note that we assume the expected power of the uniform clutter is equal to the power of background noise, so the CNR of the uniform clutter is 1. In fact, since the background noise is unknown, we estimate the CNR of the uniform clutter as well.
Then, $\tilde{\bm{x}}^{\rm c}_{\tau,k} =\{{\bm{x}} _{\tau,k} ^{\rm {c}} \ \bm{D}^{\rm c}_{\tau,k}\}$ is the spatial state of the nonuniform clutter component $\tau$, $\tau=1,\ldots,N_C$, where ${\bm{x}} _{\tau,k} ^{\rm {c}}$ and $\bm{D}^{\rm c}_{\tau,k}$ are the corresponding position and shape parameters, respectively.
At time $k$, we define $\bm{\Pi}_{k}=\{\pi_{\tau,k}\}_{\tau=0} ^{N_C}$ as the clutter joint mixing weights, where $\pi_{\tau,k}$ is the mixing weight of clutter component $\tau$ and satisfies $0\leq \pi_{\tau,k}\leq 1$ and $\sum_{\tau=0}^{N_C} \pi_{\tau,k}=1$.
We define the clutter joint spatial state as $\tilde{\bm{X}} _{k}^{\rm c}=\{\tilde{\bm{x}} _{\tau,k}^{\rm c}\}_{\tau=1} ^{N_C}$ and the clutter joint CNR as ${\bm{\sigma}} _{k}^{\rm c}=\{{\sigma} _{\tau,k}^{\rm c}\}_{\tau=1} ^{N_C}$.
We define the spatial state sequence, the mean CNR sequence, the mixing weight sequence of clutter component $\tau$ over time $1$ to time $K$ as ${\tilde{\bm{X}}} _{\tau,1:K}^{\rm t}=\{\tilde{\bm{x}} _{\tau,k}^{\rm c}\}_{k=1} ^{K}$, ${\bm{\sigma}} _{\tau,1:K}^{\rm c}=\{{\sigma} _{\tau,k}^{\rm c}\}_{k=1} ^{K}$,
${\bm{\pi}} _{\tau,1:K}^{\rm t}=\{{{\pi}} _{\tau,k}^{\rm t}\}_{k=1} ^{K}$, respectively.
We define the clutter joint augmented state sequence and the clutter joint mixing weights sequence over time $1$ to time $K$ as $\bm{X} _{1:K} ^{\rm c} = \{\bm{X} _{k} ^{\rm c}\}_{k=1} ^{K}$ and $\bm{\Pi} _{1:K} ^{\rm c} = \{\bm{\Pi} _{k} ^{\rm c}\}_{k=1} ^{K}$, respectively.

We assume the PDF of clutter spatial state $p(\tilde{\bm{x}} _{i,k}^{\rm c})$ and the PDF of clutter mean CNR $p(\sigma _{i,k} ^{\rm c})$ are independent, and the PDF of the clutter augmented state can be factorized as $p(\bm{X}^{\rm c}_{i,k}) = p(\tilde{\bm{x}} _{i,k}^{\rm c}) p(\sigma _{i,k} ^{\rm c})$.
Since the position measurement likelihood given the clutter spatial state~(will be detailed in Section~\ref{subsec: Measurement model}) is a Gaussian distribution, we define the PDF $p(\tilde{\bm{x}} _{i,k}^{\rm c})$ as a Gaussian-Wishart (GW) distribution which is the conjugate prior of the mean and covariance for a Gaussian distribution.
We refer the reader to~\cite{2006Pattern} for details of the Wishart distribution.
Therefore, we introduce the GW prior governing the spatial parameters of each nonuniform clutter component, given by
\begin{equation}\label{equ:prior2}
\begin{split}
  p (& \tilde{\bm{x}}^{\rm c}_{\tau,k}) = p (\bm{x}^{\rm c}_{\tau,k}, \bm{D}^{\rm c}_{\tau,k}) = p(\bm{x}^{\rm c}_{\tau,k}|\bm{D}^{\rm c}_{\tau,k}) p(\bm{D}^{\rm c}_{\tau,k}) = \\
  & \mathcal{N}({\bm{x}} ^{\rm c} _{\tau,k} ; \hat{\bm{x}} ^{\rm c} _{\tau,k}, (\beta^{\rm c} _{\tau,k} {\bm{D}} ^{\rm c} _{\tau,k})^{-1})
  \mathcal{W}({\bm{D}} ^{\rm c} _{\tau,k} ; {\bm{W}} ^{\rm c} _{\tau,k}, {{\upsilon}} ^{\rm c} _{\tau,k}).
\end{split}
\end{equation}
Assuming that the clutter and targets have the same likelihood function on signal strength, the PDF of clutter mean CNR $p(\sigma _{i,k} ^{\rm c})$ is also chosen to be the IG distribution.
In addition, since the likelihood of the data association event given the clutter mixing weights~(will be detailed in Section~\ref{subsec: Measurement model}) is a multinomial distribution, we choose a Dirichlet distribution on the clutter mixing weights ${\rm Dir}(\bm{\Pi}_{k}|\bm{\alpha}_k) = C(\bm{\alpha}_k) \prod _{\tau=0} ^{N_C} {\pi_{\tau,k}} ^ {\alpha_{\tau,k} }$, where $C(\bm{\alpha}_k)$ is the normalization constant, and $\bm{\alpha}_k =\{{\alpha} _{\tau,k}\} _{\tau=0} ^{N_C}$ with ${\alpha}_{\tau,k}$ as the prior number of points associated with clutter component $\tau$.

By assuming that each clutter augmented state transits independently based on the Markov process, the PDF of $\bm{X} _{1:K} ^{\rm c}$ can be written as
\begin{equation}\label{equ:clutter transition X}
p(\bm{X} _{1:K} ^{\rm c})= \prod_{\tau = 1}^{N_T} p(\bm{X}_{\tau,1}^{\rm {c}}) \prod_{k = 2}^{K} p(\bm{X}_{\tau,k}^{\rm {c}}|\bm{X}_{\tau,k-1}^{\rm {c}}), \\
\end{equation}
where $p(\bm{X}_{\tau,1}^{\rm {c}})$ is the prior PDF at time $1$ and $p(\bm{X}_{\tau,k}^{\rm {c}}|\bm{X}_{\tau,k-1}^{\rm {c}})$ is the transition PDF of clutter augmented state.
Assume that the clutter spatial transition PDFs $p(\tilde{\bm{x}}^{\rm c}_{\tau,k}| \tilde{\bm{x}}^{\rm c}_{\tau,k-1})$ and the clutter mean CNR transition PDFs $p(\sigma ^{\rm c} _{\tau,k}| \sigma ^{\rm c} _{\tau,k-1})$ are independent, one has $p({\bm{X}}^{\rm c}_{\tau,k}| {\bm{X}}^{\rm c}_{\tau,k-1}) = p(\tilde{\bm{x}}^{\rm c}_{\tau,k}| \tilde{\bm{x}}^{\rm c}_{\tau,k-1}) p(\sigma ^{\rm c} _{\tau,k}| \sigma ^{\rm c} _{\tau,k-1})$.
In some real applications, compared with the high speed of targets, the dynamics of the interference mainly changes on shape rather than on spatial movement.
For example, the atmosphere moves slowly while its shape varies as the meteorological conditions;
the chaff cloud hovers in the air while its scattering volume spreads~\cite{Richards2010}.
Thus, we assume that the position of the nonuniform clutter changes within a relatively small region and the shape of the nonuniform clutter varies over time.
Accordingly, we define the transition PDF of the clutter spatial state as the GW distribution as well, that is,
\begin{equation}\label{equ:Transition GW}
\begin{split}
  p(\!\tilde{\bm{x}}^{\rm c}_{\tau,k}| \tilde{\bm{x}}^{\rm c}_{\tau,k-1}) \!= & \mathcal{N}({\bm{x}} ^{\rm c} _{\tau,k} ; \hat{\bm{x}} ^{\rm c} _{\tau,k|k-1}, (\beta^{\rm c} _{\tau,k|k-1} {\bm{D}} ^{\rm c} _{\tau,k})^{-1}) \\
  & \times \mathcal{W}({\bm{D}} ^{\rm c} _{\tau,k} ; {\bm{W}} ^{\rm c} _{\tau,k|k-1}, {{\upsilon}} ^{\rm c} _{\tau,k|k-1}),
\end{split}
\end{equation}
where $\hat{\bm{x}} ^{\rm c} _{\tau,k|k-1}=\hat{\bm{x}} ^{\rm c} _{\tau,k-1}$, ${\beta} ^{\rm c} _{\tau,k|k-1}= \beta ^{\rm c} _{\tau, k-1}$, ${\bm{W}} ^{\rm c} _{\tau,k|k-1}=\xi {\bm{W}} ^{\rm c} _{\tau,k-1}$, ${{\upsilon}} ^{\rm c} _{\tau,k|k-1}=\xi ({{\upsilon}} ^{\rm c} _{\tau,k-1}-m-1)+m+1$, with $\hat{\bm{x}} ^{\rm c} _{\tau,k-1}$, $\beta ^{\rm c} _{\tau, k-1}$, ${\bm{W}} ^{\rm c} _{\tau,k-1}$, ${{\upsilon}} ^{\rm c} _{\tau,k-1}$ being the parameters of $p(\tilde{\bm{x}}^{\rm c}_{\tau,k-1})$, $m=2$ being the dimension of the position measurement, and $\xi$ being a forgetting factor~\cite{Huang2018}.
Again, we assume that the clutter spatial state can be predicted in reverse by the same transition PDFs as the forward prediction.
The transition PDF $p(\sigma ^{\rm c} _{\tau,k}| \sigma ^{\rm c} _{\tau,k-1})$ and the inverse transition PDF $p(\sigma ^{\rm c} _{\tau,k-1}| \sigma ^{\rm c} _{\tau,k})$ of clutter mean CNR are the same as those of target mean SNR with corresponding forgetting factor $u^{\rm c}$.
By assuming that each clutter component appears and disappears independently and transits based on the Markov process, the PDF of $\bm{\Pi} _{1:K} ^{\rm t}$ can be represented as
\begin{equation}\label{equ:clutter transition P}
p(\bm{\Pi} _{1:K} ^{\rm t}) =\prod_{\tau = 1}^{N_T} p({\pi}_{\tau,1}) \prod_{k = 2}^{K} p({\pi}_{\tau,k}|\pi_{\tau,k-1}), \\
\end{equation}
where $p({\pi}_{\tau,1})$ is the prior PDF at time $1$ and $p({\pi}_{\tau,k}|\pi_{\tau,k-1})$ is transition PDF of clutter mixing weight,
which is also defined as the Dirichlet distribution, that is $p(\pi_{\tau,k}| \pi_{\tau,k-1}) = {\rm Dir}({\pi}_{\tau,k}|{\alpha}_{\tau,k|k-1})$.
Assuming that the mixing weights change slowly and $\alpha_{\tau,k-1} / \sum_{\tau'=0}^{N_C} \alpha_{\tau',k-1} = \alpha_{\tau,k} / \sum_{\tau'=0}^{N_C} \alpha_{\tau',k}$.
We define $\alpha_{\tau,k|k-1} = \kappa M_{k-1} \alpha_{\tau,k-1} / \sum_{\tau'=0}^{N_C} \alpha_{\tau',k-1}$ by a heuristic approach, where $\kappa$ is a balance parameter tuning the effects of the prior knowledge and $M_{k-1}$ is the total number of clutter at time $k-1$.
We assume that the clutter mixing weights can be predicted in reverse by the same transition PDFs as the forward prediction.

\subsection{Measurement Modelling}\label{subsec: Measurement model}
At time $k$, let $\bm{Y}_k=\{\bm{Y}_{j,k}\}_{j=1}^{N_{M,k}}$ be the measurements, which are generated from radar range-azimuth-strength map using the constant false alarm detector, followed by a peak extraction scheme and detection with a threshold $d>0$, and $N_{M,k}$ is the number of measurements.
Each measurement $\bm{Y}_{j,k} =[{\bm{y}}_{j,k}^ {\rm{T}} \ m_{j,k}]^{\rm{T}}$ consists of two elements: (i) the spatial information in the polar coordinates ${\bm{y}}_{j,k}=[r_{j,k} \ \xi_{j,k}]^{\rm T}$; (ii) the measured signal strength $m_{j,k} = S_{j,k}/N_0>d$, where $S_{j,k}$ is received signal power.
We define the measurement sequence over time $1$ to time $K$ as $\bm{Y} _{1:K} = \{\bm{Y} _{k}\}_{k=1} ^{K}$.

For the target-originated measurement, the spatial measurement likelihood of $\bm{y}_{j,k}$ given $\bm{x}_{i,k}$ is denoted as $p({\bm{y}}_{j,k}| {\bm{x}}_{i,k})$, which can be determined by the spatial measurement equation~\cite{Bar1995Multitarget}.
The RCS fluctuations of target can be captured by the Swerling-I and Swerling-III models~\cite{Richards2010}, and the corresponding PDF of strength $m_{j,k}$ in noise background given the target mean SNR $\sigma _{i,k} ^{\rm t}$ can be represented by the general Rayleigh distribution~\cite{Skolnik2002}
\begin{equation}\label{equ:Rayleigh probability density}
  \mathcal{R}(m_{j,k};\sigma_{i,k} ^{\rm t},n) =\frac{2m_{j,k} ^{2n-1}} {(\sigma_{i,k} ^{\rm t} +1)^n} \exp \left( -n\frac{m_{j,k}^2} {\sigma_{i,k} ^{\rm t}} \right),
\end{equation}
where $n=1$ and $n=2$ denotes the Swerling-I and Swerling-III models, respectively.
Using the approximate expression, the corresponding detection probability is~\cite{Yang2018}
\begin{equation}\label{equ:detection probability}
  p_D^d=\exp\left( -n\frac{d^2}{\sigma_{i,k} ^{\rm t}+1} \right).
\end{equation}
The general Rayleigh PDF after thresholding becomes
\begin{equation}\label{equ:pdf after thresholding1}
  \mathcal{R}^d(m_{j,k};\sigma_{i,k} ^{\rm t},n) \!=\! \frac{2m_{j,k}^{2n-1}}{(\sigma_{i,k} ^{\rm t}+1)^n}\exp\! \left(\! -n\frac{m_{j,k}^2-d^2}{\sigma_{i,k} ^{\rm t}+1} \right).
\end{equation}

Likewise in \cite{Mertens2016,Yang2018}, we assume that the measured position and the measured strength information are independent of each other; the corresponding likelihood function can be decomposed as
\begin{equation}\label{equ:clutter distribution X}
  p(\bm{Y}_{j,k}|\bm{X}_{i,k}^{\rm t})=p({\bm{y}}_{j,k} | {\bm{x}}_{i,k}^{\rm t}) \mathcal{R}^d(m_{j,k};\sigma_{i,k}^{\rm t},n).
\end{equation}

We use the finite mixture model to model the spatial distribution of clutter, given by
\begin{equation}\label{equ:finite mixture distribution}
p({\bm{y}}_{j,k}|\tilde{\bm{X}} _{k}^{\rm c},\bm{\Pi}_k) = \pi_{0,k}U(V_G) + \sum _{\tau=1} ^{N_C} \pi_{\tau,k} p({\bm{y}}_{j,k}|\tilde{\bm{x}} _{\tau,k}^{\rm c}),
\end{equation}
where $U(V_G)=1/V_G$ is the uniform distribution representing the spatial distribution of uniform clutter component $\tau=0$; $p({\bm{y}}_{j,k} |\tilde{\bm{x}} _{\tau,k}^{\rm c}) =\mathcal{N} ({\bm{y}} _{j,k}; {\bm{x}} _{\tau,k}^{\rm c}, ({\bm{D}} _{\tau,k}^{\rm c})^{-1})$ is the Gaussian distribution with mean ${\bm{x}} _{\tau,k} ^{\rm c}$ and precision matrix ${\bm{D}} _{\tau,k}^{\rm c}$ representing the spatial distribution of nonuniform clutter component $\tau$.
Note that we use precision matrix rather than covariance matrix as this somewhat simplifies the mathematics.
By assuming that the nonuniform clutter component is caused by the volumetric scattering interference and the clutter point is dominated by the underlying scattering body, we interpret ${\bm{x}} _{\tau,k} ^{\rm c}$ and $({\bm{D}} _{\tau,k} ^{\rm c})^{-1}$ as the centroid and shape of each corresponding nonuniform clutter region, respectively.

We assume that the clutter and targets have the same power fluctuates and the PDF of strength likelihood function is also a Rayleigh distribution after thresholding,
\begin{equation}\label{equ:pdf after thresholding2}
  \mathcal{R}^d(m_{j,k};\sigma_{i,k} ^{\rm c},n) \!=\! \frac{2m_{j,k}^{2n-1}}{(\sigma_{i,k} ^{\rm c}+1)^n}\exp\!\! \left( -n\frac{m_{j,k}^2-d^2}{\sigma_{i,k} ^{\rm c}+1} \right).
\end{equation}

As noted in Eq.~\eqref{equ:finite mixture distribution}, the measurement likelihood is dependent on both clutter augmented states and clutter mixing weights; this coupling issue will make MTT difficult to solve.
In the following, the measurement likelihood is decoupled by introducing a data association event.

\subsection{Data Association}\label{subsec: Data Association}
Let $\bm{A}_k= \bm{A}_k^{\rm t} \cup \bm{A}_k^{\rm c}$ be the joint data association events at time $k$, where $\bm{A}_k^{\rm t}=\{a_{i,j,k}^{\rm t}\}_ {i=1} ^{N_T} \ \! _{j=0} ^{N_{M,k}}$ is the joint data association events between measurements and targets, $\bm{A}_k^{\rm c}=\{a_{\tau,j,k}^{\rm c}\}_ {\tau=0} ^{N_C} \ \! _{j=0} ^{N_{M,k}}$ is the joint data association events between measurements and clutter.
The binary association variable $a^{\rm t}_{i,j,k}$ denotes that the measurement $\bm{Y}_{j,k}$ is generated by target $i$ if $a^{\rm t}_{i,j,k} = 1$; likewise, the binary association variable $a^{\rm c}_{\tau,j,k}$ denotes that the measurement $\bm{Y}_{j,k}$ belongs to clutter component $\tau$ if $a^{\rm c}_{\tau,j,k} = 1$, given as
\begin{equation}\label{equ:data association model of target}
a^{\rm t}_{i,j,k} =
\begin{cases}
1, & \text{if $\bm{Y}_{j,k}$ is generated by target $i$,} \\
0, & \text{otherwise,}
\end{cases}
\end{equation}
\begin{equation}\label{equ:data association model of clutter}
a^{\rm c}_{\tau,j,k} =
\begin{cases}
1, & \text{if $\bm{Y}_{j,k}$ belongs to clutter component $\tau$,} \\
0, & \text{otherwise.}
\end{cases}
\end{equation}
In particular, $a^{\rm t}_{i,0,k}$ denotes that the target $i$ is missed and $a^{\rm c}_{\tau,0,k}$ represents that the clutter component $\tau$ disappears.
For the convenience of description, the following data association event sets are defined, ${\bm{A} _{j,k}} = \{ a_{i,j,k}^{\rm{t}}\} _{i = 1}^{N_T} \cup \{ a_{\tau,j,k}^{\rm{c}}\} _{\tau = 0}^{N_C} $, $\bm{A} _{i,k}^{\rm{t}} = \{ a_{i,j,k}^{\rm{t}}\} _{j = 0}^{N_{M,k}}$, and $\bm{A} _{\tau,k}^{\rm{c}} = \{ a_{\tau,j,k}^{\rm{c}}\} _{j = 0}^{N_{M,k}}$.
We define the joint data association sequence over time $1$ to time $K$ as $\bm{A} _{1:K} = \{\bm{A} _{k}\}_{k=1} ^{K}$.

A valid joint data association event $\bm{A}_k$ satisfies the following three constraints: (a) Each measurement is originated from at most one target or belongs to at most one clutter component, denoted as ${I_{j,k}}({\bm{A} _{j,k}})$;
(b) Each target can generate at most one measurement, denoted as $E_{i,k}^{\rm{t}}(\bm{A} _{i,k}^{\rm{t}})$; (c) Each clutter component either generates measurements or not, denoted as $E_{\tau,k}^{\rm{c}}(\bm{A} _{\tau,k}^{\rm{c}})$.
According to the above three constrains, the following constraint equations are obtained.
\begin{equation}\label{equ:C1}
{I_{j,k}}({\bm{A} _{j,k}}) =
\begin{cases}
1, & \text{if $ \sum_{{a_{i,j,k}} \in {\bm{A} _{j,k}}} {{a_{i,j,k}}}  = 1 $,} \\
0, & \text{otherwise.}
\end{cases}
\end{equation}
\begin{equation}\label{equ:C2}
  E_{i,k}^{\rm{t}}(\bm{A} _{i,k}^{\rm{t}})=
  \begin{cases}
  1, & \text{if $\sum_{a_{i,j,k}^{\rm{t}} \in \bm{A} _{i,k}^{\rm{t}}}{a_{i,j,k}^{\rm{t}}} = 1$,} \\
  0, & \text{otherwise.} \\
  \end{cases}
\end{equation}
\begin{equation}\label{equ:C3}
E_{\tau,k}^{\rm{c}}(\bm{A} _{\tau,k}^{\rm{c}}) =
\begin{cases}
0, & \begin{aligned}
       \text{if} & \text{ $\exists j>0$ such that} \\
       & \text{$a_{\tau,0,k}^{\rm{c}} = 1$ and $a_{\tau,j,k}^{\rm{c}} > 0$,}
     \end{aligned}
\\
1, & \text{otherwise.}
\end{cases}
\end{equation}
Define the following set of constraints.
\begin{equation}\label{equ:set of constraints1}
  \begin{aligned}
    & \bm{I}({\bm{A} _{1:K}}) = \prod _{k=1}^{K} {\bm{I}_{k}}({\bm{A} _{k}}) = \prod _{k=1}^{K} \prod _{j=1}^{N_{M,k}} {I_{j,k}}({\bm{A} _{j,k}}), \\
  \end{aligned}
\end{equation}
\begin{equation}\label{equ:set of constraints2}
  \begin{aligned}
    \bm{E}(\bm{A} _{1:K}) & = \prod _{k=1}^{K} {\bm{E}_{k}}({\bm{A} _{k}}) = \prod _{k=1}^{K} {\bm{E}_{k}^{\rm t}}({\bm{A} _{k}^{\rm t}}) {\bm{E}_{k}^{\rm c}}({\bm{A} _{k}^{\rm c}}) \\
    & = \prod _{k=1}^{K} \prod _{i=1}^{N_T} E_{i,k}^{\rm{t}}(\bm{A} _{i,k}^{\rm{t}}) \prod _{\tau=0}^{N_C} E_{\tau,k}^{\rm{c}}(\bm{A} _{\tau,k}^{\rm{c}}).
  \end{aligned}
\end{equation}

The joint prior probability of data association sequence $\bm{A}_{1:k}$ given the joint target visibility state sequence $\bm{S}_{1:k}$ is
\begin{equation}\label{pirAonS}
\begin{split}
  p(\bm{A}_{1:K}& |\bm{S}_{1:K}) = \prod_{k=1}^{K} p(\bm{A}_{k}^{\rm t}|\bm{S}_{k}) \\
 & = \prod_{k=1}^{K} \prod_{i=1}^{N_T} P_{\rm d} (s_{i,k}) ^{1 - a^{\rm t}_{i,0,k}} (1 - P_{\rm d}(s_{i,k}))^{a^{\rm t}_{i,0,k}},
\end{split}
\end{equation}
where $P_{\rm d} (s_{i,k})$ represents the detection probability of target $i$ given $s_{i,k}$, i.e., $P_{\rm d} (s_{i,k}=1)=\hat{p}_{{\rm D},i,k}^d$ and $P_{\rm d} (s_{i,k}=0)=\varepsilon$ ($0 < \varepsilon \ll 1$)~\cite{Lan2020}, where $\hat{p}_{{\rm D},i,k}^d$ is estimated detection probability of target $i$ calculated by Eq.~\eqref{equ:detection probability} with the estimated target mean SNR $\mathds{E}(\sigma_{i,k}^{\rm t})$ and detection threshold $d$.

The probability that the measurement $\bm{Y}_{j,k}$ belongs to the clutter component $\tau$ can be represented by the clutter mixing weights $\pi_{\tau,k}$, i.e., $p(a_{\tau,j,k} ^{\rm c}=1)= \pi_{\tau,k}$.
The conditional distribution of the data association event given the clutter mixing weights can be written as a multinomial distribution,
\begin{equation}\label{pirAonPi}
  p(\bm{A}_{1:K}|\bm{\Pi}_{1:K}) = \prod_{k=1}^{K} p(\bm{A}_{k}^{\rm c}|\bm{\Pi}_{k}) = \prod_{k=1}^{K} \prod_{\tau=0}^{N_C} \prod_{j=1}^{N_M} \pi_{\tau,k} ^{a_{\tau,j,k}^{\rm c}}.
\end{equation}
Similarly, the conditional PDF of the spatial measurement $\bm{y}_{j,k}$ given the clutter joint spatial state $\tilde{\bm{X}} _{k}^{\rm c}$ and the joint data association event $\bm{A}_{\tau,k}^{\rm c}$ can be represented by
\begin{equation}\label{equ:measurement likelihood decoupe}
p({\bm{y}}_{j,k}|\tilde{\bm{X}} _{k}^{\rm c},\bm{A}_{\tau,k}^{\rm c}) = \prod_{\tau = 0}^{N_C} p(\bm{y}_{j,k}| \tilde{\bm{x}} _{\tau,k} ^{\rm c}) ^{{a_{\tau,j,k}^{\rm c}}}.
\end{equation}
We have therefore found an equivalent formulation of the finite mixture model (as Eq.~\eqref{equ:finite mixture distribution}) involving a data association event, leading to significant simplifications of our MP method that will be presented in Section~\ref{sec:MP-Based Approach}.
We can represent the conditional distribution of the measurement sequence $\bm{Y}_{1:k}$ given the target joint augmented state sequence $\bm{X}_{1:K}^{\rm{t}}$, clutter joint augmented state sequence $\bm{X}_{1:K}^{\rm{c}}$, and data association sequence $\bm{A}_{1:K}$ as
\begin{equation}\label{equ:measurement}
\begin{split}
& p(\bm{Y}_{1:K}|\bm{X}_{1:K}^{\rm{t}}, \bm{X}_{1:K}^{\rm{c}}, \bm{A}_{1:K}) = \prod_{k = 1}^{K} p(\bm{Y}_{k}|\bm{X}_{k}^{\rm{t}}, \bm{X}_{k}^{\rm{c}}, \bm{A}_{k}) \\
& = \prod_{k = 1}^{K} \prod_{j = 0}^{N_{M,k}} \prod_{\tau = 0}^{N_C} p(\bm{y}_{j,k}|\tilde{\bm{x}} _{\tau,k} ^{\rm c}) ^{{a_{\tau,j,k}^{\rm c}}} \prod_{i = 1}^{N_T} p(\bm{y}_{j,k}|\bm{x} _{i,k} ^{\rm t}) ^{a ^{\rm t} _{i,j,k}}.
\end{split}
\end{equation}
Note that, by introducing the data association $\bm{A}_{k}$, $\bm{X}_k^{\rm t}$ is conditionally independent of $\bm{E}_k$ and $\bm{X}_k^{\rm c}$ is conditionally independent of $\bm{\Pi}_k$, and we obtain a new measurement likelihood function as in Eq.~\eqref{equ:measurement}.
Furthermore, the new measurement likelihood function can simplify the derivation of message passing rules~(The details will be given in Section \ref{sec:MP-Based Approach}).

\subsection{The Joint PDF and Problem Statement}\label{subsec:Problem Statement}
Let ${\bm{\Theta}}_{1 : K} = \{\bm{X}_{1:K}^{\rm{t}}, \bm{X}_{1:K}^{\rm{c}}, \bm{S}_{1:K}, \bm{\Pi}_{1:K}, \bm{A}_{1:K}\}$ denote the collection of all the latent variables.
The joint posterior PDF $\mathcal{L}(\bm{\Theta}_{1:K})$ can be factorized as
\begin{equation}\label{equ:joint PDF1}
\begin{split}
\mathcal{L}& ( \bm{\Theta}_{1:K}) = \frac{p(\bm{X}_{1:K}^{\rm{t}}, \bm{X}_{1:K}^{\rm{c}}, \bm{S}_{1:K}, \bm{\Pi}_{1:K}, \bm{A}_{1:K}, \bm{Y}_{1:K})}{p(\bm{Y}_{1:K})} \\
& \propto p(\bm{Y}_{1:K}|\bm{X}_{1:K}^{\rm{t}}, \bm{X}_{1:K}^{\rm{c}}, \bm{A}_{1:K}) \\
& \ \times p(\bm{X}_{1:K}^{\rm{t}}) p(\bm{S}_{1:K}) p(\bm{X}_{1:K}^{\rm{c}}) p(\bm{\Pi}_{1:K}) \\
& \ \times p(\bm{A}_{1:K}|\bm{S}_{1:K}) p(\bm{A}_{1:K}|\bm{\Pi}_{1:K}) \bm{E}(\bm{A}_{1:K})  \bm{I}(\bm{A}_{1:K}).
\end{split}
\end{equation}
Insert Eq.~\eqref{equ:target transition} for $p(\bm{X}_{1:K}^{\rm{t}})$, Eq.~\eqref{equ:target visibility} for $p(\bm{S}_{1:K})$, Eq.~\eqref{equ:clutter transition X} for $p(\bm{X}_{1:K}^{\rm{c}})$, Eq.~\eqref{equ:clutter transition P} for $p(\bm{\Pi}_{1:K})$, Eq.~\eqref{equ:set of constraints1} for $\bm{I}(\bm{A}_{1:K})$, Eq.~\eqref{equ:set of constraints2} for $\bm{E}(\bm{A}_{1:K})$, Eq.~\eqref{pirAonS} for $p(\bm{A}_{1:K}|\bm{S}_{1:K})$, Eq.~\eqref{pirAonPi} for $p(\bm{A}_{1:K}|\bm{\Pi}_{1:K})$, and Eq.~\eqref{equ:measurement} for $p(\bm{Y}_{1:K}|\bm{X}_{1:K}^{\rm{t}}, \bm{X}_{1:K}^{\rm{c}}, \bm{A}_{1:K})$, yielding the factorization of Eq.~\eqref{equ:joint PDF1} as Eq.~\eqref{equ:joint PDF}.
\begin{figure}[!htbp]
\centering
\begin{equation}\label{equ:joint PDF}
\begin{split}
& \mathcal{L}(\bm{\Theta}_{1:K}) \\
\propto & {\prod_{k = 1}^{K} \prod_{j = 0}^{N_{M,k}} \prod_{\tau = 0}^{N_C} p(\bm{y}_{j,k}|\tilde{\bm{x}} _{\tau,k} ^{\rm c}) ^{{a_{\tau,j,k}^{\rm c}}} \prod_{i = 1}^{N_T} p(\bm{y}_{j,k}|\bm{x} _{i,k} ^{\rm t}) ^{a ^{\rm t} _{i,j,k}}} \\
\times & {\prod_{i = 1}^{N_T} p(\bm{x}_{i,1}^{\rm {t}}) p(\bm{s}_{i,1}) \prod_{k = 2}^{K} p(\bm{x}_{i,k}^{\rm {t}}|\bm{x}_{i,k-1}^{\rm {t}}) p(\bm{s}_{i,k}|\bm{s}_{i,k - 1})} \\
\times & \! {\prod_{\tau = 0}^{N_C} p(\bm{x}_{\tau,1}^{\rm {c}}) p({\pi}_{\tau,1}) \prod_{k = 2}^{K} \! p(\bm{x}_{\tau,k}^{\rm {c}}|\bm{x}_{\tau,k-1}^{\rm {c}}) p({\pi}_{\tau,k}|{\pi}_{\tau,k - 1})} \\
\times & {\prod_{k = 1}^{K}\bm{E}_k(\bm{A}_k) \bm{I}_k (\bm{A}_k)}
    {\prod_{k = 1}^{K} p(\bm{A}_k |\bm{S}_k) p(\bm{A}_k |\bm{\Pi}_k)}.
\end{split}	
\end{equation}
\end{figure}

The aim of RMTT is to simultaneously estimate $\bm{X}_{1:K}^{\rm{t}}$ (target augmented state estimation), $\bm{S}_{1:K}$ (target detection), $\bm{X}_{1:K}^{\rm{c}}$ (clutter augmented state estimation) and  $\bm{\Pi}_{1:K}$ (clutter mixing weights estimation), given measurements $\bm{Y}_{1:K}$ with unknown $\bm{A}_{1:K}$.
The posterior PDFs of $\bm{X}_{1:K}^{\rm{t}}$, $\bm{S}_{1:K}$, $\bm{X}_{1:K}^{\rm{c}}$ and $\bm{\Pi}_{1:K}$ can be obtained by marginalizing $\mathcal{L} (\bm{\Theta} _{1:K})$ in the Bayesian framework.
Unfortunately, the marginalizing of data association is exponentially complex, making exact solution computationally prohibitive.
Therefore, the combined BP-MF approximation is adopted in the next section.

\section{MP-Based Approach for RMTT}\label{sec:MP-Based Approach}
In this section, we first present the framework for solving the RMTT problem using the combined BP-MF MP algorithm.
Then, the message update rules and the approximate beliefs of all hidden variables, including target joint augmented states, clutter joint augmented states, target visibility states, clutter mixing weights and data association, are derived.
Finally, we present the initialization, implementation and computational complexity for the proposed RMTT algorithm.

\subsection{Combined BP-MF MP Approach for RMTT}\label{subsec:Combined BP-MF MP approach for RMTT}
The combined BP-MF MP approach performs inference on a probabilistic graphical model~(typically, a factor graph) by exchanging messages.
The first step to apply the combined BP-MF approach is to design the factor graph corresponding to the joint PDF of the problem to be solved, for instance, Eq.~\eqref{equ:joint PDF}.
This is accomplished by associating each random variable or vector in the joint PDF to a variable node in the factor graph and associating each functions defined on random variables~(including PDFs, conditional PDFs, constrains etc.) to a factor node in the factor graph, and connect a variable node with a factor node if the latter is a function of the former.
The factor graph is then split into an MF region and a BP region to maximize their advantages and circumvent their disadvantages.
Specifically, BP can deal effectively with hard constraints and has good approximations to marginal PDFs, but is not suitable for situations containing both continuous and discrete hidden variables.
MF approximation can guarantee convergence and is straightforward to derive for conjugate-exponential models, but is incompatible with hard constraint.
The message update rules and the beliefs of each hidden variables are eventually derived by solving the constrained minimum region-based free energy problem on the factor graph via the Lagrangian relaxation method.
For the detailed derivation of the combined BP-MF MP approach, the reader is referred to~\cite{Riegler2013}.
Following the ideas of the combined MP approach, next we present our proposed MP-based method for solving the RMTT problem.

\begin{figure}[!htbp]
\centering
\includegraphics[scale = 0.55]{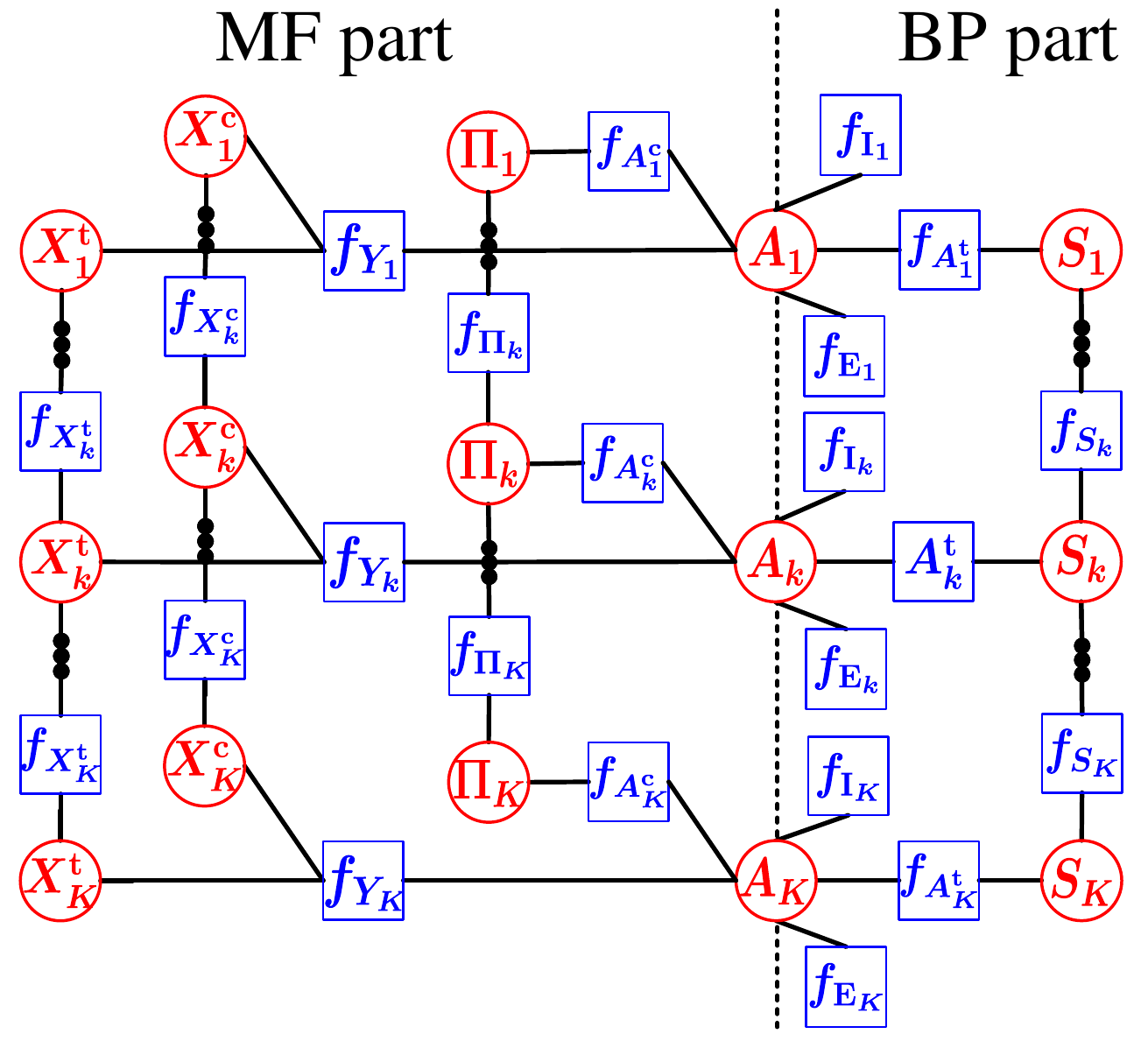}
\caption{Factor graph represented by Eq.~\eqref{equ:joint PDF}.}
\label{fig:1-FG}
\end{figure}

By observing Eq.~\eqref{equ:joint PDF}, we define the factor nodes $f_{\bm{Y}_k} \triangleq p({\bm{Y}_k | \bm{X}_k ^{\rm{t}}, \bm{X} _k ^{\rm{c}}, \bm{A} _k})$, $f_{\bm{X}_k^{\text{t}}} \triangleq \prod_{i=1}^{N_T} p(\bm{X}_{i,k}^{\rm {t}}|\bm{X}_{i,k-1}^{\rm {t}})$, $f_{\bm{X}_k^{\text{c}}} \triangleq \prod_{\tau=1}^{N_C} p(\bm{X}_{\tau,k}^{\rm {c}}|\bm{X}_{\tau,k-1}^{\rm {c}})$, $f_{\bm{S}_k} \triangleq \prod_{i=1}^{N_T} p(s_{i,k}|s_{i,k-1})$, $f_{\bm{\Pi}_k} \triangleq \prod_{\tau=1}^{N_C} p(\pi_{\tau,k}|\pi_{\tau,k-1})$, $f_{\bm{A}_{k}^{\rm{t}}} \triangleq p ({\bm{A} _{k} ^{\rm{t}}| \bm{S} _{k}})$, $f_{\bm{A}_{k}^{\rm{c}}} \triangleq p ({\bm{A} _{k} ^{\rm{c}} | \bm{\Pi}_{k}})$,
$f_{\bm{E}_{k}} \triangleq {\bm {E} _{k}}(\bm{A} _{k})$, $f_{\bm{I}_{k}} \triangleq {\bm {I} _{k}}(\bm{A} _{k})$,
the set of variable nodes $\mathcal{I} \triangleq \{ \bm{X}_k^{\text{t}}, \bm{X}_k^{\text{c}}, \bm{S}_k, \bm{\Pi}_k, \bm{A}_k \}_{k=1}^K$,
and the set of factor nodes
$\mathcal{F} \triangleq \{ f_{\bm{Y}_k}, f_{\bm{X}_k^{\text{t}}}, f_{\bm{X}_k^{\text{c}}}, f_{\bm{S}_k}, f_{\bm{\Pi}_k}, f_{\bm{A}_{k}^{\rm{t}}}, f_{\bm{A}_{k}^{\rm{c}}}, f_{\bm{E}_{k}}, f_{\bm{I}_{k}}\}_{k=1}^K$.
The corresponding factor graph is illustrated in Fig.~\ref{fig:1-FG}.

We next describe how the factor graph in Fig.~\ref{fig:1-FG} is divided into a BP part and an MF part.
BP is used to estimate the target visibility state in order to obtain a good approximation for the corresponding posterior PDFs.
BP is also used for data association, which contains hard constraints as in Eq.~\eqref{equ:C1}-Eq.~\eqref{equ:C3}.
The MF approximation is used for the estimation of target joint augmented state, clutter joint augmented state and clutter mixture weight, as these hidden variables are conjugate-exponential models and therefore simple MP rules can be obtained.
Accordingly, we split $\mathcal{F}$ into $\mathcal{F}_{\mathrm{BP}}$ and $\mathcal{F}_{\mathrm{MF}}$ with
\begin{equation}
\begin{split}
\mathcal{F}_{\text{BP}} = & \big\{f_{\bm{S}_k}, f_{\bm{A}_{k}^{\rm{t}}}, f_{\bm{E}_{k}}, f_{\bm{I}_{k}}\big\}_{k=1}^K, \\
\mathcal{F}_{\text{MF}} = & \big\{f_{\bm{Y}_k}, f_{\bm{X}_k^{\text{t}}}, f_{\bm{X}_k^{\text{c}}}, f_{\bm{\Pi}_k}, f_{\bm{A}_{k}^{\rm{c}}}\big\}_{k=1}^K.
\end{split}
\end{equation}
We have
\begin{equation}
\begin{split}
\mathcal{I}_{\text{BP}} = \big\{ \bm{S}_k, \bm{A}_k \big\}_{k=1}^K,\ \mathcal{I}_{\text{MF}} =  \big\{\bm{X}_k^{\text{t}}, \bm{X}_k^{\text{c}}, \bm{\Pi}_k, \bm{A}_k \big\}_{k=1}^K.
\end{split}
\end{equation}
Then, the joint posterior PDF $\mathcal{L}(\Theta_{1:K})$ can be written as
\begin{equation}\label{RFG}
\mathcal{L}(\Theta_{1:K}) = \overbrace{\prod_{k=1}^K f_{\bm{Y}_k} f_{\bm{X}_k^{\text{t}}} f_{\bm{X}_k^{\text{c}}} f_{\bm{\Pi}_k} f_{\bm{A}_{k}^{\rm{c}}}}^{\text{MF region}} \times \overbrace{\prod_{k = 1}^K f_{\bm{S}_k} f_{\bm{A}_{k}^{\rm{t}}} f_{\bm{E}_{k}} f_{\bm{I}_{k}}}^{\text{BP region}}.
\end{equation}

The beliefs of hidden variables are approximated by~\cite{Riegler2013}
\begin{equation}\label{equ:approximated1}
b_{\bm{X}}(\bm{X}_{1:K}^{\text{t}}) \propto \prod \limits_{\alpha \in \mathcal{S}_{\text{MF}}(\bm{X}_{1:K}^{\text{t}})} m_{{\alpha} \rightarrow \bm{X}_{1:K}^{\text{t}}}^{\text{MF}}(\bm{X}_{1:K}^{\text{t}}),
\end{equation}
\begin{equation}\label{equ:approximated2}
b_{\bm{X}}(\bm{X}_{1:K}^{\text{c}}) \propto \prod \limits_{\alpha \in \mathcal{S}_{\text{MF}}(\bm{X}_{1:K}^{\text{c}})} m_{{\alpha} \rightarrow \bm{X}_{1:K}^{\text{c}}}^{\text{MF}}(\bm{X}_{1:K}^{\text{c}}),
\end{equation}
\begin{equation}\label{equ:approximated3}
b_{\bm{S}}(\bm{S}_{1:K}) \propto \prod \limits_{\alpha \in \mathcal{S}_{\text{BP}}(\bm{S}_{1:K})} m_{{\alpha} \rightarrow \bm{S}_{1:K}}^{\text{BP}}(\bm{S}_{1:K}),
\end{equation}
\begin{equation}\label{equ:approximated4}
b_{\bm{\Pi}}(\bm{\Pi}_{1:K}) \propto \prod \limits_{\alpha \in \mathcal{S}_{\text{MF}}(\bm{\Pi}_{1:K})} m_{{\alpha} \rightarrow \bm{\Pi}_{1:K}}^{\text{MF}}(\bm{\Pi}_{1:K}),
\end{equation}
\begin{equation}\label{equ:approximated5}
b_{\bm{A}}(\bm{A}_{1:K}) \propto \!\!\!\!\!\!\!\!\!\!\!\! \prod \limits_{{\alpha} \in \mathcal{S}_{\text{BP}}(\bm{A}_{1:K})} \!\!\!\!\!\!\!\!\!\!\!\! m_{{\alpha} \rightarrow \bm{A}_{1:K}}^{\text{BP}}(\bm{A}_{1:K}) \!\!\!\!\!\!\!\!\!\!\! \prod \limits_{{\alpha} \in \mathcal{S}_{\text{MF}}(\bm{A}_{1:K})} \!\!\!\!\!\!\!\!\!\!\! m_{{\alpha} \rightarrow \bm{A}_{1:K}}^{\text{MF}}(\bm{A}_{1:K}).
\end{equation}

Next, we present the detailed derivations of the beliefs for each hidden variables and the messages in Eqs.~\eqref{equ:approximated1}-Eqs.~\eqref{equ:approximated5}.
Before we present the details, we introduce the following proposition~\cite{Riegler2013, Lan2020}, which is required in the derivations of the beliefs.
\begin{proposition}
For all messages passed from variable nodes $i$ to factor nodes $\alpha$ in the MF region, $n_{i \rightarrow \alpha} =b _i(\bm{x}_i)$.
\label{cor-1}
\end{proposition}

\subsection{Derivations of Belief $b_{\bm{X}}(\bm{X}_{1:K}^{\rm t})$}\label{subsec:Derivations of target state Belief}
\begin{figure}[!htbp]
\centering
\includegraphics[scale = 0.45]{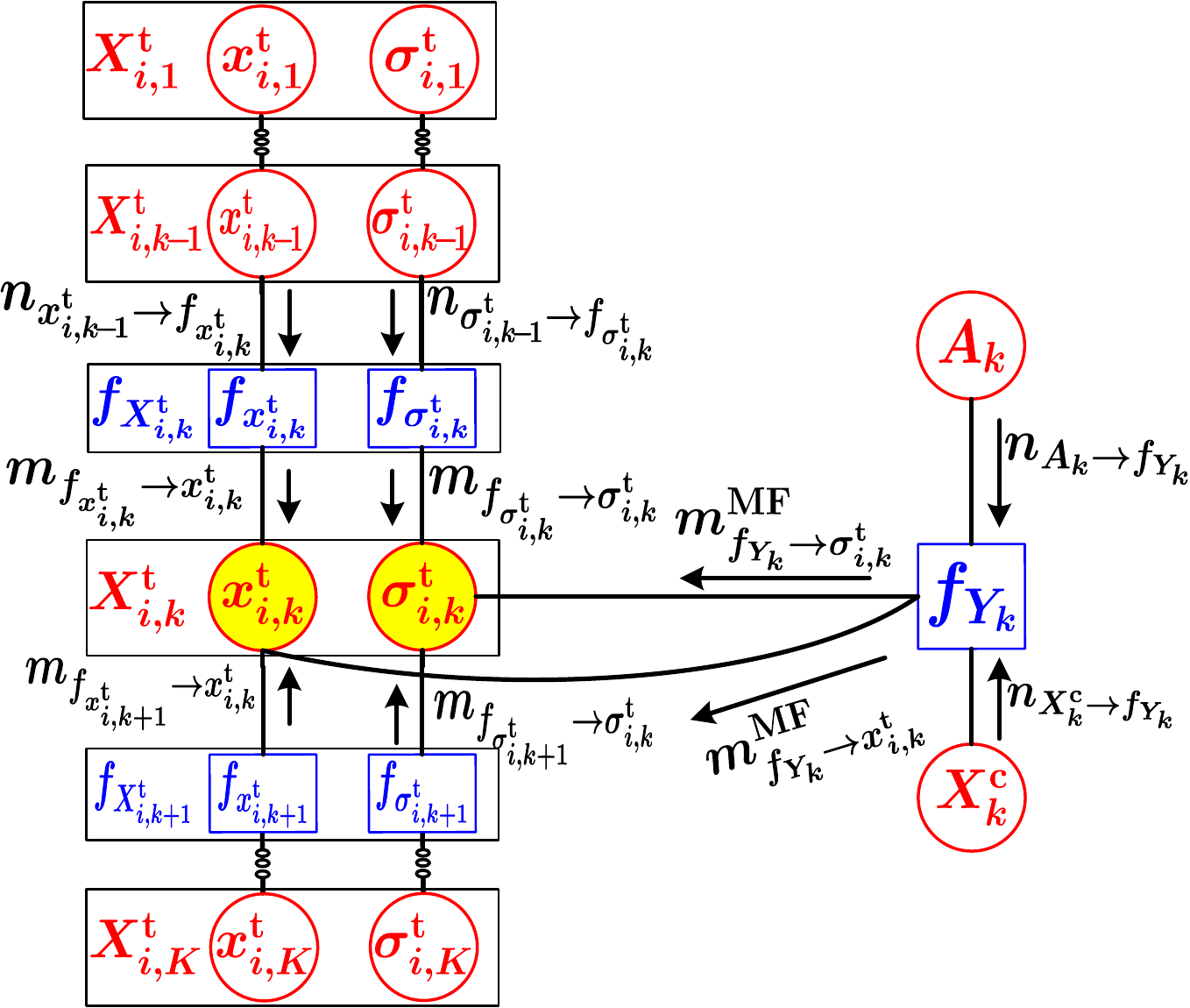}
\caption{The subgraph for target state estimation.}
\label{fig:Subgraph-x}
\end{figure}

Fig.~\ref{fig:Subgraph-x} shows the target tracking and mean SNR estimation subgraph corresponding to the belief $b_{\bm{x}}({\bm{x}}_{1:K} ^{\rm t})$ and $b_{\bm{\sigma}}(\bm{\sigma} _{i,1:K} ^{\rm t})$, $i=1,\ldots,N_T$, which contains the variable nodes ${\bm{x}} ^{\rm t} _{i,k}$ and $\sigma ^{\rm t}_{i,k}$, $k=1,\ldots,K$.

\subsubsection{The belief of target kinematic state} The belief $b_{{\bm{x}}}({\bm{x}} ^{\rm t} _{i,k})$ can be derived as
\begin{equation}\label{eq:bKSxX}
b_{{\bm{x}}} ({\bm{x}} ^{\rm t} _{i,k} ) \propto
    \underbrace{ m^{\text{MF}}_{f_{{\bm{x}} ^{\rm t} _{i,k}  } \rightarrow {\bm{x}} ^{\rm t} _{i,k}} m^{\text{MF}}_{f_{\bm{Y}_k} \rightarrow {\bm{x}} ^ {\rm t} _{i,k}}}_{\overrightarrow{b}_{{\bm{x}}} ({\bm{x}} ^{\rm t} _{i,k})} \times \underbrace{m^{\text{MF}}_{f_{{\bm{x}} ^{\rm t} _{i,k+1}  } \rightarrow {\bm{x}} ^{\rm t} _{i,k}}}_ {\overleftarrow{b}_{{\bm{x}}} ({\bm{x}} ^{\rm t} _{i,k} )}.
\end{equation}

Initialize $\overrightarrow{b}_{{\bm{x}}} ({\bm{x}} ^{\rm t} _{i,1}) = \mathcal{N}({\bm{x}} ^{\rm t} _{i,1}; \hat{\bm{x}} ^{\rm t} _{i,1|1}, {\bm{P}} ^{\rm t} _{i,1|1})$ as a Gaussian distribution at time 1, the forward messages $\overrightarrow{b}_{{\bm{x}}} ({\bm{x}} ^{\rm t} _{i,k})$ then can be calculated from time $2$ to time $K$.
The factor-to-variable messages corresponding to $\overrightarrow{b} _{{\bm{x}}} ({\bm{x}} ^{\rm t} _{i,k})$ include
\begin{equation}\label{equ:F2VM}
\begin{split}
   & m^{\text{MF}}_{f_{{\bm{x}} ^{\rm t} _{i,k}} \rightarrow {\bm{x}} ^{\rm t} _{i,k}} \\
  =  & \exp\big(\int n_{{\bm{x}} ^{\rm t} _{i,k-1}  \rightarrow f_{{\bm{x}} ^{\rm t} _{i,k}}}
       \ln p\left({\bm{x}} ^{\rm t} _{i,k} |{\bm{x}} ^{\rm t} _{i,k-1} \right) d{{\bm{x}} ^{\rm t} _{i,k-1} } \big),
\end{split}
\end{equation}
\begin{equation}\label{equ:F2VM2XT}
\begin{split}
    m^{\text{MF}}_{f_{\bm{Y}_k} \rightarrow {\bm{x}} ^{\rm t} _{i,k} } \!\!\! = \exp \big( \sum_{j = 1}^{N_{M,k}} n_{\bm{A}_k \rightarrow f_{\bm{Y}_k}}
     \ln p( {\bm{y}}_{j,k}|{\bm{x}} ^{\rm t} _{i,k} , \bm{A}_k)\big).
\end{split}
\end{equation}

According to Eq.~\eqref{eq:bKSxX}, we have $n_{{\bm{x}} ^{\rm t} _{i,k-1} \rightarrow f_{{\bm{x}} ^{\rm t} _{i,k}}} = \overrightarrow{b} _{{\bm{x}}} ({\bm{x}} ^{\rm t} _{i,k-1})$ since the backward message $\overleftarrow{b} _{{\bm{x}}} ({\bm{x}} ^{\rm t} _{i,k-1})$ is not available yet, and $n_{\bm{A}_k \rightarrow f_{\bm{Y}_k}} = b_{\bm{A}} (a_{i,j,k}^{\rm t})$.
Note that the belief $\overrightarrow{b} _{{\bm{x}}} ({\bm{x}} ^{\rm t} _{i,k-1}) = \mathcal{N}({\bm{x}} ^{\rm t} _{i,k-1}; \hat{\bm{x}} ^{\rm t} _{i,k-1|k-1}, {\bm{P}} ^{\rm t} _{i,k-1|k-1})$ is also a Gaussian distribution.
Then, Eq.~\eqref{equ:F2VM} and Eq.~\eqref{equ:F2VM2XT} can be derived as
\begin{equation}\label{equ:F2VM1}
\begin{split}
  m^{\text{MF}}_{f_{{\bm{x}} ^{\rm t} _{i,k}} \rightarrow {\bm{x}} ^{\rm t} _{i,k}}
  &  \propto \mathcal{N}({\bm{x}} ^{\rm t} _{i,k} ; \hat{\bm{x}} ^{\rm t} _{i,k|k-1}, {\bm{P}} ^{\rm t} _{i,k|k-1}),
\end{split}
\end{equation}
\begin{equation}\label{equ:F2VMXT21}
\begin{split}
m^{\text{MF}}_{f_{\bm{Y}_k} \rightarrow {\bm{x}} ^{\rm t} _{i,k} }
 = & \prod_{j = 1}^{N_{M,k}} \mathcal{N} ({\bm{y}}_{j,k} ; h({\bm{x}} ^{\rm t} _{i,k}), \bm{R}_k) ^ {\hat{a}_{i,j,k}^{\rm t}},
\end{split}
\end{equation}
where $\hat{\bm{x}} ^{\rm t} _{i,k|k-1}$ and ${\bm{P}} ^{\rm t} _{i,k|k-1}$ are calculated accorded to the transition PDF $p({\bm{x}} ^{\rm t} _{i,k+1} |{\bm{x}} ^{\rm t} _{i,k})$; $\hat{a}_{i,j,k}^{\rm t}$ is the expectation of ${a}_{i,j,k}^{\rm t}$ taken over $b_a({a}_{i,j,k}^{\rm t})$.
Eq.~\eqref{equ:F2VMXT21} can be derived as $m^{\text{MF}}_{f_{\bm{Y}_k} \rightarrow {\bm{x}} ^{\rm t} _{i,k} } = \mathcal{N} (\bar{\bm{y}}_{i,k} ; h({\bm{x}} ^{\rm t} _{i,k}), \bar{\bm{R}}_{i,k})$ with~\cite{Lan2020}
\begin{equation}\label{equ:synthetic}
\bar{\bm{y}}_{i,k}=\frac{\sum_{j = 1}^{N_{M,k}} \hat{a}_{i,j,k}^{\rm t} {\bm{y}}_{j,k}}{1- \hat{a}_{i,0,k}^{\rm t}},\quad \bar{\bm{R}}_{i,k}= \frac{\bm{R}_k}{1- \hat{a}_{i,0,k}^{\rm t}}.
\end{equation}
Substitute Eq.~\eqref{equ:F2VM1} and Eq.~\eqref{equ:F2VMXT21} into Eq.~\eqref{eq:bKSxX}, yielding
\begin{equation}\label{eq:bKSx}
\begin{split}
  \overrightarrow{b}& _{{\bm{x}}} ({\bm{x}} ^{\rm t} _{i,k} ) = \mathcal{N}({\bm{x}} ^{\rm t} _{i,k} ; \hat{\bm{x}} ^{\rm t} _{i,k|k}, {\bm{P}} ^{\rm t} _{i,k|k}),
\end{split}
\end{equation}
where $\hat{\bm{x}} ^{\rm t} _{i,k|k}$ and ${\bm{P}} ^{\rm t} _{i,k|k}$ are calculated by the Unscented Kalman Filter (UKF)~\cite{sarkka2013}.

Next, the backward message ${\overleftarrow{b}_{{\bm{x}}} ({\bm{x}} ^{\rm t} _{i,k})}$ is calculated from time $K$ to time $1$.
Again, we need initial message ${\overleftarrow{b}_{{\bm{x}}} ({\bm{x}} ^{\rm t} _{i,K})}$.
We take ${\overleftarrow{b}_{{\bm{x}}} ({\bm{x}} ^{\rm t} _{i,K})}=1$ for all ${\bm{x}} ^{\rm t} _{i,K}$.
Then, ${\overleftarrow{b}_{{\bm{x}}} ({\bm{x}} ^{\rm t} _{i,k})}$ from time $K-1$ to time $1$ can be derived as
\begin{equation}\label{equ:Back X c}
\begin{split}
& {\overleftarrow{b}_{{\bm{x}}} ({\bm{x}} ^{\rm t} _{i,k})} = m^{\text{MF}}_{f_{{\bm{x}} ^{\rm t} _{i,k+1}  } \rightarrow {\bm{x}} ^{\rm t} _{i,k}} \\
= & \exp\big(\int n_{{\bm{x}} ^{\rm t} _{i,k+1}  \rightarrow f_{{\bm{x}} ^{\rm t} _{i,k}}} \ln p\left({\bm{x}} ^{\rm t} _{i,k+1} |{\bm{x}} ^{\rm t} _{i,k} \right) d{{\bm{x}} ^{\rm t} _{i,k+1} } \big).
\end{split}
\end{equation}

We multiply ${\overleftarrow{b}_{{\bm{x}}} ({\bm{x}} ^{\rm t} _{i,k})}$ by $\overrightarrow{b}_{{\bm{x}}} ({\bm{x}} ^{\rm t} _{i,k})$ and use a nonlinear fixed-interval smoother, deriving that  ${b}_{{\bm{x}}} ({\bm{x}} ^{\rm t} _{i,k}) \propto \overrightarrow{b}_{{\bm{x}}} ({\bm{x}} ^{\rm t} _{i,k}) \overleftarrow{b}_{{\bm{x}}} ({\bm{x}} ^{\rm t} _{i,k}) \propto \mathcal{N}({\bm{x}} ^{\rm t} _{i,k} ; \hat{\bm{x}} ^{\rm t} _{i,k|K}, {\bm{P}} ^{\rm t} _{i,k|K})$ with $\hat{\bm{x}} ^{\rm t} _{i,k|K}$ and ${\bm{P}} ^{\rm t} _{i,k|K}$ being calculated by the Unscented Rauch-Tung-Striebel Smoother (URTSS)~\cite{sarkka2013}.

\subsubsection{The belief of target mean SNR}
The belief $b_{\sigma}(\sigma ^{\rm t} _{i,k} )$ can be calculated as
\begin{equation}\label{equ:bKSsigmaS}
b_{\sigma} (\sigma ^{\rm t} _{i,k} ) \propto
    \underbrace{m^{\text{MF}}_{f_{\sigma ^{\rm t} _{i,k}} \rightarrow \sigma ^{\rm t} _{i,k}} m^{\text{MF}}_{f_{\bm{Y}_k} \rightarrow \sigma ^{\rm t} _{i,k} }}_{\overrightarrow{b}_{\sigma} (\sigma ^{\rm t} _{i,k})} \times \underbrace{m^{\text{MF}}_{f_{\sigma ^{\rm t} _{i,k+1}} \rightarrow \sigma ^{\rm t} _{i,k}}}_{\overleftarrow{b}_{\sigma} (\sigma ^{\rm t} _{i,k} )}.
\end{equation}

Initializing $\overrightarrow{b}_{\sigma} (\sigma ^{\rm t} _{i,1}) = \mathcal{I}(\sigma ^{\rm t} _{i,1} ;\alpha ^{\rm t} _{i,1|1} ,\beta ^{\rm t} _{i,1|1} )$ as an IG distribution at time 1, the forward messages $\overrightarrow{b}_{\sigma} (\sigma ^{\rm t} _{i,k} )$ then can be calculated from time $2$ to time $K$.
The factor-to-variable messages corresponding to $\overrightarrow{b}_{\sigma} (\sigma ^{\rm t} _{i,k})$ include
\begin{equation}\label{equ:F2VM_sigma}
\begin{split}
    & m^{\text{MF}}_{f_{\sigma ^{\rm t} _{i,k}} \rightarrow \sigma ^{\rm t} _{i,k}} \\
  = & \exp\big(\int  n_{\sigma ^{\rm t} _{i,k-1}  \rightarrow f_{\sigma ^{\rm t} _{i,k} }}
\ln p(\sigma ^{\rm t} _{i,k} |\sigma ^{\rm t} _{i,k-1} ) d{\sigma ^{\rm t} _{i,k-1} } \big),
\end{split}
\end{equation}
\begin{equation}\label{equ:F2VM211S}
m^{\text{MF}}_{f_{\bm{Y}_k} \rightarrow \sigma ^{\rm t} _{i,k} } \!\!= \! \exp \Big( \sum_{j = 1}^{N_{M,k}} \! n_{\bm{A}_{k} \rightarrow f_{\bm{Y}_k}}
\ln p(m_{j,k} | \sigma ^{\rm t} _{i,k} , \bm{A}_{k})\Big).
\end{equation}

We have $n_{\sigma ^{\rm t} _{i,k-1}  \rightarrow f_{\sigma ^{\rm t} _{i,k}}} = \overrightarrow{b}_{\sigma} (\sigma ^{\rm t} _{i,k-1} )$ according to Proposition~\ref{cor-1}, $\overrightarrow{b}_{\sigma} (\sigma ^{\rm t} _{i,k-1} ) = \mathcal{I}(\sigma ^{\rm t} _{i,k-1} ;\alpha ^{\rm t} _{i,k-1|k-1} ,\beta ^{\rm t} _{i,k-1|k-1} )$ is an IG distribution, and $n_{\bm{A}_k \rightarrow f_{\bm{Y}_k}} = b_{\bm{A}} (a_{i,j,k}^{\rm t})$.
According to $p(\sigma ^{\rm t} _{i,k} |\sigma ^{\rm t} _{i,k-1} )$, we get $m^{\text{MF}}_{f_{{\sigma ^{\rm t} _{i,k}} } \rightarrow \sigma ^{\rm t} _{i,k}} \propto \mathcal{I} (\sigma ^{\rm t} _{i,k} ; \alpha ^{\rm t} _{i,k|k-1} , \beta ^{\rm t} _{i,k|k-1} )$, where $\beta ^{\rm t} _{i,k|k-1} ={\beta ^{\rm t} _{i,k-1|k-1}}/ {u^{\rm t}}$ and $\alpha ^{\rm t} _{i,k} = ({{\alpha ^{\rm t} _{i,k-1|k-1} +u^{\rm t}-1}})/ {u^{\rm t}}$.
Eq.~\eqref{equ:F2VM211S} is derived as
\begin{equation}\label{equ:F2VM21}
\begin{split}
m^{\text{MF}}_{f_{\bm{Y}_k} \rightarrow \sigma ^{\rm t} _{i,k} }
 = & \prod_{j = 1}^{N_{M,k}} \mathcal{R}^d (m_{j,k};\sigma ^{\rm t} _{i,k}, n) ^ {\hat{a}_{i,j,k}^{\rm t}},
\end{split}
\end{equation}
with $\hat{a}_{i,j,k}^{\rm t}$ being the expectation of ${a}_{i,j,k}^{\rm t}$ taken over belief $b_{\bm{A}} (a_{i,j,k}^{\rm t})$.
Then the belief $\overrightarrow{b}_{\sigma}(\sigma ^{\rm t} _{i,k} )$ can be derived as
\begin{equation}\label{equ:bKSsigma1}
\begin{split}
\overrightarrow{b}_{\sigma} (\sigma ^{\rm t} _{i,k} ) &  \propto \
  \mathcal{I} (\sigma ^{\rm t} _{i,k} ; \alpha ^{\rm t} _{i,k|k-1} , \beta ^{\rm t} _{i,k|k-1} ) \\
  & \quad \times \prod_{j = 1}^{N_{M,k}} \mathcal{R}^d (m_{j,k};\sigma ^{\rm t} _{i,k}, n) ^ {\hat{a}_{i,j,k}^{\rm t}} \\
  & = \prod_{j = 1}^{N_{M,k}} \mathcal{I} (\sigma ^{\rm t} _{i,k}; \alpha ^{\rm t} _{i,j,k|k}, \beta ^{\rm t} _{i,j,k|k}) ^ {\hat{a}_{i,j,k}^{\rm t}},
\end{split}
\end{equation}
where $\beta ^{\rm t} _{i,j,k|k} = \beta ^{\rm t} _{i,k|k-1} + n {m_{j,k}^2} - n d ^ 2$ and  $\alpha ^{\rm t} _{i,j,k|k} = \alpha ^{\rm t} _{i,k|k-1} + n$.
By applying the product of the IG distributions, Eq.~(\ref{equ:bKSsigma1}) is derived as $\overrightarrow{b}_{\sigma} (\sigma ^{\rm t} _{i,k} ) = \mathcal{I} (\sigma ^{\rm t} _{i,k}; \alpha ^{\rm t} _{i,k|k}, \beta ^{\rm t} _{i,k|k})$ with
\begin{equation}\label{equ:posterior IG parameters4}
\begin{split}
\beta ^{\rm t} _{i,k|k}   = \frac{\sum_{j = 1}^{N_{M,k}} {\hat{a}_{i,j,k}^{\rm t}} \beta  ^{\rm t} _{i,j,k|k}} {1 - {\hat{a}_{i,0,k}^{\rm t}}},
%\\
\alpha ^{\rm t} _{i,k|k}  = \frac{\sum_{j = 1}^{N_{M,k}} {\hat{a}_{i,j,k}^{\rm t}} \alpha ^{\rm t} _{i,j,k|k}} {1 - {\hat{a}_{i,0,k}^{\rm t}}}.
\end{split}
\end{equation}

Next, the backward message $\overleftarrow{b}_{\sigma} (\sigma ^{\rm t} _{i,k})$ is calculated from time $K$ to time $1$, starting with $\overleftarrow{b}_{\sigma} (\sigma ^{\rm t} _{i,K})=1$ for $\sigma ^{\rm t} _{i,K}$, which can be calculated as
\begin{equation}\label{equ:Black}
\begin{split}
& \overleftarrow{b}_{\sigma} (\sigma ^{\rm t} _{i,k}) = m^{\text{MF}}_{f_{\sigma ^{\rm t} _{i,k+1}} \rightarrow \sigma ^{\rm t} _{i,k}}\\
= & \exp\big( \int m^{\text{MF}}_{f_{\sigma ^{\rm t} _{i,k+1}} \rightarrow \sigma ^{\rm t} _{i,k}} \ln p(\sigma ^{\rm t} _{i,k+1} |\sigma ^{\rm t} _{i,k} ) d{\sigma ^{\rm t} _{i,k+1} } \big).
\end{split}
\end{equation}
Note that $m^{\text{MF}}_{f_{\sigma ^{\rm t} _{i,k+1}} \rightarrow \sigma ^{\rm t} _{i,k}} = b_{\sigma} (\sigma ^{\rm t}_ {i,k+1})$, which is calculated by Eq.~\eqref{equ:bKSsigmaS} and must be an IG distribution, i.e., $b_{\sigma} (\sigma ^{\rm t}_ {i,k+1})= \mathcal{I} (\sigma ^{\rm t} _{i,k+1} ; \alpha ^{\rm t} _{i,k+1|K}, \beta ^{\rm t} _{i,k+1|K})$.
By the definition of $p(\sigma ^{\rm t} _{i,k+1} |\sigma ^{\rm t} _{i,k} )$, we have
$\overleftarrow{b}_{\sigma} (\sigma ^{\rm t} _{i,k} ) \propto \mathcal{I} (\sigma ^{\rm t} _{i,k} ; \alpha ^{\rm t} _{i,k|k+1}, \beta ^{\rm t} _{i,k|k+1}) $, which is also an IG distribution, where $\beta ^{\rm t} _{i,k|k+1} ={\beta ^{\rm t} _{i,k+1|K}}/ {u^{\rm t}}$ and $\alpha ^{\rm t} _{i,k|k+1} = ({{\alpha ^{\rm t} _{i,k+1|K} +u^{\rm t}-1}})/ {u^{\rm t}}$.

Finally, the belief $b_{\sigma} (\sigma ^{\rm t} _{i,k})$ is calculated by an IG smoother as $b_{\sigma} (\sigma ^{\rm t} _{i,k}) = \overrightarrow{b}_{\sigma} (\sigma ^{\rm t} _{i,k} ) \overleftarrow{b}_{\sigma} (\sigma ^{\rm t} _{i,k})$.
By applying the product of IG, we have $b_{\sigma} (\sigma ^{\rm t} _{i,k})= \mathcal{I} (\sigma ^{\rm t} _{i,k} ; \alpha ^{\rm t} _{i,k|K}, \beta ^{\rm t} _{i,k|K})$, with $\alpha ^{\rm t} _{i,k|K}=\alpha ^{\rm t} _{i,k|k}+\alpha ^{\rm t} _{i,k|k+1}+1$ and $\beta ^{\rm t} _{i,k|K}=\beta ^{\rm t} _{i,k|k}+\beta ^{\rm t} _{i,k|k+1}$.

\subsection{Derivations of Belief $b_{\bm{X}}(\bm{X}_{1:K}^{\rm c})$}\label{subsec:Derivations of clutter state Belief}
\begin{figure}[!htbp]
\centering
\includegraphics[scale = 0.45]{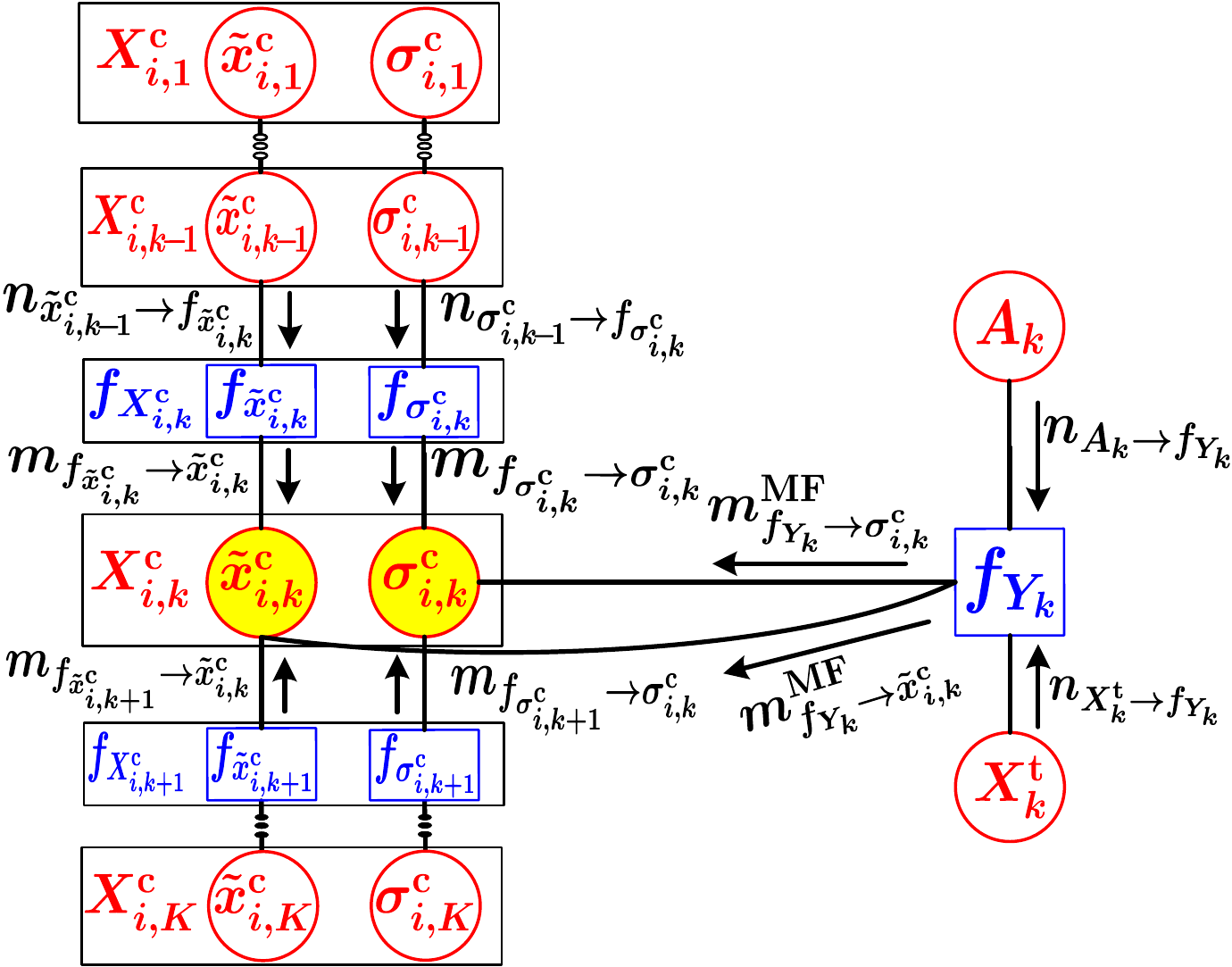}
\caption{The clutter estimation subgraph of $b_{{\bm{x}} ^{\rm c} _{\tau,k}}({\bm{x}} ^{\rm c} _{\tau,k})$.}
\label{fig:Subgraph-C}
\end{figure}

Fig.~\ref{fig:Subgraph-C} shows the clutter spatial state and mean CNR estimation subgraph corresponding to the belief $b_{{\bm{X}}}(\tilde{\bm{X}} ^{\rm c} _{\tau,1:K})$ and $b_{\bm{\sigma}}(\bm{\sigma} ^{\rm c}_{\tau,1:K})$, $\tau=0,\ldots,N_C$, which contains the variable nodes $\tilde{\bm{x}} ^{\rm c} _{\tau,k}$ and $\sigma ^{\rm c}_{\tau,k}$, $k=1,\ldots,K$.

\subsubsection{The belief of clutter spatial state} The belief $b_{{\bm{x}}}(\tilde{\bm{x}} ^{\rm c} _{\tau,k})$ can be calculated as
\begin{equation}\label{eq:bKSxXC}
b_{{\bm{x}}} (\tilde{\bm{x}} ^{\rm c} _{\tau,k}) \propto
    \underbrace{m^{\text{MF}}_{f_{\tilde{\bm{x}} ^{\rm c} _{\tau,k} \rightarrow \tilde{\bm{x}} ^{\rm c} _{\tau,k}}} m^{\text{MF}}_{f_{\bm{Y}_k} \rightarrow \tilde{\bm{x}} ^ {\rm c} _{\tau,k}}}_{\overrightarrow{b}_{{\bm{x}}} (\tilde{\bm{x}} ^{\rm c} _{\tau,k})} \times \underbrace{m^{\text{MF}}_{f_{\tilde{\bm{x}} ^{\rm c} _{\tau,k+1} \rightarrow \tilde{\bm{x}} ^{\rm c} _{\tau,k}}}}_{\overleftarrow{b}_{{\bm{x}}} (\tilde{\bm{x}} ^{\rm c} _{\tau,k})}.
\end{equation}

We initialize $\overrightarrow{b}_{{\bm{x}}} (\tilde{\bm{x}} ^{\rm c} _{\tau,1})$ as a GW distribution at time 1, i.e., $ \overrightarrow{b}_{{\bm{x}}}  (\tilde{\bm{x}} ^{\rm c} _{\tau,1}) \propto \mathcal{W}({\bm{D}} ^{\rm c} _{\tau,1} ; {\bm{W}} ^{\rm c} _{\tau,1|1}, {{\upsilon}} ^{\rm c} _{\tau,1|1}) \times \mathcal{N}({\bm{x}} ^{\rm c} _{\tau,1} ; \hat{\bm{x}} ^{\rm c} _{\tau,1|1}, (\beta^{\rm c} _{\tau,1|1} {\bm{D}} ^{\rm c} _{\tau,1})^{-1})$.
Then the forward messages $\overrightarrow{b}_{{\bm{x}}} (\tilde{\bm{x}} ^{\rm c} _{\tau,k})$ can be calculated from time $2$ to time $K$, which consist of factor-to-variable messages as follows.
\begin{equation}\label{equ:F2VMC}
\begin{split}
   & m^{\text{MF}}_{f_{\tilde{\bm{x}} ^{\rm c} _{\tau,k} \rightarrow \tilde{\bm{x}} ^{\rm c} _{\tau,k}}} \\
 = & \exp\big(\int n_{\tilde{\bm{x}} ^{\rm c} _{\tau,k-1}  \rightarrow f_{\tilde{\bm{x}} ^{\rm c} _{\tau,k}}} \ln p (\tilde{\bm{x}} ^{\rm c} _{\tau,k}| \tilde{\bm{x}} ^{\rm c} _{\tau,k-1} ) d{\tilde{\bm{x}} ^{\rm c} _{\tau,k-1} } \big),
\end{split}
\end{equation}
\begin{equation}\label{equ:F2VM2XTC}
\begin{split}
    & m^{\text{MF}}_{f_{\bm{Y}_k} \rightarrow \tilde{\bm{x}} ^{\rm c} _{\tau,k} } \!\!
  = \exp \big( \sum_{j = 1}^{N_{M,k}} \!\! n_{\bm{A}_k \rightarrow f_{\bm{Y}_k}}
     \ln p( {\bm{y}}_{j,k}|\tilde{\bm{x}} ^{\rm c} _{\tau,k}, \bm{A}_k)\big).
\end{split}
\end{equation}
According to Proposition~\ref{cor-1}, we have $n_{\tilde{\bm{x}} ^{\rm c} _{\tau,k-1}  \rightarrow f_{\tilde{\bm{x}} ^{\rm c} _{\tau,k}}} = \overrightarrow{b}_{{\bm{x}}} (\tilde{\bm{x}} ^{\rm c} _{\tau,k-1})$ and $n_{\bm{A}_k \rightarrow f_{\bm{Y}_k}} = b_{\bm{A}} (a_{\tau,j,k}^{\rm c})$.
Similarly, the belief $\overrightarrow{b}_{{\bm{x}}} (\tilde{\bm{x}} ^{\rm c} _{\tau,k-1})$ is also a GW distribution, given by,
\begin{equation}\label{equ:prior1}
\begin{split}
  \overrightarrow{b}&_{{\bm{x}}} (\tilde{\bm{x}} ^{\rm c} _{\tau,k-1}) \propto \mathcal{W}({\bm{D}} ^{\rm c} _{\tau,k-1} ; {\bm{W}} ^{\rm c} _{\tau,k-1|k-1}, {{\upsilon}} ^{\rm c} _{\tau,k-1|k-1}) \\
 & \times \mathcal{N}({\bm{x}} ^{\rm c} _{\tau,k-1} ; \hat{\bm{x}} ^{\rm c} _{\tau,k-1|k-1}, (\beta^{\rm c} _{\tau,k-1|k-1} {\bm{D}} ^{\rm c} _{\tau,k-1})^{-1}).
\end{split}
\end{equation}
By the definition of $p(\tilde{\bm{x}} ^{\rm c} _{\tau,k}| \tilde{\bm{x}} ^{\rm c} _{\tau,k-1})$, we get
\begin{equation}\label{equ:priorCTrans}
\begin{split}
  m&^{\text{MF}}_{f_{\tilde{\bm{x}} ^{\rm c} _{\tau,k} \rightarrow \tilde{\bm{x}} ^{\rm c} _{\tau,k}}} \propto \mathcal{W}({\bm{D}} ^{\rm c} _{\tau,k} ; {\bm{W}} ^{\rm c} _{\tau,k|k-1}, {{\upsilon}} ^{\rm c} _{\tau,k|k-1})\\
 & \times \mathcal{N}({\bm{x}} ^{\rm c} _{\tau,k} ; \hat{\bm{x}} ^{\rm c} _{\tau,k|k-1}, (\beta^{\rm c} _{\tau,k|k-1} {\bm{D}} ^{\rm c} _{\tau,k})^{-1}),
\end{split}
\end{equation}
where $\hat{\bm{x}} ^{\rm c} _{\tau,k|k-1}=\hat{\bm{x}} ^{\rm c} _{\tau,k-1|k-1}$, ${\beta} ^{\rm c} _{\tau,k|k-1}= \beta ^{\rm c} _{\tau, k-1|k-1}$, ${\bm{W}} ^{\rm c} _{\tau,k|k-1}=\xi {\bm{W}} ^{\rm c} _{\tau,k-1|k-1}$, ${{\upsilon}} ^{\rm c} _{\tau,k|k-1}=\xi ({{\upsilon}} ^{\rm c} _{\tau,k-1|k-1}-m-1)+m+1$.
Eq.~(\ref{equ:F2VM2XTC}) is calculated as
\begin{equation}\label{equ:F2VMXTC21}
\begin{split}
m^{\text{MF}}_{f_{\bm{Y}_k} \rightarrow \tilde{\bm{x}} ^{\rm c} _{\tau,k} }
 = & \prod_{j = 1}^{N_{M,k}} \mathcal{N} ({\bm{y}}_{j,k} ; {\bm{x}} ^{\rm c} _{\tau,k}, ({\bm{D}} ^{\rm c} _{\tau,k})^{-1}) ^ {\hat{a}_{\tau,j,k}^{\rm c}},
\end{split}
\end{equation}
where $\hat{a}_{\tau,j,k}^{\rm c}$ is the expectation of ${a}_{\tau,j,k}^{\rm c}$ taken over belief $b_{\bm{A}} (a_{\tau,j,k}^{\rm c})$.
Substituting Eq.~\eqref{equ:priorCTrans} and Eq.~\eqref{equ:F2VMXTC21} into Eq.~\eqref{eq:bKSxXC} and according to \cite{2006Pattern}, $\overrightarrow{b}_{{\bm{x}}} (\tilde{\bm{x}} ^{\rm c} _{\tau,k})$ is calculated as
\begin{equation}\label{equ:Postrior}
\begin{split}
  \overrightarrow{b}_{{\bm{x}}} ({\bm{x}} ^{\rm c} _{\tau,k}) \propto \ & \mathcal{N}({\bm{x}} ^{\rm c} _{\tau,k} ; \hat{\bm{x}} ^{\rm c} _{\tau,k|k}, (\beta^{\rm c} _{\tau,k|k} {\bm{D}} ^{\rm c} _{\tau,k})^{-1}) \\
  & \times \mathcal{W}({\bm{D}} ^{\rm c} _{\tau,k} ; {\bm{W}} ^{\rm c} _{\tau,k|k}, {{\upsilon}} ^{\rm c} _{\tau,k|k}),
\end{split}
\end{equation}
where we define
\begin{equation}\label{equ:PostriorE1}
\begin{split}
  & \beta ^{\rm c} _{\tau,k|k} =\beta ^{\rm c} _{\tau,k|k-1} + N ^{\rm c}_{\tau, k}, \quad {{\upsilon}} ^{\rm c} _{\tau,k|k} =  {{\upsilon}} ^{\rm c} _{\tau,k|k-1} + N ^{\rm c}_{\tau, k}, \\
  & \hat{\bm{x}} ^{\rm c} _{\tau,k|k} = \frac{1}{\beta ^{\rm c} _{\tau,k}} (\beta ^{\rm c} _{\tau,k|k-1} \hat{\bm{x}} ^{\rm c} _{\tau,k|k-1} + N ^{\rm c}_{\tau, k} \bar{\bm{x}} ^{\rm c} _{\tau,k} ),\\
  & ({\bm{W}} ^{\rm c} _{\tau,k|k})^{-1} = ({\bm{W}} ^{\rm c} _{\tau,k|k-1})^{-1} + N ^{\rm c}_{\tau, k} \bm{\Xi}^{\rm c}_{\tau, k} + \\
  & \frac{\beta ^{\rm c} _{\tau,k|k-1} N ^{\rm c}_{\tau, k}}{\beta ^{\rm c} _{\tau,k|k-1}+N ^{\rm c}_{\tau, k}} (\bar{\bm{x}} ^{\rm c} _{\tau,k} - \hat{\bm{x}} ^{\rm c} _{\tau,k|k-1}) (\bar{\bm{x}} ^{\rm c} _{\tau,k} - \hat{\bm{x}} ^{\rm c} _{\tau,k|k-1})^{\rm{T}}.
\end{split}
\end{equation}
In Eq.~(\ref{equ:PostriorE1}), we use three statistics of the measurement data set, given by
\begin{equation}\label{equ:PostriorE}
\begin{split}
  & N ^{\rm c}_{\tau, k} =  \sum_{j=1}^{N_{M,k}} \hat{a}_{\tau,j,k}^{\rm c}, \quad \bar{\bm{x}} ^{\rm c} _{\tau,k} = \frac{1}{N ^{\rm c}_{\tau, k} } \sum_{j=1}^{N_{M,k}} \hat{a}_{\tau,j,k}^{\rm c} {\bm{y}}_{j,k}, \\
  & {\bm{\Xi}} ^{\rm c} _{\tau,k} = \frac{1}{N ^{\rm c}_{\tau, k} } \sum_{j=1}^{N_{M,k}} \hat{a}_{\tau,j,k}^{\rm c} ({\bm{y}}_{j,k}-\bar{\bm{x}} ^{\rm c} _{\tau,k})({\bm{y}}_{j,k}-\bar{\bm{x}} ^{\rm c} _{\tau,k})^{\rm{T}}.
\end{split}
\end{equation}

Then, the backward message $\overleftarrow{b}_{{\bm{x}}} (\tilde{\bm{x}} ^{\rm c} _{\tau,k})$ is calculated from time $K$ to time $1$, starting with $\overleftarrow{b}_{{\bm{x}}} (\tilde{\bm{x}} ^{\rm c} _{\tau,K})=1$ for all $\tilde{\bm{x}} ^{\rm c} _{\tau,K}$ at time $K$, which can be calculated as
\begin{equation}\label{equ:Back X}
\begin{split}
& \overleftarrow{b}_{{\bm{x}}} (\tilde{\bm{x}} ^{\rm c} _{\tau,k}) = m^{\text{MF}}_{f_{\tilde{\bm{x}} ^{\rm c} _{\tau,k+1} \rightarrow \tilde{\bm{x}} ^{\rm c} _{\tau,k}}}\\
= & \exp\big(\int n_{\tilde{\bm{x}} ^{\rm c} _{\tau,k+1}  \rightarrow f_{\tilde{\bm{x}} ^{\rm c} _{\tau,k}}} \ln p (\tilde{\bm{x}} ^{\rm c} _{\tau,k+1}| \tilde{\bm{x}} ^{\rm c} _{\tau,k} ) d{\tilde{\bm{x}} ^{\rm c} _{\tau,k+1} } \big),
\end{split}
\end{equation}
where $n_{\tilde{\bm{x}} ^{\rm c} _{\tau,k+1}  \rightarrow f_{\tilde{\bm{x}} ^{\rm c} _{\tau,k}}} = b_{{\bm{x}}} (\tilde{\bm{x}} ^{\rm c} _{\tau,k+1})$ is a GW distribution, i.e., $\overleftarrow{b}_{{\bm{x}}} (\tilde{\bm{x}} ^{\rm c} _{\tau,k+1}) = \mathcal{N} ({\bm{x}} ^{\rm c} _{\tau,k+1} ; $ $\hat{\bm{x}} ^{\rm c} _{\tau,k+1|K}, (\beta^{\rm c} _{\tau,k+1|K} {\bm{D}} ^{\rm c} _{\tau,k+1})^{-1}) \times $ $ \mathcal{W}({\bm{D}} ^{\rm c} _{\tau,k+1} ; {\bm{W}} ^{\rm c} _{\tau,k+1|K}, {{\upsilon}} ^{\rm c} _{\tau,k+1|K})$.
Via the definition of $p (\tilde{\bm{x}} ^{\rm c} _{\tau,k+1}| \tilde{\bm{x}} ^{\rm c} _{\tau,k} )$,
$\overleftarrow{b}_{{\bm{x}}} (\tilde{\bm{x}} ^{\rm c} _{\tau,k})$ is also a GW distribution
\begin{equation}\label{eq:backword of clutter x}
\begin{split}
\overleftarrow{b}_{{\bm{x}}} (\tilde{\bm{x}} ^{\rm c} _{\tau,k}) = & \mathcal{N} ({\bm{x}} ^{\rm c} _{\tau,k} ; \hat{\bm{x}} ^{\rm c} _{\tau,k|k+1}, (\beta^{\rm c} _{\tau,k|k+1} {\bm{D}} ^{\rm c} _{\tau,k})^{-1}) \\
& \times \mathcal{W}({\bm{D}} ^{\rm c} _{\tau,k} ; {\bm{W}} ^{\rm c} _{\tau,k|k+1}, {{\upsilon}} ^{\rm c} _{\tau,k|k+1}) ,
\end{split}
\end{equation}
where $\hat{\bm{x}} ^{\rm c} _{\tau,k|k+1}=\hat{\bm{x}} ^{\rm c} _{\tau,k+1|K}$, ${\beta} ^{\rm c} _{\tau,k|k+1}= \beta ^{\rm c} _{\tau, k+1|K}$, ${\bm{W}} ^{\rm c} _{\tau,k|k+1}=\xi {\bm{W}} ^{\rm c} _{\tau,k+1|K}$, ${{\upsilon}} ^{\rm c} _{\tau,k|k+1}=\xi ({{\upsilon}} ^{\rm c} _{\tau,k+1|K}-m-1)+m+1$.

Finally, the belief ${b}_{{\bm{x}}} (\tilde{\bm{x}} ^{\rm c} _{\tau,k})$ is calculated by a GW smoother as $b_{{\bm{x}}} (\tilde{\bm{x}} ^{\rm c} _{\tau,k}) \propto {\overrightarrow{b}_{{\bm{x}}} (\tilde{\bm{x}} ^{\rm c} _{\tau,k})} {\overleftarrow{b}_{{\bm{x}}} (\tilde{\bm{x}} ^{\rm c} _{\tau,k})}$.
By applying the product of GW distribution, we have
\begin{equation}\label{eq:Smooth of clutter x 1}
\begin{split}
{b}_{{\bm{x}}} (\tilde{\bm{x}} ^{\rm c} _{\tau,k}) = & \mathcal{N} ({\bm{x}} ^{\rm c} _{\tau,k} ; \hat{\bm{x}} ^{\rm c} _{\tau,k|K}, (\beta^{\rm c} _{\tau,k|K} {\bm{D}} ^{\rm c} _{\tau,k})^{-1}) \\
& \times \mathcal{W}({\bm{D}} ^{\rm c} _{\tau,k} ; {\bm{W}} ^{\rm c} _{\tau,k|K}, {{\upsilon}} ^{\rm c} _{\tau,k|K}),
\end{split}
\end{equation}
where $\beta^{\rm c} _{\tau,k|K} = \beta^{\rm c} _{\tau,k|k} + \beta^{\rm c} _{\tau,k|k+1}$, $\hat{\bm{x}} ^{\rm c} _{\tau,k|K}=(\beta^{\rm c} _{\tau,k|k} \hat{\bm{x}} ^{\rm c} _{\tau,k|k} + \beta^{\rm c} _{\tau,k|k+1} \hat{\bm{x}} ^{\rm c} _{\tau,k|k+1}) / \beta^{\rm c} _{\tau,k|K}$, $({\bm{W}} ^{\rm c} _{\tau,k|K})^{-1} =  ({\bm{W}} ^{\rm c} _{\tau,k|k})^{-1} + ({\bm{W}} ^{\rm c} _{\tau,k|k+1})^{-1}$ and ${{\upsilon}} ^{\rm c} _{\tau,k|K} = {{\upsilon}} ^{\rm c} _{\tau,k|k} + {{\upsilon}} ^{\rm c} _{\tau,k|k+1}-m-1$.

\subsubsection{The belief of clutter mean CNR} The belief $b_{\sigma}(\sigma ^{\rm c} _{\tau,k} )$ is derived by
\begin{equation}\label{equ:bKSsigma11C}
b_{\sigma} (\sigma ^{\rm c} _{\tau,k} ) \propto
    \underbrace{m^{\text{MF}}_{f_{\sigma ^{\rm c} _{\tau,k} \rightarrow \sigma ^{\rm c} _{\tau,k} }} m^{\text{MF}}_{f_{\bm{Y}_k} \rightarrow \sigma ^{\rm c} _{\tau,k} }}_{\overrightarrow{b}_{\sigma} (\sigma ^{\rm c} _{\tau,k})} \times \underbrace{m^{\text{MF}}_{f_{\sigma ^{\rm c} _{\tau,k+1}} \rightarrow \sigma ^{\rm c} _{\tau,k}}}_{\overleftarrow{b}_{\sigma} (\sigma ^{\rm c} _{\tau,k} )}.
\end{equation}

Using the same initialisation and MP relus as target mean SNR estimation, the forward message $\overrightarrow{b}_{\sigma}(\sigma ^{\rm c} _{\tau,k} )$ can be calculated as $\overrightarrow{b}_{\sigma} (\sigma ^{\rm c} _{\tau,k} ) \propto \mathcal{I} (\sigma ^{\rm c} _{\tau,k}; \alpha ^{\rm c} _{\tau,k|k}, \beta ^{\rm c} _{\tau,k|k})$, where
\begin{equation}\label{equ:posterior IG parameters2}
\begin{split}
\beta ^{\rm c} _{\tau,k|k} = \frac{\sum_{j = 1}^{N_{M,k}} {\hat{a}_{\tau,j,k}^{\rm c}} \beta ^{\rm c} _{\tau,j,k|k}} {1 - \hat{a}_{\tau,0,k}^{\rm c}},
% \\
\alpha ^{\rm c} _{i,k|k}   = \frac{\sum_{j = 1}^{N_{M,k}} {\hat{a}_{\tau,j,k}^{\rm c}} \alpha ^{\rm c} _{\tau,j,k|k}} {1 - \hat{a}_{\tau,0,k}^{\rm c}},
\end{split}
\end{equation}
where $\beta ^{\rm c} _{\tau,j,k|k} \!=\!\! \beta ^{\rm c} _{\tau,k|k-1} + nm_{j,k} ^ {2} - n d ^ {2},  \alpha ^{\rm c} _{\tau,j,k|k} \!=\!\! \alpha ^{\rm c} _{\tau,k|k-1} + n$.

Again, referring to target mean SNR estimation, the backward message $\overleftarrow{b}_{\sigma} (\sigma ^{\rm c} _{\tau,k})$ can be calculated as $\overleftarrow{b}_{\sigma} (\sigma ^{\rm c} _{\tau,k} ) \propto \mathcal{I} (\sigma ^{\rm c} _{\tau,k} ; \alpha ^{\rm c} _{\tau,k|k+1}, \beta ^{\rm c} _{\tau,k|k+1})$, where $\beta ^{\rm c} _{\tau,k|k+1} ={\beta ^{\rm c} _{\tau,k+1|K}}/ {u^{\rm c}}$ and $\alpha ^{\rm c} _{\tau,k|k+1} = ({{\alpha ^{\rm c} _{\tau,k+1|K} +u^{\rm c}-1}})/ {u^{\rm c}}$.

Finally, we have $b_{\sigma} (\sigma ^{\rm c} _{\tau,k})= \mathcal{I} (\sigma ^{\rm c} _{\tau,k} ; \alpha ^{\rm c} _{\tau,k|K}, \beta ^{\rm c} _{\tau,k|K})$ by an IG smoother, where $\alpha ^{\rm c} _{\tau,k|K}=\alpha ^{\rm c} _{\tau,k|k}+\alpha ^{\rm c} _{\tau,k|k+1}+1$ and $\beta ^{\rm c} _{\tau,k|K}=\beta ^{\rm c} _{\tau,k|k}+\beta ^{\rm c} _{\tau,k|k+1}$.

\subsection{Derivations of Belief $b_{\bm{S}}(\bm{S}_{1:K})$}\label{subsec:Derivations of detection state belief}
\begin{figure}[!htbp]
\centering
\includegraphics[scale = 0.6]{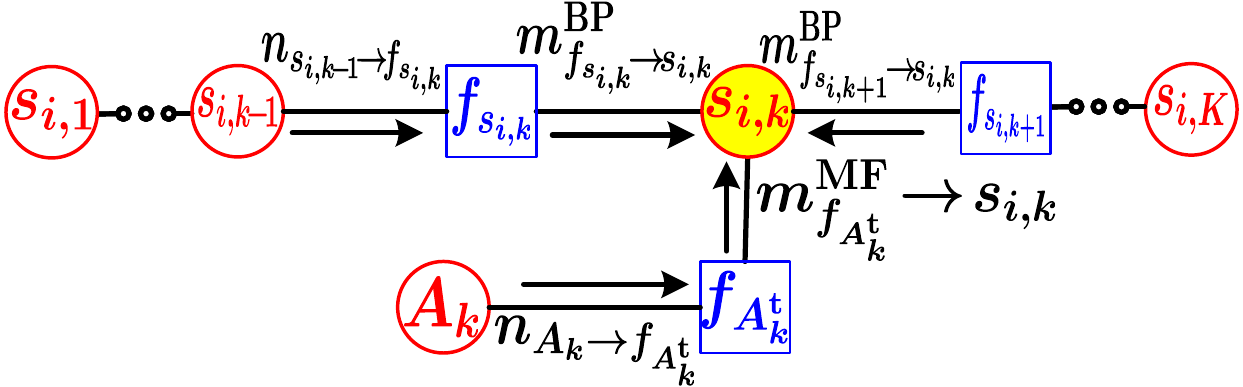}
\caption{The subgraph for target visibility state estimation.}
\label{fig:Subgraph-e}
\end{figure}

Fig.~\ref{fig:Subgraph-e} shows the subgraph for the estimation of target visibility state $s_{i,k}$, $k=1,\ldots,K$.
The belief $b_{s}(s_{i,k})$ can be calculated as
\begin{equation} \label{Equ:beFinalTS}
b_{s}(s_{i,k}) \propto \underbrace{m^{\text{BP}}_{f_{s_{i,k}} \rightarrow s_{i,k}} m^{\text{BP}}_{f_{\bm{A}_k} \rightarrow s_{i,k}}}_{\overrightarrow{b} _{s}(s_{i,k})} \times \underbrace{m^{\text{BP}}_{f_{s_{i,k+1}} \rightarrow s_{i,k}}}_{\overleftarrow{b} _{s}(s_{i,k})}.
\end{equation}

Initializing $\overrightarrow{b} _{s}(s_{i,1})$ as a Bernoulli distribution at time $1$, the forward messages $\overrightarrow{b} _{s}(s_{i,k})$ then can be calculated from time $2$ to time $K$, which consist of factor-to-variable messages
\begin{equation}\label{equ:A2e11T}
\begin{split}
  m^{\text{BP}}_{f_{s_{i,k}} \rightarrow s_{i,k}}
  = & \sum_{s_{i,k-1}=0}^{1} p ({s}_{i,k}|s_{i,k-1}) n_{{s}_{i,k-1}^{\rm t} \rightarrow f_{s_{i,k}}} \\
  = & \bm{T}_k \overrightarrow{b} _s({s}_{i,k-1}),
\end{split}
\end{equation}
\begin{equation}\label{equ:A2e12T}
\begin{split}
  m^{\text{BP}}_{f_{\bm{A}_k} \rightarrow s_{i,k}}
  \!\! = \!\!\!\! \sum_{a_{i,0,k} ^{\rm t} =0}^{1} p(a_{i,0,k} ^{\rm t}| s_{i,k}) n_{a_{i,0,k} ^{\rm t} \rightarrow f_{\bm{A}_k}}
  \!\! = \bm{\xi}_k(s_{i,k}),
\end{split}
\end{equation}
where $n_{a_{i,0,k} ^{\rm t} \rightarrow f_{\bm{A}_k}}=b_a(a_{i,0,k})$.
Substituting Eq.~\eqref{equ:A2e11T} and Eq.~\eqref{equ:A2e12T} into Eq.~\eqref{Equ:beFinalTS}, the belief $\overrightarrow{b} _{s}(s_{i,k})$ is written as $ \overrightarrow{b}_{s}(s_{i,k}) = \bm{T}_k b_s({s}_{i,k-1}) \bm{\xi}_k(s_{i,k})$.
The belief $\overrightarrow{b}_{\bm{s}}(\bm{s}_{i,1:K})$ from time $1$ to time $K$ can be derived as
\begin{equation}\label{eq:bee1k}
\overrightarrow{b}_{\bm{s}}(\bm{s}_{i,1:K}) =  \prod_{k = 1}^K \overrightarrow{b}_{s}(s_{i,k}) = \overrightarrow{b} _{s}(s_{i,1}) \bm{\xi}_{1}({s}_{i,1}) \prod_{k = 2}^K  \bm{T}_k \bm{\xi}_k(s_{i,k}).
\end{equation}

Next, the backward message $\overleftarrow{b} _{s}(s_{i,k})$ is calculated from time $K$ to time $1$, starting with $\overleftarrow{b} _{s}(s_{i,K})=1$ for all $s_{i,K}$, which is given by
\begin{equation}\label{equ:Blck message S}
\begin{split}
& \overleftarrow{b} _{s}(s_{i,k}) = m^{\text{BP}}_{f_{s_{i,k+1}} \rightarrow s_{i,k}} \\
= & \sum_{s_{i,k+1}=0}^1 \!\! m^{\text{BP}}_{f_{s_{i,k+2}} \rightarrow s_{i,k+1}} m^{\text{BP}}_{f_{\bm{A}_{k+1}} \rightarrow s_{i,k+1}} p ({s}_{i,k+1}|s_{i,k}).
\end{split}
\end{equation}

Thus, the belief $b_{\bm{s}}(\bm{s}_{i,1:K})$ can be recognized as a hidden Markov model (HMM) with observation sequence $\{\bm{\xi}_{1}({s}_{i,1}), \ldots, \bm{\xi}_{1}({s}_{i,K})\}$, transition matrix $\bm{T}_k$ and initial probability $\overrightarrow{b} _{s}(s_{i,1})$.
We adopt the forward-backward algorithm~\cite{Rabiner1989} to estimate the belief $b_{\bm{s}}(\bm{s}_{i,1:K})$, resulting an HMM smoother.
The track management can be achieved by comparing the belief $b_{s}(s_{i,k})$ with the track confirmation and deletion thresholds.

\subsection{Derivations of Belief $b_{\bm{\Pi}}(\bm{\Pi}_{1:K})$}\label{subsec:Derivations of clutter mixing weights}
\begin{figure}[!htbp]
\centering
\includegraphics[scale = 0.6]{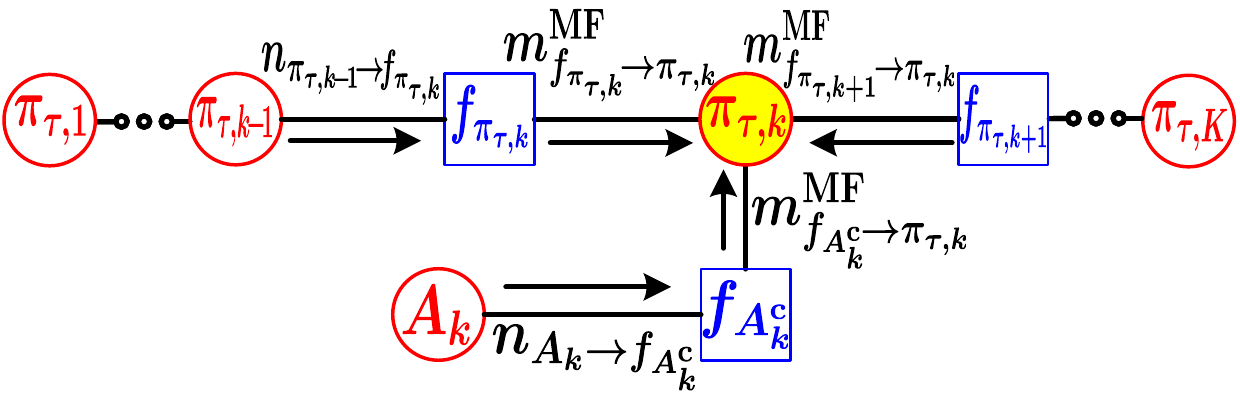}
\caption{The mixing weights estimation subgraph of $b_{\bm{\Pi}}(\bm{\Pi}_{1:K})$.}
\label{fig:Subgraph-p}
\end{figure}

Fig.~\ref{fig:Subgraph-p} shows the clutter mixing weights estimation subgraph corresponding to the belief $b_{\bm{\pi}}(\bm{\pi}_{\tau,1:K})$, $\tau=0,\ldots,N_T$, which contains the variable nodes ${\pi}_{\tau,k}$, $k=1,\ldots,K$.
The belief $b_{{\pi}}({\pi}_{\tau,k})$ is given by
\begin{equation} \label{Equ:beFinalT}
b_{{\pi}}({\pi}_{\tau,k}) \propto \underbrace{m^{\text{MF}}_{f_{{\pi}_{\tau,k}} \rightarrow {\pi}_{\tau,k}} m^{\text{MF}}_{f_{\bm{A}_k} \rightarrow {\pi}_{\tau,k}}}_{\overrightarrow{b}_{{\pi}}({\pi}_{\tau,k})} \times \underbrace{m^{\text{MF}}_{f_{{\pi}_{\tau,k+1}} \rightarrow {\pi}_{\tau,k}}} _{\overleftarrow{b}_{{\pi}}({\pi}_{\tau,k})}.
\end{equation}

Initializing $\overrightarrow{b}_{{\pi}}({\pi}_{\tau,1}) \propto {\rm Dir} ({\pi}_{\tau,1}; \alpha_{\tau,1|1})$ as a Dirichlet distribution at time $1$, the forward messages $\overrightarrow{b}_{{\pi}}({\pi}_{\tau,k})$ then can be calculated from time $2$ to time $K$, which consist of factor-to-variable messages
\begin{equation}\label{equ:A2e11TC}
  m^{\text{MF}}_{f_{{\pi}_{\tau,k}} \rightarrow {\pi}_{\tau,k}}
  \!\!=\!\! \int n_{{\pi}_{\tau,k-1} \rightarrow f_{{\pi}_{\tau,k}}} \ln p ({\pi}_{\tau,k}|{\pi}_{\tau,k-1}) d {\pi}_{\tau,k},
\end{equation}
\begin{equation}\label{equ:F2VMDir1}
\begin{split}
  m ^{\text{MF}}_{f_{\bm{A}_k^{\rm c}} \rightarrow {\pi}_{\tau,k}} =   \exp \Big( \sum_{j = 1}^{N_{M,k}}  n_{\bm{A}_{k} \rightarrow f_{\bm{A}_{k}^{\rm c}}}
  \ln p( a_{\tau,j,k}^{\rm c} | {\pi}_{\tau,k} )\Big).
\end{split}
\end{equation}
Note that $n_{{\pi}_{\tau,k-1} \rightarrow f_{{\pi}_{\tau,k}}} = \overrightarrow{b}_{{\pi}}({\pi}_{\tau,k-1})$ is a Dirichlet distribution, i.e., $\overrightarrow{b}_{{\pi}}({\pi}_{\tau,k-1}) = {\rm Dir} ({\pi}_{\tau,k-1}; \alpha_{\tau,k-1|k-1})$.
By the definition of $p ({\pi}_{\tau,k}|{\pi}_{\tau,k-1})$, the predicted belief also follows a Dirichlet distribution, i.e., $m^{\text{MF}}_{f_{{\pi}_{\tau,k}} \rightarrow {\pi}_{\tau,k}} \propto {\rm Dir} ({\pi}_{\tau,k}; \alpha_{\tau,k|k-1})$, where $\alpha_{\tau,k|k-1} = \kappa M_{k-1} \alpha_{\tau,k-1|k-1} / \sum_{\tau'=0}^{N_C} \alpha_{\tau',k-1|k-1}$ and $M_{k-1}$ is the estimated total number of clutter at time $k-1$.
Then, the message $m^{\text{MF}}_{f_{\bm{A}_k^{\rm c}} \rightarrow {\pi}_{\tau,k}}$ is calculated as
\begin{equation}\label{equ:F2VMDir2}
\begin{split}
  m ^{\text{MF}}_{f_{\bm{A}_k^{\rm c}} \rightarrow {\pi}_{\tau,k}} = & \exp \Big( \sum_{j = 1}^{N_{M,k}} \hat{a}_{\tau,j,k}^{\rm c} \ln {\pi}_{\tau,k} \Big) \\
   = & ({{\pi}_{\tau,k}}) ^ {\sum_{j = 1}^{N_{M,k}} \hat{a}_{\tau,j,k}}.
\end{split}
\end{equation}
Thus, the forward message $\overrightarrow{b}_{{\pi}}({\pi}_{\tau,k})$ is calculated as
\begin{equation}\label{equ:A2e11TCF}
\begin{split}
  \overrightarrow{b}_{{\pi}}({\pi}_{\tau,k}) \propto & {\rm Dir} ({\pi}_{\tau,k-1}; \alpha_{\tau,k|k-1}) ({{\pi}_{\tau,k}}) ^ {\sum_{j = 1}^{N_{M,k}} \hat{a}_{\tau,j,k}} \\
   \propto & ({{\pi}_{\tau,k}}) ^ {\sum_{j = 1}^{N_{M,k}} \hat{a}_{\tau,j,k} + \alpha_{\tau,k|k-1}}.
\end{split}
\end{equation}
We recognize $\overrightarrow{b}_{{\pi}}({\pi}_{\tau,k}) = {\rm{Dir}} ({\pi}_{\tau,k}| {\alpha}_{\tau,k|k})$ as a Dirichlet distribution with ${\alpha}_{\tau,k|k}=\sum_{j = 1}^{N_{M,k}} \hat{a}_{\tau,j,k} + \alpha_{\tau,k|k-1}$.

Next, the backward message $\overleftarrow{b}_{{\pi}}({\pi}_{\tau,k})$ is calculated from time $K$ to time $1$, starting with $\overleftarrow{b} _{{\pi}} ({\pi}_{\tau,K})=1$ for all ${\pi}_{\tau,K}$, which is given by
\begin{equation}\label{equ:Blck message pi}
\begin{split}
& \overleftarrow{b} _{{\pi}} ({\pi}_{\tau,k}) = m^{\text{MF}} _{f_{{\pi}_{\tau,k+1}} \rightarrow {\pi}_{\tau,k}}\\
= & \exp\big( \int m^{\text{MF}}_{f_{\pi _{\tau,k+1}} \rightarrow \pi _{\tau,k}} \ln p(\pi _{\tau,k+1} | \pi _{\tau,k} ) d{\pi _{\tau,k+1} } \big).
\end{split}
\end{equation}
Note that $m^{\text{MF}}_{f_{\pi _{\tau,k+1}} \rightarrow \pi _{\tau,k}} = {b}_{\pi} (\pi_ {\tau,k+1})$ is a Dirichlet distribution, given by ${b}_{\pi} (\pi_ {\tau,k+1}) = {\rm Dir} ({\pi}_{\tau,k+1}; \alpha_{\tau,k+1|K})$.
By the definition of $p(\pi _{\tau,k+1} | \pi _{\tau,k} )$, we have
$\overleftarrow{b} _{{\pi}} ({\pi}_{\tau,k}) \propto {\rm{Dir}} ({\pi}_{\tau,k}| {\alpha}_{\tau,k|k+1})$, which is also a Dirichlet distribution, where $\alpha_{\tau,k|k+1} = \kappa M_{k+1} \alpha_{\tau,k+1|K} / \sum_{\tau'=0}^{N_C} \alpha_{\tau',k+1|K}$, where $M_{k+1}$ is the estimated total number of clutter at time $k+1$.

Finally, the belief $b_{{\pi}}({\pi}_{\tau,k})$ is calculated by a Dirichlet smoother as $b_{{\pi}}({\pi}_{\tau,k}) = \overrightarrow{b}_{{\pi}}({\pi}_{\tau,k}) \overleftarrow{b}_{{\pi}}({\pi}_{\tau,k})$.
By applying the product of Dirichlet distribution, we have $b_{{\pi}}({\pi}_{\tau,k}) = {\rm{Dir}} ({\pi}_{\tau,k}| {\alpha}_{\tau,k|K})$, where ${\alpha}_{\tau,k|K}={\alpha}_{\tau,k|k} +{\alpha}_{\tau,k|k+1}$.
The estimated number of clutter component $\tau$ is $M_{\tau,k}=M_k {\alpha}_{\tau,k|K}$ with $M_k=\sum_{\tau = 0}^{N_{C}} \sum_{j = 1}^{N_{M,k}} \hat{a}_{\tau,j,k}$ being the estimated total number of clutter.

\subsection{Derivations of Beliefs $b_{\bm{A}}(\bm{A}_{1:K})$}\label{subsec:Derivations of data association state belief}
\begin{figure}[!htbp]
\centering
\includegraphics[scale = 0.56]{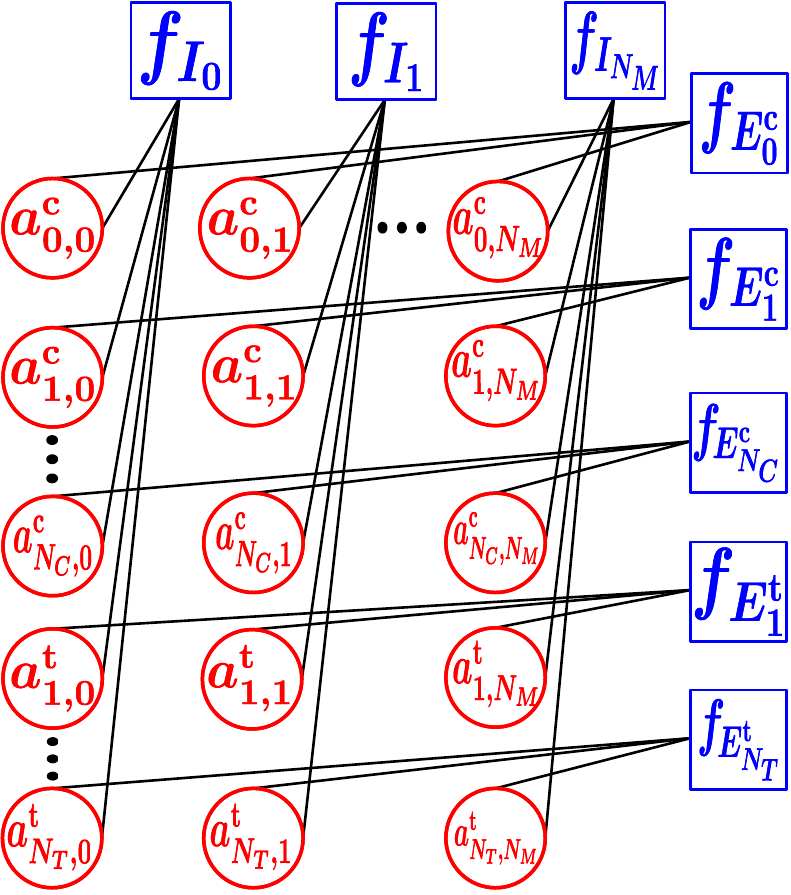}
\caption{The global data association subgraph of $b_{\bm{A}}(\bm{A}_{k})$. The time index $k$ is omitted for the simplicity. In addition, each variable node $a_{i,j,k}^{\text{t}} \ (a_{\tau,j,k}^{\text{c}})$ also connects with the factor nodes $f_{\bm{y}_{j,k}}$ and $f_{\bm{A}_{k}^{\rm t}} \ (f_{\bm{A}_{k}^{\rm c}})$ as in Fig.~\ref{fig:Subgraph-a}, which are not shown here for the simplicity.}
\label{fig:Subgraph-a1}
\end{figure}

\begin{figure}[!htbp]
\centering
\includegraphics[scale = 0.6]{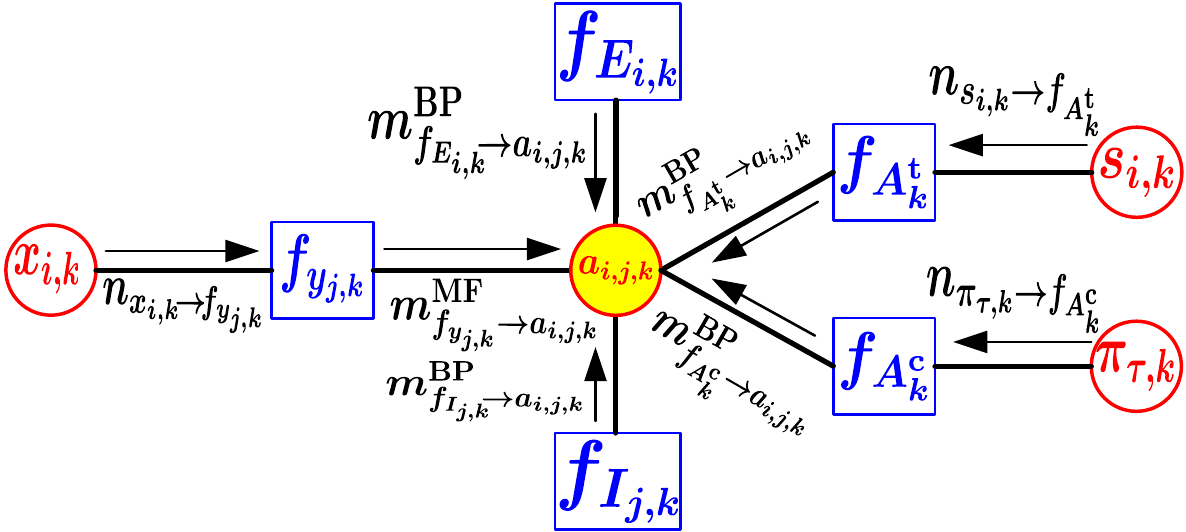}
\caption{The local subgraph of data association. The $\bm{x}_{i,k}$, $\bm{E}_{i,k}$ and $a_{i,j,k}$ represent $\bm{x}_{i,k}^{\text{t}} \ (\bm{x}_{i,k}^{\text{c}})$, $\bm{E}_{i,k}^{\text{t}} \ (\bm{E}_{i,k}^{\text{c}})$ and $a_{i,j,k}^{\text{t}} \ (a_{\tau,j,k}^{\text{c}})$ when calculate the association between measurements and targets (clutter).}
\label{fig:Subgraph-a}
\end{figure}

Fig.~\ref{fig:Subgraph-a1} and Fig.~\ref{fig:Subgraph-a} show the data association global subgraph and the corresponding local MP subgraph for the belief of  $\bm{A}_{k}$, respectively, which contains the variables nodes $a_{i,j,k}^{\rm t}$ and $a_{\tau,j,k}^{\rm c}$, $i=1,\ldots,N_T$, $\tau=0,\ldots,N_C$, $j=0,\ldots,N_{M,k}$, $k=1,\ldots,K$.
The belief $b_{\bm{A}}(\bm{A}_{k})$ can be computed as
\begin{equation} \label{Eq::beFinal}
\begin{split}
b_{\bm{A}}(\bm{A}_{k}) \propto & m^{\text{MF}}_{f_{\bm{Y}_k} \rightarrow \bm{A}_k} \times m^{\text{BP}}_{f_{\bm{A}_k^{\text{t}}} \rightarrow \bm{A}_k} \times m^{\text{BP}}_{f_{\bm{A}_k^{\text{c}}} \rightarrow \bm{A}_k} \\
& \times m^{\text{BP}}_{f_{\bm{I}_k} \rightarrow \bm{A}_k} \times m^{\text{BP}}_{f_{\bm{E}_k} \rightarrow \bm{A}_k}.
\end{split}
\end{equation}
The belief $m^{\text{MF}}_{f_{\bm{Y}_k} \rightarrow \bm{A}_k}$ is computed as
\begin{equation}\label{equ:Yk2Ak}
  m^{\text{BP}}_{f_{\bm{Y}_k} \rightarrow \bm{A}_k} = \prod _{j=0} ^ {N_{M,k}} \prod _{i=1} ^ {N_{T}} m^{\text{MF}}_{f_{\bm{Y}_k} \rightarrow a_{i,j,k}^{\rm t}} \prod _{\tau=0} ^ {N_{C}} m^{\text{MF}}_{f_{\bm{Y}_k} \rightarrow a_{\tau,j,k}^{\rm c}}.
\end{equation}

The message $m^{\text{MF}}_{f_{\bm{Y}_k} \rightarrow a_{i,j,k}^{\rm t}}$ is calculated as
\begin{equation}\label{equ:Yk2Ak0}
\begin{split}
    & m^{\text{MF}}_{f_{\bm{Y}_k} \rightarrow a_{i,j,k}^{\rm t}} \\
  = & \exp \int n_{\bm{X}_{i,k}^{\rm t} \rightarrow f_{\bm{Y}_k}} \ln p(\bm{Y}_{j,k}|\bm{X}_{i,k}^{\rm t}) ^{a_{i,j,k}^{\rm t}} {d} \bm{X}_{i,k}^{\rm t} \\
  = & \exp \big( \int n_{ {\bm{x}} _{i,k}^{\rm t} \rightarrow f_{\bm{Y}_k}} \ln p({\bm{y}}_{j,k}| {\bm{x}} _{i,k}^{\rm t}) ^{a_{i,j,k}^{\rm t}} {d} \bm{x}_{i,k}^{\rm t} + \\
    & \qquad \int n_{{\sigma}_{i,k}^{\rm t} \rightarrow f_{\bm{Y}_k}} \ln p(m_{j,k};{\sigma}_{i,k}^{\rm t}) ^{a_{i,j,k}^{\rm t}} {d} {\sigma}_{i,k}^{\rm t} \big).
\end{split}
\end{equation}
Since $n_{ {\bm{x}} _{i,k}^{\rm t} \rightarrow f_{\bm{Y}_k}}=b( {\bm{x}} _{i,k}^{\rm t})$ and $n_{{\sigma}_{i,k}^{\rm t} \rightarrow f_{\bm{Y}_k}}=b({\sigma}_{i,k}^{\rm t})$, Eq.~\eqref{equ:Yk2Ak0} can be calculated as
\begin{equation}\label{equ:Yk2Ak2}
\begin{split}
& m^{\text{MF}}_{f_{\bm{Y}_k} \rightarrow a_{i,j,k}^{\rm t}} \\
  = & \exp \! \big[ a_{i,j,k}^{\rm t} \big( \mathds{E} [\ln p({\bm{y}}_{j,k}| {\bm{x}} _{i,k}^{\rm t})] \!+\! \mathds{E} [\ln p(m_{j,k};{\sigma}_{i,k}^{\rm t})] \big) \big],
\end{split}
\end{equation}
where,
\begin{equation}\label{equ:Yk2Ak01}
\begin{split}
  & \mathds{E} [\ln p({\bm{y}}_{j,k}| {\bm{x}} _{i,k}^{\rm t})] \overset{\rm{c}}{=} \mathds{E} \big[ ({\bm{y}}_{j,k}- \bm{H}  {\bm{x}} _{i,k}^{\rm t})^{\rm{T}} \bm{R}^{-1}_{k} ({\bm{y}}_{j,k}-\bm{H}  {\bm{x}} _{i,k}^{\rm t}) \big] \\
  & \overset{\rm{c}}{=} {\rm{Tr}} \Big( \bm{R}^{-1}_{k} \big( \bm{H} \bm{P} _{i,k}^{\rm t} \bm{H}^{\rm{T}} + (\bm{H}\hat{\bm{x}} _{i,k}^{\rm t}-{\bm{y}}_{j,k}) (\bm{H}\hat{\bm{x}} _{i,k}^{\rm t}-{\bm{y}}_{j,k})^{\rm{T}} \big) \Big),
\end{split}
\end{equation}
and $\mathds{E} [\ln p(m_{j,k};{\sigma}_{i,k}^{\rm t})]$ can be calculated as~(The details are shown in Appendix~\ref{sec:APP1}),
\begin{equation}\label{equ:Yk2Ak02}
\begin{split}
    & \mathds{E} [\ln p(m_{j,k};{\sigma}_{i,k}^{\rm t})] \\
  = & (2n-1) \ln m_{j,k} - n m_{j,k}^2 \frac{\alpha_{i,k}^{\rm t}}{\beta_{i,k}^{\rm t}} +\frac{n }{2\alpha_{i,k}^{\rm t}-4}.
\end{split}
\end{equation}

The message $m^{\text{MF}}_{f_{\bm{Y}_k} \rightarrow a_{\tau,j,k}^{\rm c}}$ is calculated as,
\begin{equation}\label{equ:Yk2Ak0C}
\begin{split}
    & m^{\text{MF}}_{f_{\bm{Y}_k} \rightarrow a_{\tau,j,k}^{\rm c}} \\
  =& \exp \int n_{\bm{X}_{\tau,k}^{\rm c} \rightarrow f_{\bm{Y}_k}} \ln p(\bm{Y}_{j,k}|\bm{X}_{\tau,k}^{\rm c}) ^{a_{\tau,j,k}^{\rm c}} {d} \bm{X}_{\tau,k}^{\rm c} \\
  = & \exp \big(\int n_{ \tilde{\bm{x}} _{\tau,k}^{\rm c} \rightarrow f_{\bm{Y}_k}} \ln p({\bm{y}}_{j,k}| \tilde{\bm{x}} _{\tau,k}^{\rm c}) ^{a_{\tau,j,k}^{\rm c}} {d} \tilde{\bm{x}}_{\tau,k}^{\rm c} + \\
    & \qquad \int n_{{\sigma}_{\tau,k}^{\rm c} \rightarrow f_{\bm{Y}_k}} \ln p(m_{j,k};{\sigma}_{\tau,k}^{\rm c}) ^{a_{\tau,j,k}^{\rm c}} {d} {\sigma}_{\tau,k}^{\rm c} \big).
\end{split}
\end{equation}
With $n_{ \tilde{\bm{x}} _{\tau,k}^{\rm c} \rightarrow f_{\bm{Y}_k}}=b( \tilde{\bm{x}} _{\tau,k}^{\rm c})$ and $n_{{\sigma}_{\tau,k}^{\rm c} \rightarrow f_{\bm{Y}_k}}=b({\sigma}_{\tau,k}^{\rm c})$, Eq.~\eqref{equ:Yk2Ak0C} can be derived as
\begin{equation}\label{equ:Yk2Ak0C5}
\begin{split}
& m^{\text{MF}}_{f_{\bm{Y}_k} \rightarrow a_{\tau,j,k}^{\rm c}} \\
= & \exp \! \big[ a_{\tau,j,k}^{\rm c} \big( \mathds{E} [\ln p({\bm{y}}_{j,k}| \tilde{\bm{x}} _{\tau,k}^{\rm c})] \!\!+\!\! \mathds{E} [\ln p(m_{j,k};\!{\sigma}_{\tau,k}^{\rm c})] \big) \big],
\end{split}
\end{equation}
where $\mathds{E} [\ln p({\bm{y}}_{j,k}| \tilde{\bm{x}} _{\tau,k}^{\rm c})] = \ln (1/V_G)$ if $\tau=0$; otherwise,
\begin{equation}\label{equ:Yk2Ak01C1}
\begin{split}
\mathds{E} & [\ln p({\bm{y}}_{j,k}| \tilde{\bm{x}} _{\tau,k}^{\rm c})] = -\frac{D}{2}\ln 2 \pi + \frac{1}{2}\mathds{E}[\ln |\bm{D}_{\tau,k}^{\rm c}|] \\
& -\frac{1}{2} \mathds{E} [({\bm{y}}_{j,k}-\hat {\bm{x}} _{\tau,k}^{\rm c})^{\rm T} \bm{D}_{\tau,k}^{\rm c} ({\bm{y}}_{j,k}-\hat {\bm{x}} _{\tau,k}^{\rm c})],
\end{split}
\end{equation}
where,
\begin{equation}\label{equ:Yk2Ak01CLam}
\begin{split}
\mathds{E}[\ln |\bm{D}_{\tau,k}^{\rm c}|] \!\! =\!\! \sum_{j=1}^{m}\psi(\frac{{{\upsilon}} ^{\rm c} _{\tau,k}\!+1\!-j}{2}) \!+\! m \ln 2 \!+\! \ln | {\bm{W}} ^{\rm c} _{\tau,k} |,
\end{split}
\end{equation}
\begin{equation}\label{equ:Yk2Ak01CE}
\begin{split}
  & \mathds{E} [({\bm{y}}_{j,k}-\hat {\bm{x}} _{\tau,k}^{\rm c})^{\rm T} \bm{D}_{\tau,k}^{\rm c} ({\bm{y}}_{j,k}-\hat {\bm{x}} _{\tau,k}^{\rm c})] \\
= & m {\beta _{\tau,k}^{\rm c}}^{-1} + {{\upsilon}} ^{\rm c} _{\tau,k} ({\bm{y}}_{j,k}-\hat {\bm{x}} _{\tau,k}^{\rm c})^{\rm T} \bm{W}_{\tau,k}^{\rm c} ({\bm{y}}_{j,k}-\hat {\bm{x}} _{\tau,k}^{\rm c}),
\end{split}
\end{equation}
where $\psi(a)=d \ln \Gamma(a) / da$ is the digamma function~\cite{2006Pattern}.
Then, $\mathds{E} [\ln p(m_{j,k};{\sigma}_{\tau,k}^{\rm c})]$ is derived as~(The details are given in Appendix~\ref{sec:APP1}),
\begin{equation}\label{equ:Yk2Ak01C2}
\begin{split}
    & \mathds{E} [\ln p(m_{j,k};{\sigma}_{\tau,k}^{\rm c})] \\
  = & (2n-1) \ln m_{j,k} - n m_{j,k}^2 \frac{\alpha_{\tau,k}^{\rm c}}{\beta_{\tau,k}^{\rm c}} +\frac{n }{2\alpha_{\tau,k}^{\rm c}-4}.
\end{split}
\end{equation}

The belief $m^{\text{BP}}_{f_{\bm{A}_k^{\text{t}}} \rightarrow \bm{A}_k}$ is calculated as
\begin{equation}\label{equ:Yk2AkA1}
  m^{\text{BP}}_{f_{\bm{A}_k^{\text{t}}} \rightarrow \bm{A}_k} = \prod _{i=1} ^ {N_{T}} m^{\text{BP}}_{f_{\bm{A}_k^{\text{t}}} \rightarrow a_{i,0,k}^{\rm t}},
\end{equation}
where,
\begin{equation}\label{equ:Ak2AkA1}
  \begin{split}
    & m^{\text{BP}}_{f_{\bm{A}_k^{\rm t}} \rightarrow a_{i,0,k}^{\rm t}} = \sum_{\bm{s}_{i,k}=0}^1 p(a_{i,0,k}^{\rm t}|s_{i,k}) n_{s_{i,k} \rightarrow f_{\bm{A}_k^{\rm t}}} \\
    & =
    \begin{cases}
      \sum_{s_{i,k}=0}^1 \big(1 - P_{\rm d}(s_{i,k})\big) m^{\text{BP}}_ {f_{s_{i,k}} \rightarrow s_{i,k}},\!\!\!\!\!\! & \text{if $ a_{i,0,k}^{\rm t} = 1 $,} \\
      \sum_{s_{i,k}=0}^1 P_{\rm d} (s_{i,k}) m^{\text{BP}}_{f_{s_{i,k}} \rightarrow s_{i,k}},\!\! & \text{otherwise.}
    \end{cases}
  \end{split}
\end{equation}

The belief $m^{\text{MF}}_{f_{\bm{A}_k^{\text{c}}} \rightarrow \bm{A}_k}$ is calculated as
\begin{equation}\label{equ:Yk2AkA2}
  m^{\text{MF}}_{f_{\bm{A}_k^{\text{c}}} \rightarrow \bm{A}_k} = \prod _{\tau=1} ^ {N_{C}} m^{\text{MF}}_{f_{\bm{A}_k^{\text{c}}} \rightarrow a_{\tau,0,k}^{\rm c}},
\end{equation}
where,
\begin{equation}\label{equ:Ak2AkA2}
  \begin{split}
    m^{\text{MF}}_{f_{\bm{A}_k^{\rm c}} \rightarrow a_{\tau,j,k}^{\rm c}} \!\!\! = & \exp \int \!\! n_{\pi_{\tau,k} \rightarrow f_{\bm{A}_k^{\rm c}}} \ln p(a_{\tau, j, k}^{\rm c}|\pi_{\tau,k}) d \pi_{\tau,k}. \\
\end{split}
\end{equation}
Note that $n_{\pi_{\tau,k} \rightarrow f_{\bm{A}_k^{\rm c}}}=b_{\pi}(\pi_{\tau,k})$.
We have
\begin{equation}\label{equ:Ak2AkA3}
  \begin{split}
    m^{\text{MF}}_{f_{\bm{A}_k^{\rm c}} \rightarrow a_{\tau,j,k}^{\rm c}}= & \exp \big( {a}_{\tau,j,k}^{\rm c} \mathds{E} [\ln \pi_{\tau,k}] \big) \\
    = & \exp \big( {a}_{\tau,j,k}^{\rm c} \Gamma (\alpha_{\tau,k}^{\rm c}) -\Gamma ( \sum_{\tau'=1}^{N_C} \alpha_{\tau',k}^{\rm c}) \big).
  \end{split}
\end{equation}

The belief $m^{\text{BP}}_{f_{\bm{I}_k} \rightarrow \bm{A}_k}$ is calculated as
\begin{equation}\label{equ:A2AkI}
  m^{\text{BP}}_{f_{\bm{I}_k} \rightarrow \bm{A}_k} = \prod _{j=1} ^ {N_{M,k}} \prod _{i=1} ^ {N_T} m^{\text{BP}}_{f_{\bm{I}_{j,k}} \rightarrow a_{i,j,k}^{\rm t}} \prod _{\tau=0} ^ {N_C} m^{\text{BP}}_{f_{\bm{I}_{j,k}} \rightarrow a_{\tau,j,k}^{\rm c}}.
\end{equation}
The belief $m^{\text{BP}}_{f_{\bm{I}_{j,k}} \rightarrow a_{i,j,k}^{\rm t}}$ and $m^{\text{BP}}_{f_{\bm{I}_{j,k}} \rightarrow a_{\tau,j,k}^{\rm c}}$ are derived as
\begin{equation}\label{equ:A2AkI1}
\begin{split}
  m^{\text{BP}}_{f_{\bm{I}_{j,k}} \rightarrow a_{i,j,k}^{\rm t}} & = \sum_{{\bm{A} _{j,k}} \backslash \{ a_{i,j,k}^{\rm t} \} } f_{\bm{I}_{j,k}} ({\bm{A} _{j,k}}) \\
    & \times \prod_{i'=1 \backslash i} ^{N_T} n_{{a_{i',j,k}^{\rm t} \rightarrow f_{\bm{I}_{j,k}}}} \prod_{\tau=0} ^{N_C} n_{{a_{\tau,j,k}^{\rm c} \rightarrow f_{\bm{I}_{j,k}}}},
\end{split}
\end{equation}
\begin{equation}\label{equ:A2AkI4}
\begin{split}
  m^{\text{BP}}_{f_{\bm{I}_{j,k}} \rightarrow a_{\tau,j,k}^{\rm c}} & = \sum_{{\bm{A} _{j,k}} \backslash \{ a_{\tau,j,k}^{\rm c} \} } f_{\bm{I}_{j,k}} ({\bm{A} _{j,k}}) \\
    & \times \prod_{i=1} ^{N_T} n_{{a_{i,j,k}^{\rm t} \rightarrow f_{\bm{I}_{j,k}}}} \prod_{\tau'=0\backslash \tau} ^{N_C} n_{{a_{\tau',j,k}^{\rm c} \rightarrow f_{\bm{I}_{j,k}}}},
\end{split}
\end{equation}
where $n_{{a_{i,j,k}^{\rm t} \rightarrow f_{\bm{I}_{j,k}}}}$ and $n_{{a_{\tau,j,k}^{\rm c} \rightarrow f_{\bm{I}_{j,k}}}}$ are
\begin{equation}\label{equ:A2AkI2}
\begin{split}
  & n_{{a_{i,j,k}^{\rm t} \rightarrow f_{\bm{I}_{j,k}}}} \!=\! m^{\rm MF}_{{f_{\bm{Y}_{k}} \rightarrow a_{i,j,k}^{\rm t}}} m^{\rm BP}_{{f_{\bm{A}_{k}^{\rm t}} \rightarrow a_{i,j,k}^{\rm t}}} m^{\rm BP}_{{f_{\bm{E}_{k} ^{\rm t}} \rightarrow a_{i,j,k}^{\rm t}}}, \\
  & n_{{a_{\tau,j,k}^{\rm c} \rightarrow f_{\bm{I}_{j,k}}}} \!=\! m^{\rm MF}_{{f_{\bm{Y}_{k}} \rightarrow a_{\tau,j,k}^{\rm c}}} m^{\rm MF}_{{f_{\bm{A}_{k} ^{\rm c}} \rightarrow a_{\tau,j,k}^{\rm c}}} m^{\rm BP}_{{f_{\bm{E}_{k} ^{\rm c}} \rightarrow a_{\tau,j,k}^{\rm c}}}.
\end{split}
\end{equation}
The belief $m^{\text{BP}}_{f_{\bm{E}_k} \rightarrow \bm{A}_k}$ is calculated as
\begin{equation}\label{equ:A2AkI3}
  m^{\text{BP}}_{f_{\bm{E}_k} \rightarrow \bm{A}_k} \!\!=\! \prod _{j=0} ^ {N_{M,k}} \! \prod _{i=1} ^ {N_{T}} m^{\text{BP}}_{f_{\bm{E}_{i,k}^{\rm t}} \rightarrow a_{i,j,k}^{\rm t}} \prod _{\tau=0} ^ {N_{C}} m^{\text{BP}}_{f_{\bm{E}_{\tau,k}^{\rm c}} \rightarrow a_{\tau,j,k}^{\rm c}}.
\end{equation}
The belief $m^{\text{BP}}_{f_{\bm{E}_{i,k}^{\rm t}} \rightarrow a_{i,j,k}^{\rm t}}$ and $m^{\text{BP}}_{f_{\bm{E}_{\tau,k}^{\rm c}} \rightarrow a_{\tau,j,k}^{\rm c}}$ are calculated as
\begin{equation}\label{equ:A2AkE1}
\begin{split}
  &m^{\text{BP}}_{f_{\bm{E}_{i,k}^{\rm t}} \rightarrow a_{i,j,k}^{\rm t}} = \!\!\!\!\!\! \sum_{{\bm{A}_{i,k}^{\rm{t}}} \backslash \{ a_{i,j,k}^{\rm t} \} } \!\!\!\!\!\!\!\!\! f_{\bm{E}_{i,k} ^{\rm t}} (\bm{A}_{i,k}^{\rm{t}}) \!\! \prod_{j'=1 \backslash j} ^{N_{M,k}} \!\! n_{{a_{i,j',k}^{\rm t} \rightarrow f_{\bm{E}_{i,k} ^{\rm t}}}}, \\
  & m^{\text{BP}}_{f_{\bm{E}_{\tau,k}^{\rm c}} \rightarrow a_{\tau,j,k}^{\rm c}} = \!\!\!\!\!\! \sum_{{\bm{A}_{\tau,k}^{\rm{c}}} \backslash \{ a_{\tau,j,k}^{\rm c} \} } \!\!\!\!\!\!\!\!\!\! f_{\bm{E}_{\tau,k} ^{\rm c}} (\bm{A}_{\tau,k}^{\rm{c}}) \!\!\!\! \prod_{j'=1 \backslash j} ^{N_{M,k}} \!\!\!\! n_{{a_{\tau,j',k}^{\rm c} \rightarrow f_{\bm{E}_{\tau,k} ^{\rm c}}}},
\end{split}
\end{equation}
where $n_{{a_{i,j,k}^{\rm t} \rightarrow f_{\bm{E}_{i,k} ^{\rm t}}}}$ and $n_{{a_{\tau,j,k}^{\rm c} \rightarrow f_{\bm{E}_{\tau,k} ^{\rm c}}}}$ are
\begin{equation}\label{equ:A2AkE2}
\begin{split}
  & n_{{a_{i,j,k}^{\rm t} \rightarrow f_{\bm{E}_{i,k} ^{\rm t}}}} = m^{\rm MF}_{{f_{\bm{Y}_{k}} \rightarrow a_{i,j,k}^{\rm t}}} m^{\rm BP}_{{f_{\bm{A}_{k}^{\rm t}} \rightarrow a_{i,j,k}^{\rm t}}} m^{\rm BP}_{{f_{\bm{I}_{k}} \rightarrow a_{i,j,k}^{\rm t}}}, \\
  & n_{{a_{\tau,j,k}^{\rm c} \rightarrow f_{\bm{E}_{\tau,k} ^{\rm c}}}} \!= m^{\rm MF}_{{f_{\bm{Y}_{k}} \rightarrow a_{\tau,j,k}^{\rm c}}} m^{\rm MF}_{{f_{\bm{A}_{k}^{\rm c}} \rightarrow a_{\tau,j,k}^{\rm c}}} m^{\rm BP}_{{f_{\bm{I}_{k}} \rightarrow a_{\tau,j,k}^{\rm c}}}.
\end{split}
\end{equation}

By applying some mathematical tricks in Appendix~\ref{sec:APP3}, we obtain the simplified messages as follows,
\begin{equation}\label{equ:A2AkE21F1}
\begin{split}
\beta_{i,j,k}^{\rm t} & =
\frac{\theta_{i,j,k}^{\rm t}}
{\theta_{0,j,k}^{\rm c} +\!\! \sum\limits_{i' = 1 \backslash i}^{N_T} \theta_{i',j,k}^{\rm t} \eta_{i',j,k}^{\rm t} +\!\! \sum\limits_{\tau = 1}^{N_C} \theta_{\tau,j,k}^{\rm c} \eta_{\tau,j,k}^{\rm c}}, \\
\end{split}
\end{equation}
\begin{equation}\label{equ:A2AkE21F2}
\begin{split}
\beta_{\tau,j,k}^{\rm c} & \!\!=\!\!
\frac{\theta_{\tau,j,k}^{\rm c}}
{\theta_{0,j,k}^{\rm c} \!\!+\!\! \sum\limits_{i = 1}^{N_T} \theta_{i,j,k}^{\rm t} \eta_{i,j,k}^{\rm t} \!+\!\! \sum\limits_{\tau' = 1 \backslash \tau}^{N_C} \theta_{\tau',j,k}^{\rm c} \eta_{\tau',j,k}^{\rm c}}, \\
\end{split}
\end{equation}
\begin{equation}\label{equ:A2AkE21F3}
\begin{split}
\eta_{i,j,k}^{\rm t} &\!=\! \frac{1}{{\sum_{j' = 0\backslash j}^{N_{M,k}} {\beta_{i,j',k}^{\rm t}} }},
\eta_{\tau,0,k}^{\rm c} \!=\! \frac{1}{\prod_{j' = 1}^{N_{M,k}} (1 + \beta_{\tau,j',k}^{\rm c})}, \\
\end{split}
\end{equation}
\begin{equation}\label{equ:A2AkE21F4}
\begin{split}
\eta_{\tau,j,k}^{\rm c} &= \frac{\prod_{j' = 1 \backslash j }^{N_{M,k}} {(1 + \beta_{\tau,j',k}^{\rm c})}}{\beta_{\tau,0,k}^{\rm c} +  \prod_{j' = 1 \backslash j}^{N_{M,k}} {(1 + \beta_{\tau,j',k}^{\rm c})}},
\end{split}
\end{equation}
where $\beta_{i,j,k}^{\rm t}$, $\beta_{\tau,j,k}^{\rm c}$, $\theta_{i,j,k}^{\rm t}$, $\theta_{\tau,j,k}^{\rm c}$, $\eta_{i,j,k}^{\rm t}$, $\eta_{\tau,j,k}^{\rm c}$ are defined in Eq.~\eqref{equ:APP31} and Eq.~\eqref{equ:APP33}.
The values of all messages $\beta_{i,j,k}^{\rm t}$, $\beta_{\tau,j,k}^{\rm c}$, $\eta _{i,j,k} ^{\rm t}$, and $\eta _{\tau,j,k} ^{\rm c}$ are initialized as one and updated via Eq.~\eqref{equ:A2AkE21F1}-Eq.~\eqref{equ:A2AkE21F4}.
The message-updating procedure is terminated after the difference between two successive messages is less than $\delta_T^{\rm BP}$ or a maximum number of iterations $r_{\rm max}^{\rm BP}$ is reached.
The data association may have multiple local optimal solutions corresponding to the multiple fixed points of the message update rules, which may prevent convergence.
To this end, we introduce a damping to the message update, i.e., $\mu = \gamma \mu^{\rm old}+(1-\gamma)\mu^{\rm new}$, where $\mu$ denote a message and $\gamma$ is a damping factor.
A higher value of $\gamma$ leads to a slower convergence rate but often with more stable convergence.

The proposed hybrid data association algorithm can be viewed as a generalization of existing data association algorithms solved with BP, which contain the one-to-one constraints that model one target generates at most one measurement and one-to-multi constraints that model one clutter component potentially generates a large number of measurements at one scan.
Specifically, the hybrid data associations degenerates into one-to-one data associations~\cite{Williams2014, meyer2017scalable, Meyer2018,Soldi2019} when only targets exist and degenerates into one-to-multi data associations~\cite{Meyer2020, meyer2021scalable} when only clutter exist.
Different from the algorithms in~\cite{Williams2014, meyer2017scalable, Meyer2018,Soldi2019,Meyer2020, meyer2021scalable} that use the two multimodal random variables to represent a redundant formulation of data association, instead as in~\cite{Lan2019,Lan-2020-107621,Lan2020}, the data association event is modelled as a binary random variable and the corresponding constraints are established, then the marginal association PDFs are inferred by BP.

\subsection{Initialization, Implementation, Computational Complexity}\label{subsec:Summary}
We introduce a scheme for initializing the beliefs of target joint augmented state $\bm{X}_{1:K}^{\rm{t}}$, target visibility state $\bm{S}_{1:K}$, clutter joint augmented state $\bm{X}_{1:K}^{\rm{c}}$, and clutter mixing weights $\bm{\Pi}_{1:K}$.
The target track is initialized by a two-point method.
We assume that the maximum speed of any target in  Cartesian coordinates is less than $v_{\rm max}$, and the number of consecutive missing measurements of any tracks is less than $L_{\rm max}$.
Suppose that there is a measurement generated from a target at scan $k$, then the measurement can be used to initialize a potential track with another measurement at scan $k+\delta,0 < \delta \leq L_{\rm max}$ within distance $\delta v_{\rm max}$.
We initialize the IG distributions of target mean SNR and clutter mean CNR using the maximum likelihood estimation, and let $\alpha = 3$ and $\beta =  2\sum_{n=1}^{N}m_n^2/N$, where $N$ is the total number of initial measurements generated from the corresponding target or clutter and $m_n$ is the strength of measurement $n$.
The belief of target visibility state is initialized as $b_{s}(s_{i,1}=1)=f_{\rm s}$ with $f_{\rm s}$ as the initial target visibility probability.
The parameters of the spatial distribution of the nonuniform clutter component, i.e., GW, are initialized by clustering technologies, such as variational mixture of Gaussians clustering~\cite{2006Pattern}.
Given that the number of measurements generated by each clutter component varies slowly, we initialize the clutter mixing weight parameter $\alpha_{\tau,1}$ with the initial mixing weights of clutter component $\tau$.

\begin{figure}[!htbp]
\centering
\includegraphics[scale = 0.56]{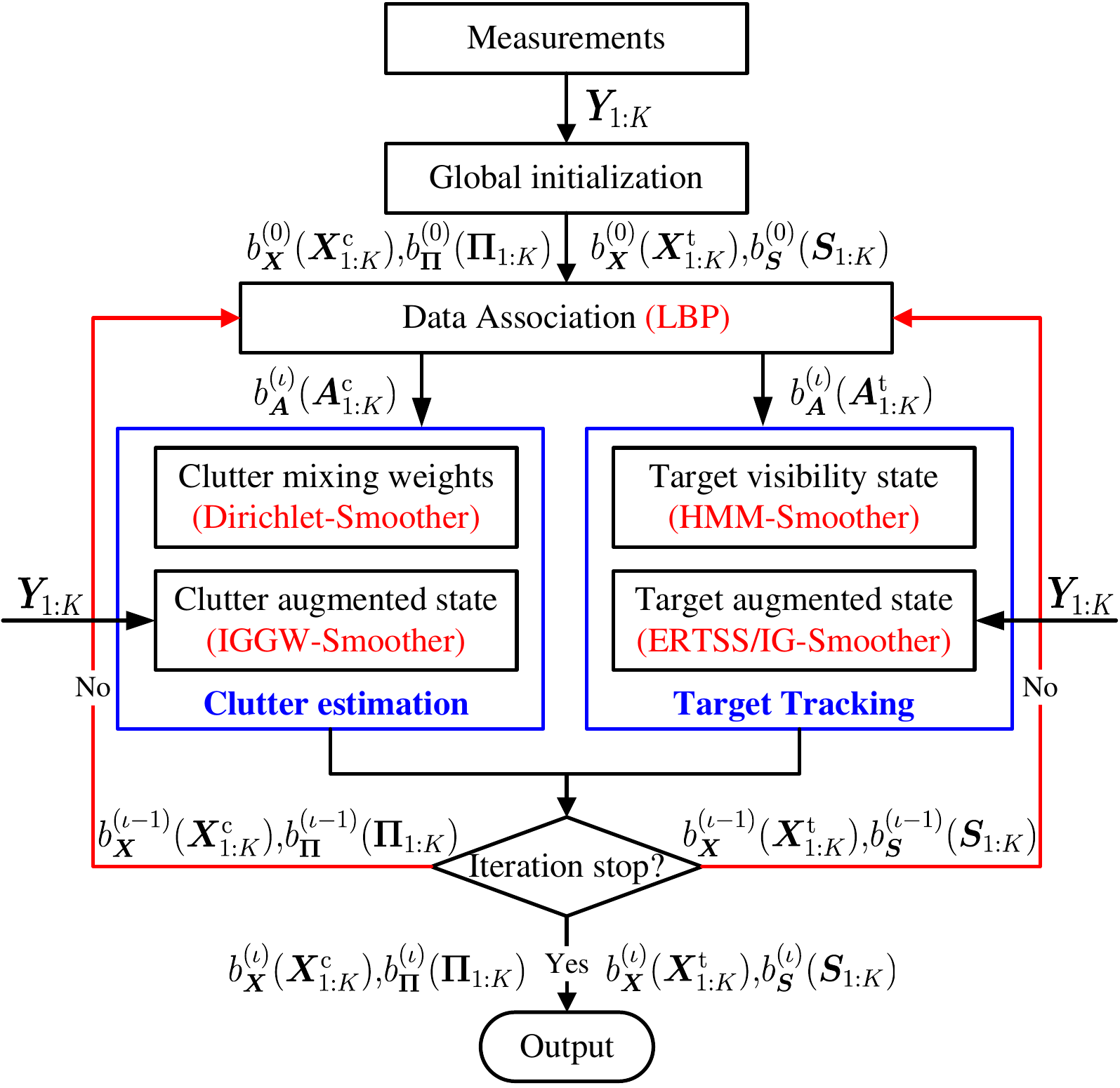}
\caption{The flowchart of MP-RMTT.}
\label{fig:Diagram}
\end{figure}

Given the initial beliefs on $\bm{X}_{1:K}^{\rm{t}}$, $\bm{S}_{1:K}$, $\bm{X}_{1:K}^{\rm{c}}$, and $\bm{\Pi}_{1:K}$, we can iteratively calculate $b_{\bm{X}}(\bm{X}_{1:K}^{\rm{t}})$, $b_{\bm{S}}(\bm{S}_{1:K})$, $b_{\bm{X}}(\bm{X}_{1:K}^{\rm{c}})$, $b_{\bm{\Pi}}(\bm{\Pi}_{1:K})$, and $b_{\bm{A}}(\bm{A}_{1:K})$ in principle by running MP described in this section.
The proposed MP algorithm is illustrated in Fig.~\ref{fig:Diagram} and summarised as \textbf{Algorithm} \ref{alg1}, which are explained as follows.
Firstly, initialize the belief $b^{(0)}_{\bm{X}}(\bm{X}_{1:K}^{\rm{t}})$, $b^{(0)}_{\bm{S}}(\bm{S}_{1:K})$, $b^{(0)}_{\bm{X}}(\bm{X}_{1:K}^{\rm{c}})$, and $b^{(0)}_{\bm{\Pi}}(\bm{\Pi}_{1:K})$.
At the $l$th iteration, the belief $b_{\bm{A}}^{(l)}(\bm{A}_{1:K}^{\rm{c}})$ is inferred using the incoming messages $b^{(l-1)}_{\bm{X}}(\bm{X}_{1:K}^{\rm{t}})$, $b^{(l-1)}_{\bm{S}}(\bm{S}_{1:K})$, $b^{(l-1)}_{\bm{X}}(\bm{X}_{1:K}^{\rm{c}})$, and $b^{(l-1)}_{\bm{\Pi}}(\bm{\Pi}_{1:K})$ calculated in the $(l-1)$th iteration, as the red lines in Fig.~\ref{fig:Diagram}.
Then $b_{\bm{A}}^{(l)}(\bm{A}_{1:K}^{\rm{c}})$ is used to infer $b^{(l)}_{\bm{X}}(\bm{X}_{1:K}^{\rm{t}})$, $b^{(l)}_{\bm{S}}(\bm{S}_{1:K})$, $b^{(l)}_{\bm{X}}(\bm{X}_{1:K}^{\rm{c}})$, and $b^{(l)}_{\bm{\Pi}}(\bm{\Pi}_{1:K})$, as the balck lines in Fig.~\ref{fig:Diagram}.
The algorithm terminates when the MP converges or when the maximum number of iterations is reached.
\begin{algorithm}[!htbp]
\caption{MP-RMTT algorithm}
\label{alg1}
\begin{algorithmic}[1]
\REQUIRE $\bm{Y} _{1:K}$, the maximum number of iterations $r_{\rm max}^{\rm MP}$;
\ENSURE $b_{\bm{X}}(\bm{X}_{1:K}^{\rm{t}})$, $b_{\bm{X}}(\bm{X}_{1:K}^{\rm{c}})$, $b_{\bm{S}}(\bm{S}_{1:K})$, $b_{\bm{\Pi}}(\bm{\Pi}_{1:K})$,  and $b_{\bm{A}}(\bm{A}_{1:K})$;
\STATE \underline{\textbf{Initialization:}} initialize $b^{(0)}_{\bm{X}}(\bm{X}_{1:K}^{\rm{t}})$, $b^{(0)}_{\bm{X}}(\bm{X}_{1:K}^{\rm{c}})$, $b^{(0)}_{\bm{S}}(\bm{S}_{1:K})$, $b^{(0)}_{\bm{\Pi}}(\bm{\Pi}_{1:K})$, the tracks number $N_T$ and clutter components number $N_C$.
\FOR{$l = 1 : r_{\rm max}^{\rm MP}$}
\STATE \underline{\textbf{Data association:}} estimate $b_{a} ^{(l)} (a_{i,j,k}^{\rm t})$ and $b_{a} ^{(l)} (a_{i,j,k}^{\rm c})$, $i=1, \ldots, N_T$, $j=1, \ldots, N_M$, $\tau=0, \ldots, N_C$, $k=1,\dots,K$ via Eq.~\eqref{equ:A2AkE21F1}-Eq.~\eqref{equ:A2AkE21F4};
\STATE \underline{\textbf{Clutter mixing weights estimation:}} calculate $b_{{\pi}}({\pi}_{\tau,k})$, $k=1,\dots,K$, $\tau=0, \ldots, N_C$ via Eq.~\eqref{Equ:beFinalT};
\STATE \underline{\textbf{Target visibility estimation:}} calculate $b_{s}(s_{i,k})$, $k=1,\dots,K$, $i=1, \ldots, N_T$, via Eq.~\eqref{Equ:beFinalTS};
\STATE \underline{\textbf{Clutter augmented state estimation:}} calculate $b_{{\bm{x}}} (\tilde{\bm{x}} ^{\rm c} _{\tau,k})$ and $b_{\sigma} (\sigma ^{\rm c} _{\tau,k} )$, $k=1,\dots,K$, $\tau=0, \ldots, N_C$, via Eq.~\eqref{eq:bKSxXC} and Eq.~\eqref{equ:bKSsigma11C}, respectively;
\STATE \underline{\textbf{Target augmented state estimation:}} calculate $b_{{\bm{x}}} ({\bm{x}} ^{\rm t} _{i,k} )$ and $b_{\sigma} (\sigma ^{\rm t} _{i,k} )$, $k=1,\dots,K$, $i=1, \ldots, N_T$, via Eq.~\eqref{eq:bKSxX} and Eq.~\eqref{equ:bKSsigmaS}, respectively.
\ENDFOR
\end{algorithmic}
\end{algorithm}

The complexity associated to the proposed MP-RMTT algorithm is discussed next.
Note that the proposed MP-RMTT is an iterative and batch process algorithm with the batch window length $K$ and the number of iterations $N_r$ among five subgraphs, and the corresponding computational cost is provided in Table.~\ref{tab:Computational Complexity} and described below.
The estimation of target joint augmented state is solved by an ERTSS and IG smoother with a computational cost $c_{x}^{\rm t}=\mathcal{O}(2KN_T)$.
The estimation of clutter joint augmented state is carried out by an IG smoother and GW smoother (referred to as IGGW smoother) with a computational cost $c_{x}^{\rm c}=\mathcal{O}(2KN_C)$.
The forward and backward algorithm is used to estimate the target visibility state and the computational cost is $c_{s}=\mathcal{O}(KN_T)$.
The estimation of clutter mixing weights is solved by a Dirichlet Smoother with a computational cost $c_{\pi}=\mathcal{O}(KN_C)$.
The data association is solved by the LBP with a computational cost $c_{a}=\mathcal{O}(KN_a(N_T+N_C)N_M)$, where $N_a$ is the maximum number of BP iterations.
Overall, the computational complexity is $c_{\rm total} = N_r(c_{x}^{\rm t}+c_{x}^{\rm c}+c_{s}+c_{\pi}+c_a)$.
\begin{table} [!htbp]
\renewcommand \arraystretch{1.19}
  \centering
  \caption{Computational complexity.}\label{tab:Computational Complexity}
\begin{tabular}{c|c|c}
  \hline
  \textbf{Hidden variables} & \textbf{Equation} & \textbf{Computational complexity} \\ \hline
  $b_{\bm{X}}(\bm{X}_{1:K}^{\rm{t}})$ & \eqref{eq:bKSxX}, \eqref{equ:bKSsigmaS} & $\mathcal{O}(2KN_rN_T)$ \\ \hline
  $b_{\bm{X}}(\bm{X}_{1:K}^{\rm{c}})$ & \eqref{eq:bKSxXC}, \eqref{equ:bKSsigma11C} & $\mathcal{O}(2KN_rN_C)$ \\ \hline
  $b_{\bm{S}}(\bm{S}_{1:K})$ & \eqref{Equ:beFinalTS} & $\mathcal{O}(KN_rN_T)$ \\ \hline
  $b_{\bm{\Pi}}(\bm{\Pi}_{1:K})$ & \eqref{Equ:beFinalT} & $\mathcal{O}(KN_rN_C)$ \\ \hline
  $b_{\bm{A}}(\bm{A}_{1:K})$ & \eqref{equ:A2AkE21F1}-\eqref{equ:A2AkE21F4} & $\mathcal{O}(KN_rN_a(N_T+N_C)N_M)$ \\ \hline
\end{tabular}
\end{table}

\section{SIMULATION AND ANALYSIS}\label{sec:EXPERIMENTAL}
\subsection{Scenario Configuration}\label{subsec:Scenario Configuration}
\subsubsection{Scenario parameters}
\begin{figure}[!htbp]
\renewcommand \arraystretch{1.25}
\centering
\includegraphics[scale = 0.6]{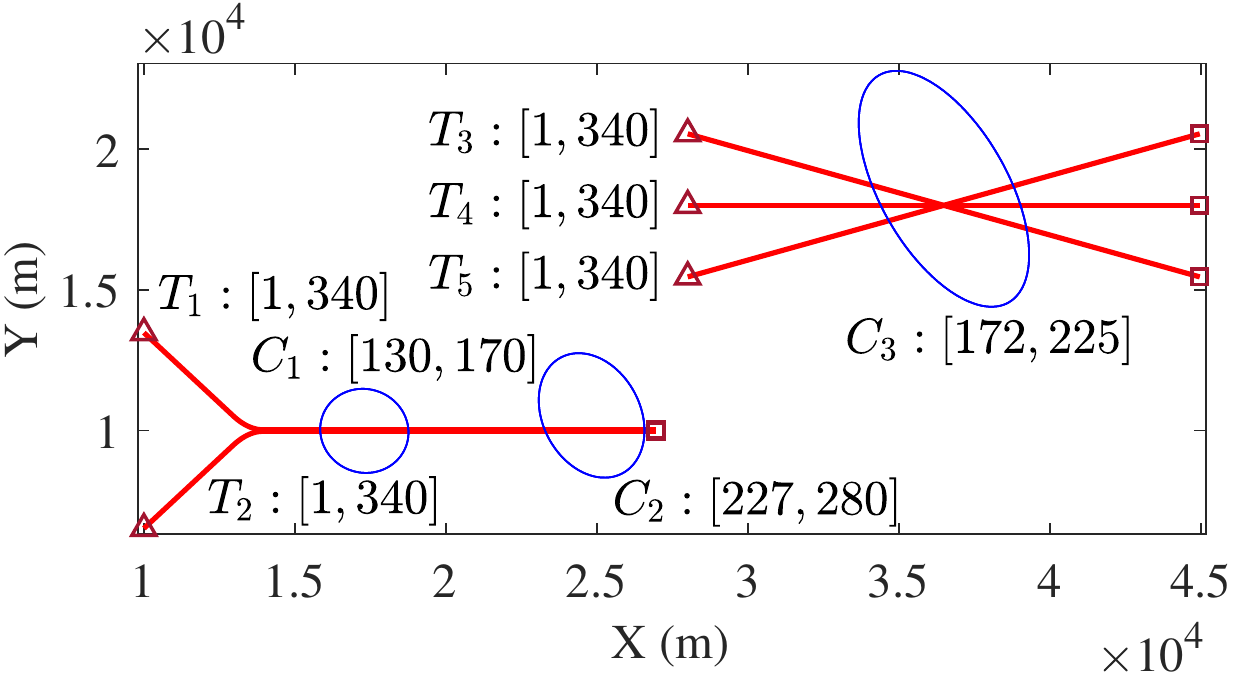}
\caption{True trajectories of targets represented by red lines and shapes of nonuniform clutter components represented by blue ellipses.
$T_i\ (C_\tau) :[k_1, k_2]$ illustrates that Target $i$ (Clutter Component $\tau$) appears and disappears at $k_1$ and $k_2$. Start/Stop positions of each target are indicated by $\vartriangle$/$\square$.}
\label{fig:True Trajectories of targets}
\end{figure}

We consider a simulation scenario with five targets together with interference from the environment and countermeasures within a two-dimensional surveillance region.
As shown in Fig.~\ref{fig:True Trajectories of targets}, all the five targets appear and disappear at $k=1$ and $k=340$, respectively.
The trajectories of Target 1 and Target 2 can be split into two intervals.
The two targets are separated spatially during the first interval.
In the second interval, both targets move in parallel at a distance of 20 m.
The other three targets (Target 3-Target 5) cross at $k=175$.
Furthermore, there are three elliptical non-uniform clutter components generated by the interference in the simulation scenario, as shown in Fig.~\ref{fig:True Trajectories of targets}.
Clutter Component 1 is generated by the chaff cloud jamming affected by Target 1 and Target 2, and Clutter Component 3 is generated by the chaff cloud jamming affected by Target 3, Target 4 and Target 5.
Clutter Component 1 appears and disappears at $k=130$ and $k=170$, respectively.
Clutter Component 3 appears and disappears at $k=172$ and $k=225$, respectively.
Clutter Component 2 is generated by the atmospheric interference, e.g., thick fog, from $k=227$ to $k=280$.
By the assumptions of high target speeds and stable atmospheric environment, the position of nonuniform clutter moves within a relatively small area and the shape of the clutter may change in different interference stages.

In the first interval of Target 1 and Target 2, both of the two targets follow the constant velocity motion moving close to each other, then follow the constant acceleration motion until the velocity in the Y-axis is zero.
In the second interval of Target 1 and Target 2, both of the two targets follow the constant velocity motion.
Since the performance evaluation focus on the second interval of Target 1 and Target 2, it is assumed  that Target 1 and Target 2 follow the constant velocity motion over all of their interval during the tracking processing for the sake of convenience.
Target 3, Target 4 and Target 5 follow constant velocity motion over all of their interval.
The parameters of the constant velocity motion are
\begin{equation}\label{equ:target parameters}
  \bm{F} = I_2 \otimes \left[
                      \begin{array}{cc}
                        1 & T \\
                        0 & 1 \\
                      \end{array}
                    \right],
  \bm{Q} = \sigma^2_v \times I_2 \otimes \left[
                      \begin{array}{cc}
                        \frac{T^4}{4} & \frac{T^3}{2} \\
                        \frac{T^3}{2} & T^2 \\
                      \end{array}
                    \right],
\end{equation}
where $T=1.25\ {\rm s}$  is the sampling period, and $\sigma^2_v = 0.01\ {\rm{m/s}}^2$ is the variance of the driving processes.
The parameters of targets are given in Table~\ref{tab:targetparam} unless noted otherwise.
\begin{table} [!htbp]
\renewcommand \arraystretch{1.25}
  \centering
  \caption{Configurations of targets.}\label{tab:targetparam}
  \begin{tabular}{c|c|c|c}
  \hline
  \textbf{ID} & \textbf{Initial state} & \textbf{Lifetime} & \textbf{SNR}\\ \hline
  $1$ & $[10000, 40, 13465, -40]^{\rm T}$  & [1, 340] & 10/3 \\ \hline
  $2$ & $[10000, 40, 6570, 40]^{\rm T}$    & [1, 340] & 50/3 \\ \hline
  $3$ & $[28000, 40, 20543, -12]^{\rm T}$   & [1, 340] & 10/3 \\ \hline
  $4$ & $[28000, 40, 18000, 0]^{\rm T}$    & [1, 340] & 50/3 \\ \hline
  $5$ & $[28000, 40, 15458, 12]^{\rm T}$    & [1, 340] & 10/3 \\ \hline
  \end{tabular}
  \begin{tablenotes}
  \footnotesize
  \item[1]$\quad\quad$ SNR$=S_{i,k}^{\rm t}/N_0$ as defined instead of being expressed in log scale.
  \end{tablenotes}
\end{table}

Each of the nonuniform clutter component is elliptical and spatially follows a two-dimensional Gaussian distribution.
The centroid and axes of the ellipse are used to determine the mean and covariance of the corresponding Gaussian distribution.
The configurations of clutter are shown in Table~\ref{tab:clutterparam}.
\begin{table} [!htbp]
  \centering
  \caption{Parameters of clutter.}\label{tab:clutterparam}
  \begin{tabular}{c|c|c|c|c|c}
  \hline
  \textbf{ID} & \textbf{Number} & \textbf{Mean} & \textbf{Covariance} & \textbf{Lifetime} & \textbf{CNR} \\ \hline
  $0$ & 30 & /                                      & /         & [1, 340] & 1\\ \hline
  $1$ & 20 & $\left[
                 \begin{array}{c}
                   \!\!\!20 \ {\rm km}\!\!\!\!\! \\
                   \!\!\!30^{\circ}\!\!\!\!\! \\
                 \end{array}
               \right]$ & $\left[\begin{array}{cc}
                    \!\!\!1 \ {\rm km}\!\!\!\! & \!\!\!0\!\!\!\!\! \\ \!\!\!0\!\!\!\! & 3^{\circ}\!\!\!\!\! \\
                \end{array}\right]^2$ & [130, 170]  & 20/3 \\ \hline
  $2$ & 20 & $\left[
                 \begin{array}{c}
                   \!\!\!27 \ {\rm km}\!\!\!\!\! \\
                   \!\!\!23^{\circ}\!\!\!\!\! \\
                 \end{array}
               \right]$ & $\left[\begin{array}{cc}
                    \!\!\!1 \ {\rm km}\!\!\!\! & \!\!\!0\!\!\!\!\! \\ \!\!\!0\!\!\!\! & 3^{\circ}\!\!\!\!\! \\
                \end{array}\right]^2$ & [227, 280]   & 10 \\ \hline
  $3$ & 30 & $\left[
                 \begin{array}{c}
                   \!\!\!41 \ {\rm km} \!\!\!\!\!\\
                   \!\!\!27^{\circ} \!\!\!\!\! \\
                 \end{array}
               \right]$ & $\left[\begin{array}{cc}
                    \!\!\!1 \ {\rm km}\!\!\!\! & \!\!\!0\!\!\!\!\! \\ \!\!\!0\!\!\!\! & 3^{\circ}\!\!\!\!\! \\
                \end{array}\right]^2$ & [172, 225]  & 20/3  \\ \hline
  \end{tabular}
  \begin{tablenotes}
  \footnotesize
  \item[1] Covariance denotes the initial covariance $\bm{{D}}_{\tau,0}^{\rm c}$ of clutter component $\tau$.
  \item[2] Number denotes the initial numbers ${\lambda_{\tau,0}}$ of clutter component $\tau$.
  \item[3] CNR$=S_{\tau,k}^{\rm c}/N_0$ as defined instead of being expressed in log scale.
  \end{tablenotes}
\end{table}

The sensor measurements consist of position information and strength information with detection threshold $d=0.715$.
The spatial measurements have zero-mean white Gaussian noise with covariance $\bm{R}=\rm{diag}(20 \ m, 0.6\degree)^2$.
The measured signal strength is sampled by the inversion method and the acceptance-rejection method for Swerling-I model and Swerling-III model~\cite{2006Pattern}, respectively.

\subsubsection{Algorithm Parameters}
The forgetting factors of target mean SNR transition PDF and clutter mean CNR transition PDF are $\mu^{\rm t} = \mu^{\rm c}=1.05$.
The forgetting factor of clutter spatial transition PDF is $\xi=0.99$.
The balance parameter of clutter mixing weight transition PDF is $\kappa=5$.
For parameters of MP, window length $K=7$ and sliding step $s=3$, $\delta_T^{\rm MP}=10^{-3}$, the maximum number of iterations $r_{\rm max}^{\rm MP}=3$.
For BP in the data association, the iterative convergence threshold $\delta_T^{\rm BP}=10^{-6}$, the maximum number of iterations $r_{\rm max}^{\rm BP}=1000$ and the damping factor $\gamma=0.9$.
The detection probability $P_{\rm d}(s_{i,k}=0)=0.01$ if the target is not visible.
The target birth probability is set to $p_{\rm b}=0.15$ and the target survival probability is set to $p_{\rm s}=0.8$.
For track initialization, the maximum speed of any target in Cartesian coordinate is $v_{\rm max}=[120 \ \rm{m/s} \ 120 \ \rm{m/s}]^{\rm T}$, and the number of consecutive missing measurements of any tracks is less than $L_{\rm max}=3$.
The initial target visibility state is $f_{\rm s}=0.5$.
The thresholds of track confirmation and deletion are set to $0.75$ and $0.5$.

\subsubsection{Performance Evaluation}
The performance metrics are given by
\begin{itemize}
  \item Correct associations rate of targets and measurements (CAR);
  \item Number of false tracks (NFT) \cite{Gorji2011};
  \item Mean optimal subpattern assignment~\cite{OSPA200} for target position estimation (MOSPA) with order and cutoff parameters as $p=2$ and $c=631$ m respectively;
  \item Relative SNR error (RSE), given by $\Delta {\sigma}_{i,k}/ {\sigma}_{i,k} $;
  \item Total number of nonuniform clutter point (TNNC);
  \item Root Mean Squares Error for nonuniform clutter position estimation (RMSE);
  \item Wasserstein distance for nonuniform clutter shape estimation (WD), given by
      $ \| \bm{x}_{k}-\bm{\hat{x}}_{k} \|^2 + {\rm tr} \Big\{ \bm{D}_{k} + \bm{\hat{D}}_{k} - 2 \sqrt{\sqrt{\bm{D}_{k}} \bm{\hat{D}}_{k} \sqrt{\bm{D}_{k}}} \Big\} $, where the square root of a matrix $\bm{D}$ is defined as the matrix $\bm{Y}$ for $\bm{D}=\bm{Y}^{\rm T}\bm{Y}$.
\end{itemize}
The values of metrics are average of 100 Monte Carlo runs.

\subsubsection{Simulation Scenarios}
We consider three scenarios to demonstrate the performance of MP-RMTT.
We outline the three scenarios in Table~\ref{tab:Simulation scenarios}.
We compare MP-RMTT against the PHD, CPHD, and CBMeMBer filter with integrated clutter estimation using measured strength information, which use the methods as in~\cite{Yang2018} with the nonuniform clutter spatial estimation method in~\cite{Liu2018}, referred to MP-RMTT, PHD, CPHD, and CBMeMBer, respectively.
We also compare the first iteration output of MP-RMTT, referred to MP-RMTT$^{\rm st}$.
In all of the algorithms, the same scenario input is used.
\begin{table} [!htbp]
\renewcommand \arraystretch{1.25}
  \centering
  \caption{Simulation scenarios.} \label{tab:Simulation scenarios}
  \begin{tabular}{c|c|c}
  \hline
  \textbf{Scenario} & \textbf{Clutter Spatial distribution} & \textbf{Nonuniform Clutter} \\ \hline
  $1$ & Uniform           & $\backslash$ \\ \hline
  $2$ & Uniform and Nonuniform & Unvarying \\ \hline
  $3$ & Uniform and Nonuniform & Varying \\ \hline
  \end{tabular}
\end{table}

\subsection{Scenario 1: Data Association Aided by AI}\label{subsec:Simulation Scenario1}
\begin{figure}[!htbp]
\centering
\includegraphics[scale = 0.5]{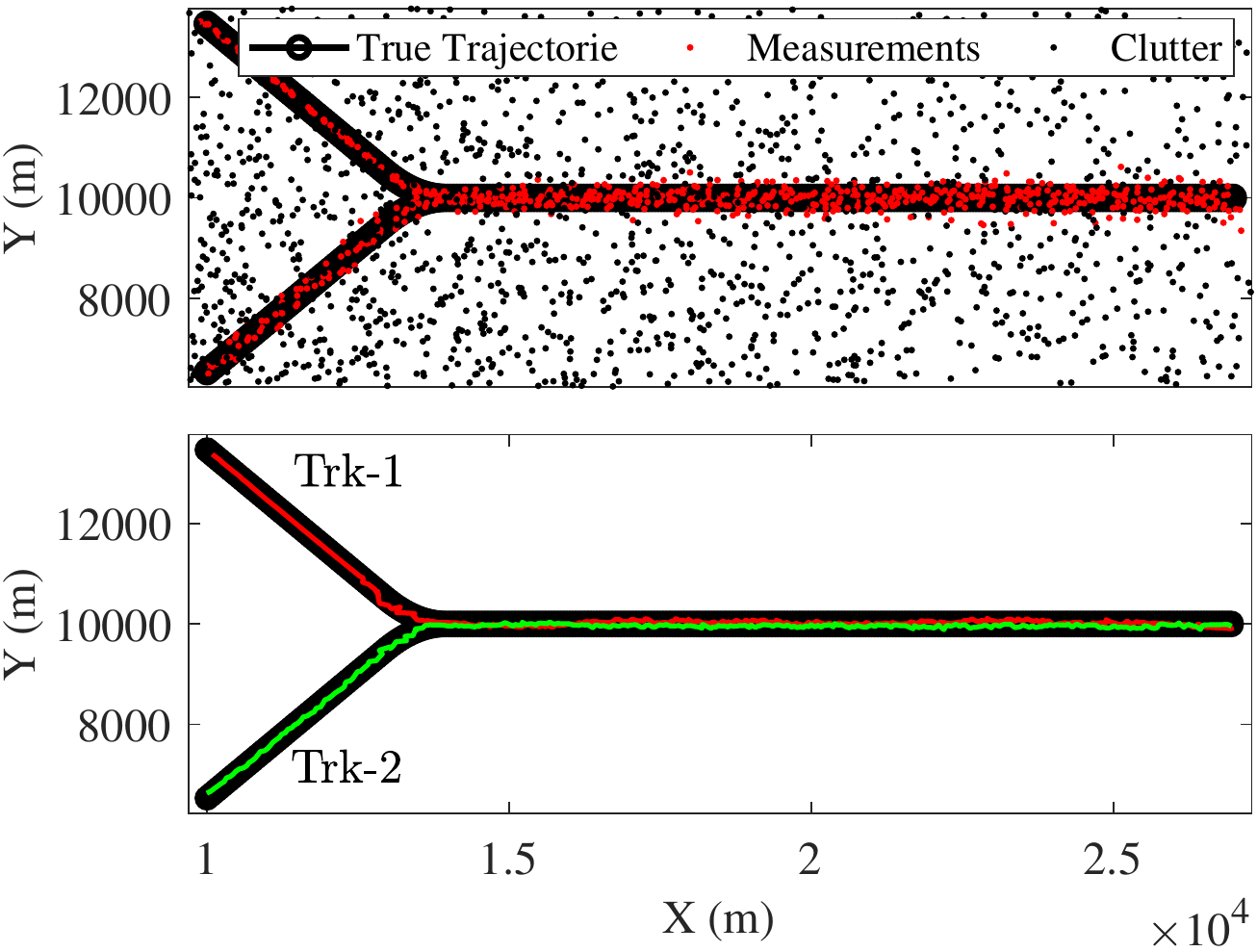}
\caption{The history of measurements generated by targets and clutter, and trajectories estimation of targets (bottom) obtained by MP-RMTT. The black and colored lines represent the target true trajectories and valid tracks, respectively. This representation is also used in the figures below.}
\label{fig:Measurement and Tracks S1}
\end{figure}

The first scenario focus on the capability of target mean SNR estimation and the discrimination ability for closely spaced targets aided by the obtained target mean SNR estimation.
There are Target 1 and Target 2, and only uniform clutter.
The history of measurements generated by targets and clutter, as well as the target trajectory estimation obtained by MP-RMTT are shown in Fig.~\ref{fig:Measurement and Tracks S1}.
During the first interval, MP-RMTT can obtain nearly stable SNR estimation for each target.
Since the trajectories of the two targets are very close in the second interval, it is obvious that the two targets cannot be distinguished by spatial measurement alone.
The tracking result in Fig.~\ref{fig:Measurement and Tracks S1} shows that the two targets can be distinguished and tracked by MP-RMTT with strength information.

\begin{figure}[!htbp]
\centering
\includegraphics[scale = 0.5]{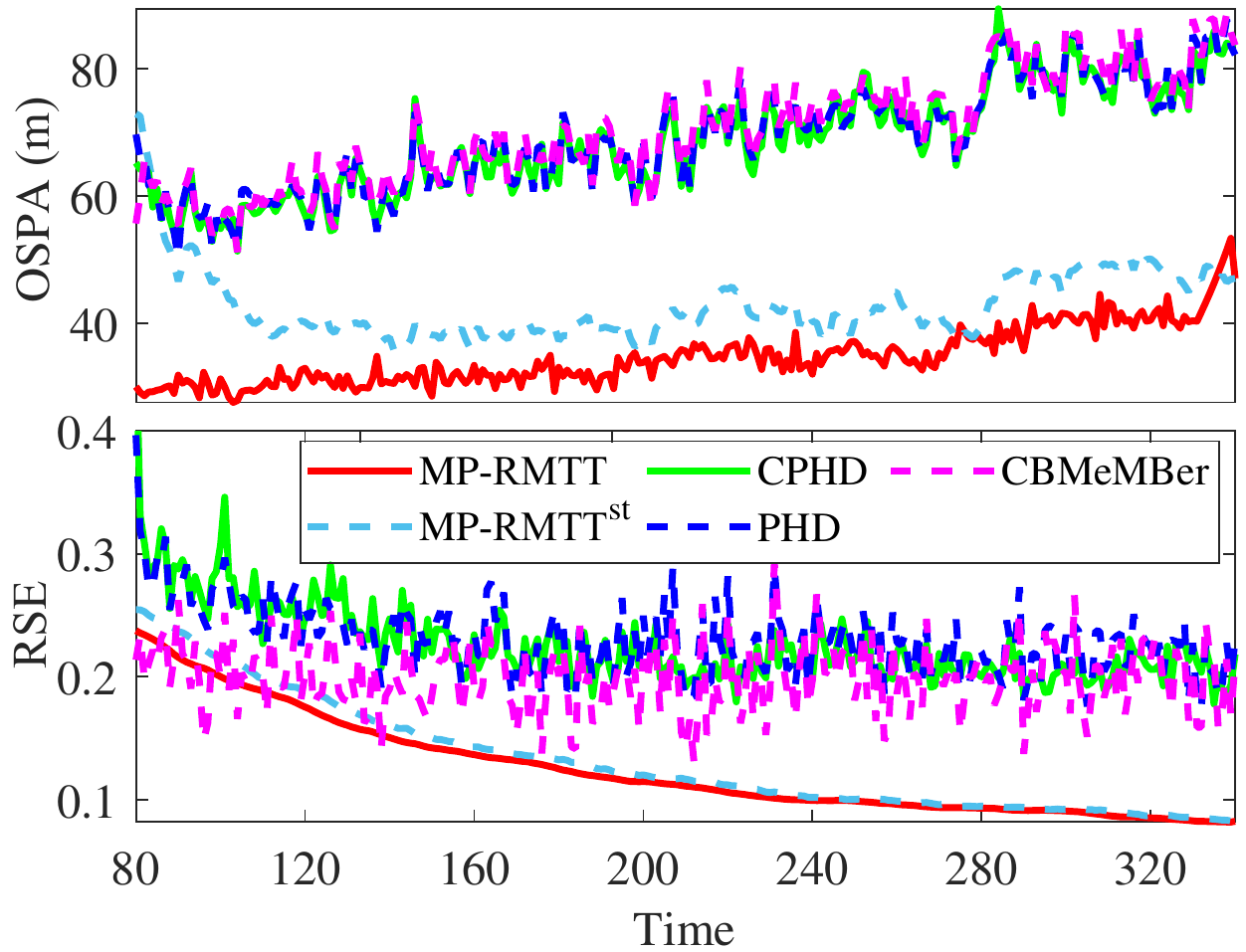}
\caption{Monte Carlo average of MOSPA distance (top) and RSE (bottom) of targets.}
\label{fig:SNR erro and OSPA S1}
\end{figure}

To analyze the performance of target mean SNR estimation and target tracking with strength information, the evolution of the MOSPA distance and RSE are evaluated, as shown in Fig.~\ref{fig:SNR erro and OSPA S1}.
It can be seen that stable SNR estimations are obtained for all algorithms as the RSE converges to a small value.
The presented results also demonstrate that MP-RMTT performs better than the PHD, CPHD, and CBMeMBer filters.

The obtained MOSPA distance and CAR results w.r.t. different target SNRs are listed in Table.~\ref{tab:table10}.
In the case of the MP-RMTT that solely exploits kinematic information (referred to MP-RMTT-KI) will inevitably lose the CAR during the second interval.
As expected, the obtained CAR take values around 50\%.
The tracking algorithm utilizing the strength information can obtain an CAR of 80\% when the SNR of Target 1 is three times larger than that of Target 2.
As the SNR of the second target increases, the ability to discriminate also improves.
The MP-RMTT with the Swerling-III model achieves almost optimal CAR with 97\% when the SNR of Target 1 is nine times larger than that of Target 2.
Note that, MP-RMTT is superior to PHD, CPHD and CBMeMBer in terms of the MOSPA metric.

\begin{table*}[!htbp]
\renewcommand \arraystretch{1.25}
\newcommand{\tabincell}[2]{\begin{tabular}{@{}#1@{}}#2\end{tabular}}
\centering
\footnotesize
\caption{\label{tab:table10} Performance comparison in different SNRs}
\begin{center}
% \resizebox{\textwidth}{10mm}
{
\begin{tabular}{ccccccccccccc}
\hline \hline
 {\multirow{3}*{\textbf{Algorithms}}} & \multicolumn{4}{c}{\tabincell{c}{$\text{SNR}=[10/3 \ 10]$ dB}} & \multicolumn{4}{c}{\tabincell{c}{$\text{SNR}=[10/3 \ 20]$ dB}} & \multicolumn{4}{c}{\tabincell{c}{$\text{SNR}=[10/3 \ 30]$ dB}} \\ \cmidrule(l){2-5}\cmidrule(l){6-9}\cmidrule(l){10-13}
 & \multicolumn{2}{c}{\tabincell{c}{Swerling-I}} & \multicolumn{2}{c}{\tabincell{c}{Swerling-II}} & \multicolumn{2}{c}{\tabincell{c}{Swerling-I}} & \multicolumn{2}{c}{\tabincell{c}{Swerling-II}} & \multicolumn{2}{c}{\tabincell{c}{Swerling-I}} & \multicolumn{2}{c}{\tabincell{c}{Swerling-II}} \\ \cmidrule(l){2-3}\cmidrule(l){4-5}\cmidrule(l){6-7}\cmidrule(l){8-9}\cmidrule(l){10-11}\cmidrule(l){12-13}
       &OSAP &CAR &OSAP&CAR &OSAP &CAR &OSAP &CAR &OSAP &CAR &OSAP&CAR \\ \hline
     MP-RMTT-KI&49.2&0.51&49.7&0.52&49.5&0.48&48.7 &0.51&51.8&0.46&48.2&0.48\\ %\hline
  MP-RMTT&{42.0}&{0.80}&\textbf{39.0}&\textbf{0.90}&{36.8}&{0.92}&\textbf{35.0} &\textbf{0.97}&{37.2} &{0.94}&\textbf{34.7}&\textbf{0.98}\\
 PHD&71.3&$\backslash$&71.6&$\backslash$&70.9&$\backslash$&70.3 &$\backslash$&69.7 &$\backslash$&70.8&$\backslash$\\
CPHD&158.5&$\backslash$&153.2&$\backslash$&159.1&$\backslash$&154.0 &$\backslash$&157.2 &$\backslash$&153.4&$\backslash$\\
CBMeMBer&80.2&$\backslash$&79.5&$\backslash$&79.3&$\backslash$&78.2 &$\backslash$&80.0 &$\backslash$&78.7&$\backslash$\\
  \hline
\end{tabular}}
\end{center}
\end{table*}

\subsection{Scenario 2: RMTT for Unvarying Clutter}\label{subsec:Simulation Scenario2}
\begin{figure}[!htbp]
\centering
\subfloat[Measurement] {\label{fig:Measurement S2} \includegraphics[scale = 0.5]{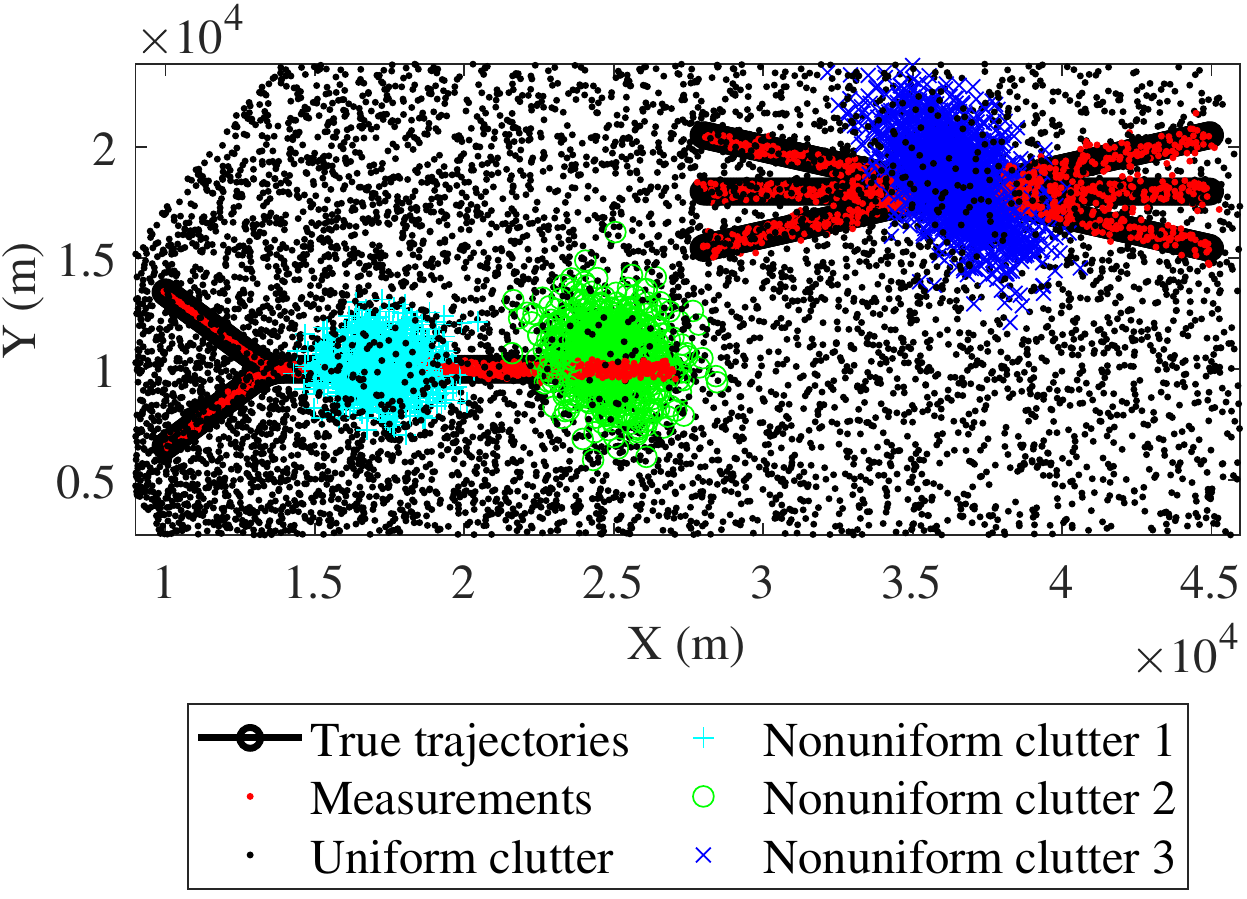}} \\
\subfloat[Trajectories and clutter estimation] {\label{fig:Tracks S2} \includegraphics[scale = 0.5]{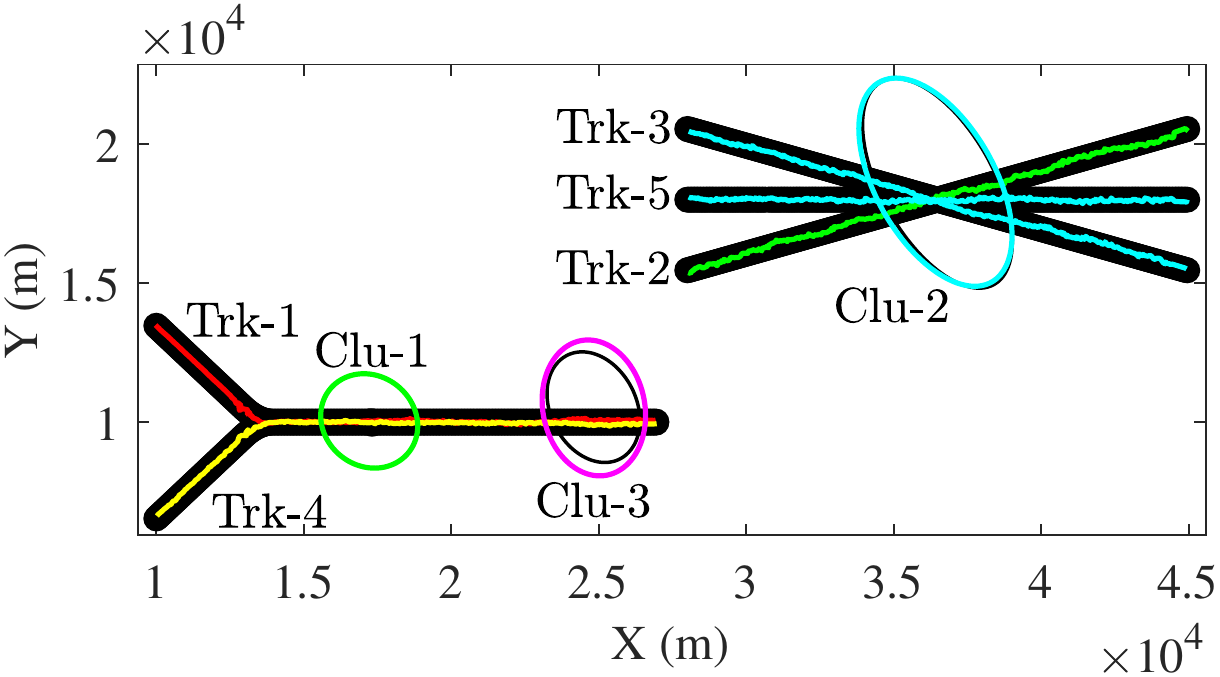}} \\
\caption{Measurements, tracks and clutter estimation obtained by MP-RMTT.}
\label{fig:Clutter points_S2}
\end{figure}

The second scenario focus on the RMTT capability in unvarying clutter background.
There are all five targets, one uniform clutter component, and three nonuniform clutter components.
The history of measurements generated by targets and clutter are shown in Fig.~\ref{fig:Measurement S2}.
The tracking and nonuniform clutter estimation results of MP-RMTT are shown in Fig.~\ref{fig:Tracks S2}.
The black and colored ellipsoids represent the covariance matrices of each nonuniform clutter component. This representation is also used in the figures below.
The presented results demonstrate that MP-RMTT tracks all five targets correctly, and no false track is generated.
Meanwhile, MP-RMTT successfully estimates the spatial and shape of the nonuniform clutter.

To reveal the performance improvements brought by the MP-RMTT, three simulation tests are performed.
In the first simulation, the spatial distribution of clutter is set to a uniform distribution where the expected number of clutter per scan is known, referred to MP-RMTT-NCE.
In the second simulation, MP-RMTT, PHD and CPHD are used for clutter estimation and target tracking, and the first iteration output of MP-RMTT is also considered.
The algorithm knows the clutter intensity exactly in the third simulation, referred to MP-RMTT-KCE.

\begin{figure}[!htbp]
\centering
\includegraphics[scale = 0.5]{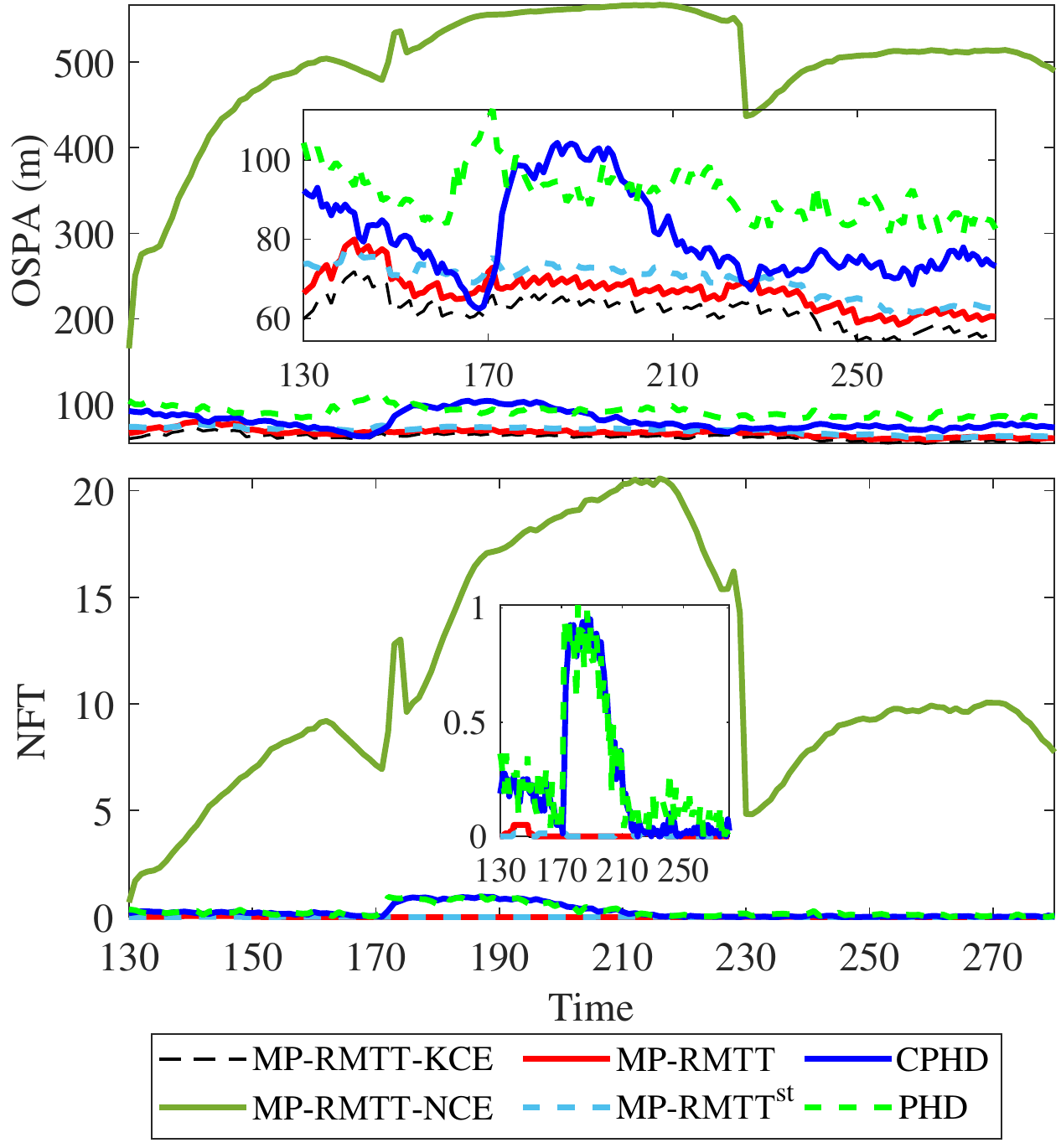}
\caption{Monte Carlo MOSPA (top) and NFT (bottom) for target tracking.}
\label{fig:target performance_S2}
\end{figure}

\begin{figure}[!htbp]
\centering
\includegraphics[scale = 0.5]{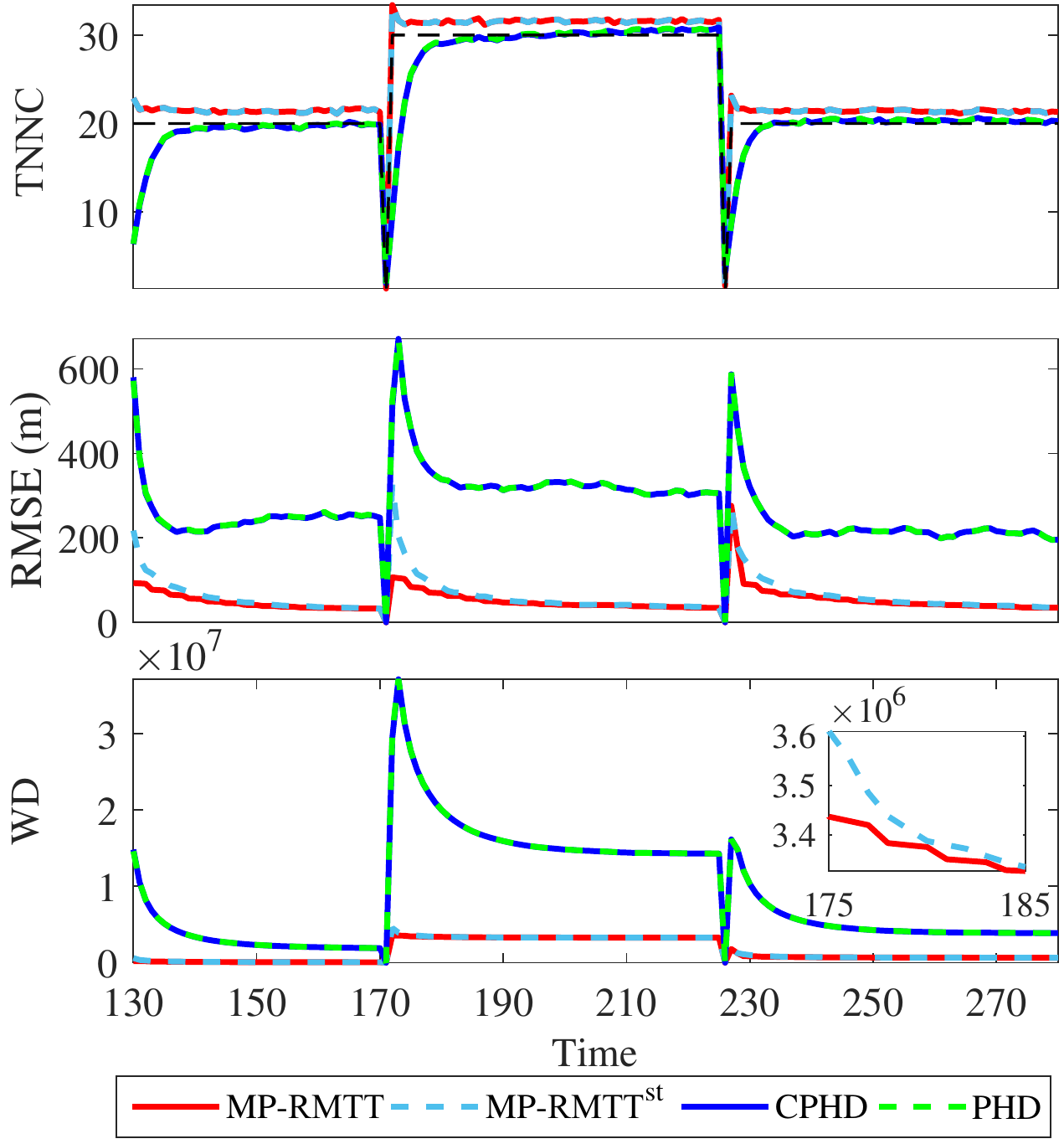}
\caption{Monte Carlo TNNC (top), RMSE (middle) and WD (bottom) for clutter estimation.}
\label{fig:clutter performance_S2}
\end{figure}

The MOSPA of target position and NFT are given in Fig.~\ref{fig:target performance_S2}.
It is clearly seen that MP-RMTT has comparable performance to algorithms with known clutter intensity perfectly.
On the contrary, the performance without clutter estimation is severely degraded.
The TNNC, RMSE and WD metrics for clutter estimation are given in Fig.~\ref{fig:clutter performance_S2}, which shows that MP-RMTT has the best clutter estimation performance.
The results indicate that the proposed MP-RMTT has comparable performance to an algorithm with a perfectly known clutter distribution, demonstrating the effectiveness and robustness of the algorithm.

\begin{figure}[!htbp]
\centering
\includegraphics[scale = 0.5]{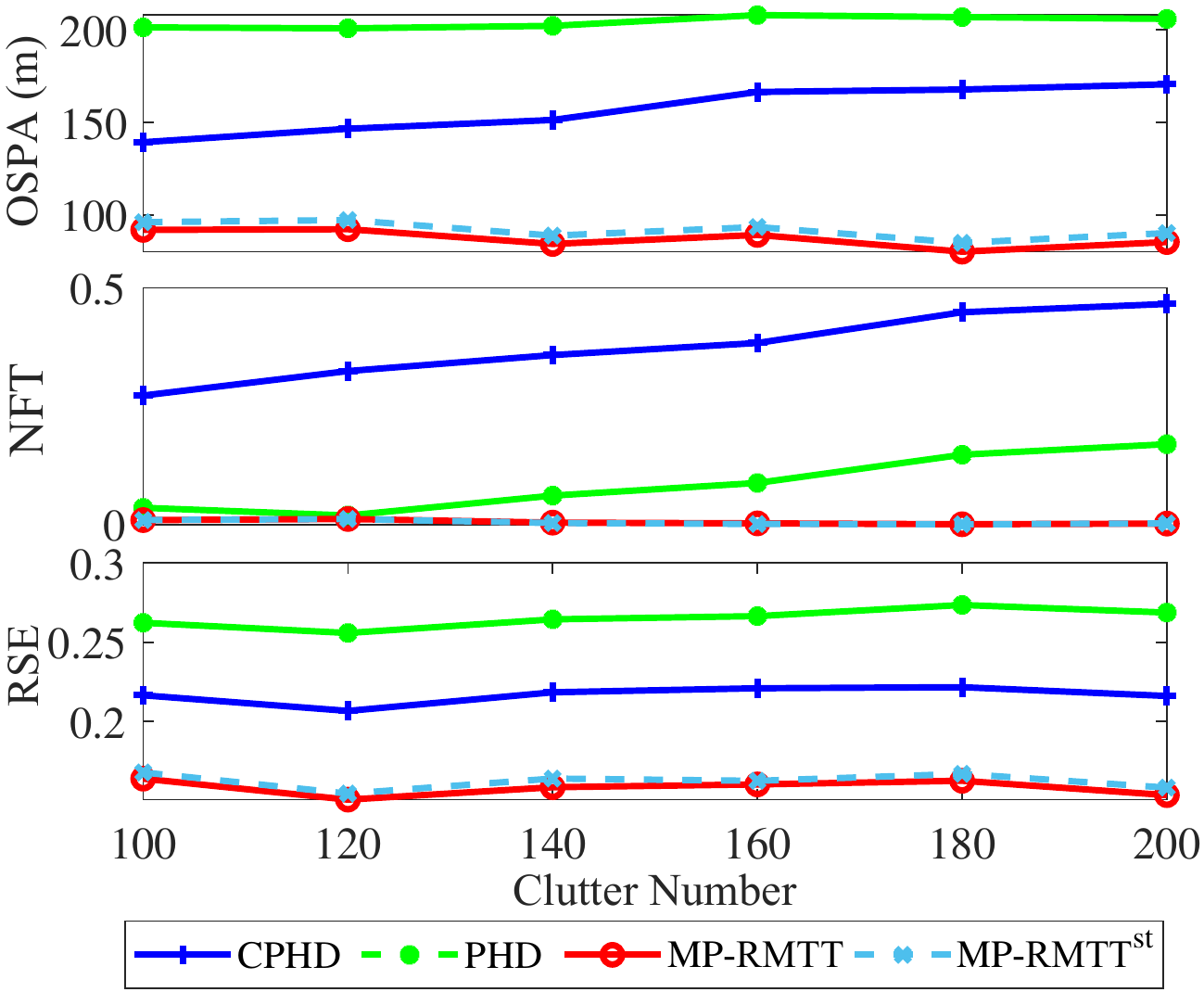}
\caption{Performance comparison of target tracking w.r.t different number of clutter.}
\label{fig:TarPer_M_S2}
\end{figure}

Fig.~\ref{fig:TarPer_M_S2} depicts the performance comparison of target tracking w.r.t. different number of clutter.
As shown in the top of Fig.~\ref{fig:TarPer_M_S2}, MP-RMTT performs better on MOSPA compared to other algorithms.
The MP-RMTT$^{\rm st}$ and MP-RMTT have comparable performance on the false track acceptance rate (as shown in the middle of Fig.~\ref{fig:TarPer_M_S2}), which are better than PHD and CPHD.
In terms of RSE (as shown in the bottom of Fig.~\ref{fig:TarPer_M_S2}), MP-RMTT is slightly better than MP-RMTT$^{\rm st}$, besides, PHD and CPHD are worst.
Meanwhile, by comparing the MOSPA and RSE of MP-RMTT$^{\rm st}$ and MP-RMTT, it is clearly seen that the performance is improved by the closed-loop process, illustrating the effectiveness of the closed-loop iterative framework.
Overall, MP-RMTT outperforms the other algorithms.
It is due to the fact that the closed-loop iterative and batch processing of MP-RMTT can improve the performance and robustness of target tracking.

\begin{figure}[!htbp]
\centering
\includegraphics[scale = 0.5]{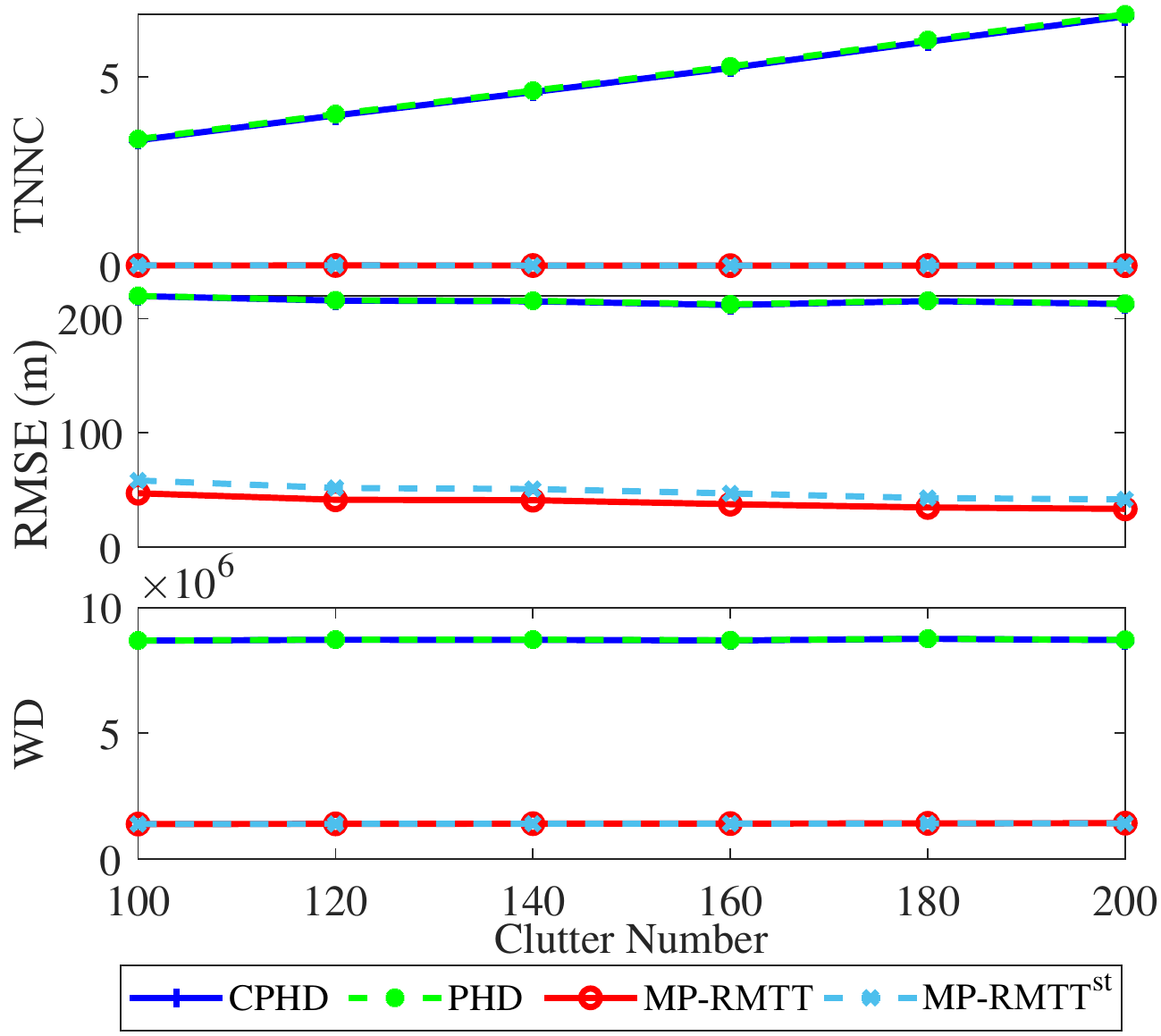}
\caption{Performance comparison of clutter estimation w.r.t different number of clutter.}
\label{fig:CltPer_M_S2}
\end{figure}

Fig.~\ref{fig:CltPer_M_S2} shows the performance comparison of clutter estimation w.r.t. different number of clutter.
The performance of MP-RMTT$^{\rm st}$ is comparable to the performance of MP-RMTT w.r.t. the number of clutter (as shown in the top of Fig.~\ref{fig:CltPer_M_S2}), which is better than PHD and CPHD.
In terms of clutter RMSE (as shown in the middle of Fig.~\ref{fig:CltPer_M_S2}), which is decreased slightly as the increase of the number of clutter, MP-RMTT outperforms the MP-RMTT$^{\rm st}$ slightly; PHD and CPHD are the worst.
The WD in the bottom of Fig.~\ref{fig:CltPer_M_S2} shows that, MP-RMTT is superior to the other two algorithms.
Because PHD and CPHD use the same clutter estimation algorithm, they have comparable performance.
MP-RMTT outperforms both PHD and CPHD.

\subsection{Scenario 3: RMTT for Varying Clutter}\label{subsec:Simulation Scenario3}
\begin{figure}[!htbp]
\centering
\subfloat[Measurement] {\label{fig:Measurement S4} \includegraphics[scale = 0.5]{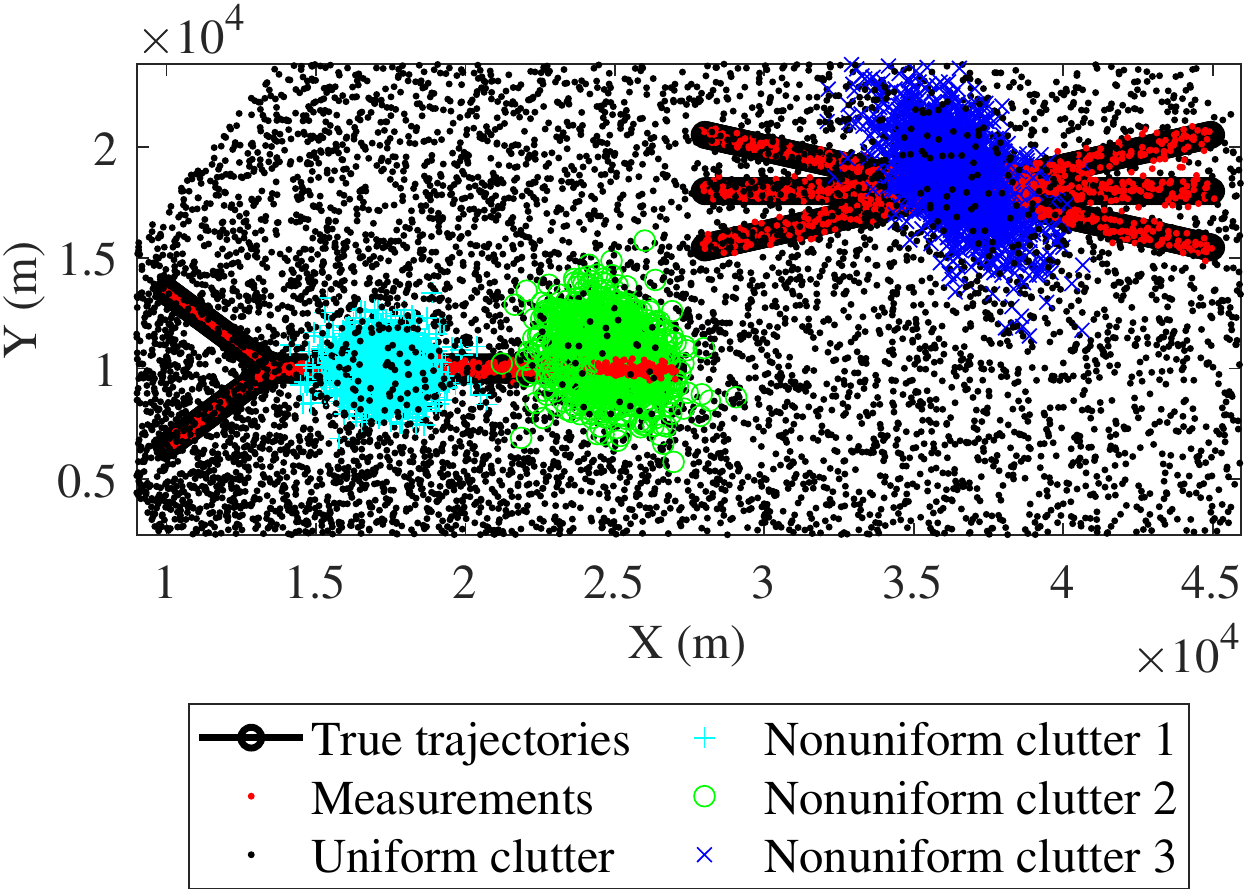}} \\
\subfloat[Trajectories and clutter estimation] {\label{fig:Tracks S4} \includegraphics[scale = 0.5]{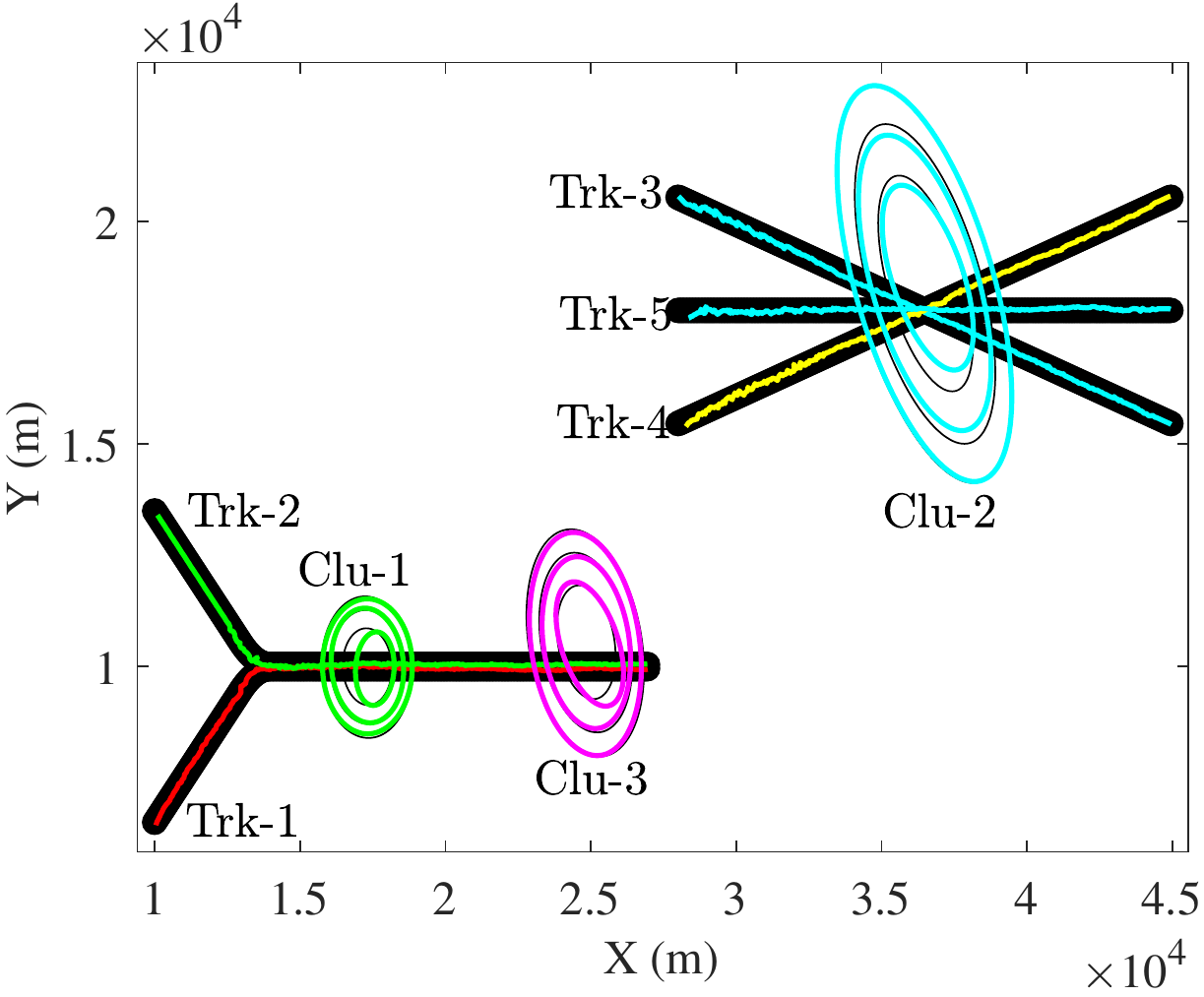}} \\
\caption{Measurements, the trajectories and nonuniform clutter estimation obtained by MP-RMTT. The three ellipses of each nonuniform clutter component from small to large represent the covariance matrix of the clutter region at $k=k^ {\tau}_{\rm start}$, $k=k^ {\tau}_{\rm start}+20$ and $k=k^ {\tau}_{\rm start}+40$ respectively, where $k^ {\tau}_{\rm start}$ is start time of the clutter component $\tau$.}
\label{fig:Clutter points_S4}
\end{figure}

The third scenario focuses on the RMTT in varying clutter background.
The scenario contains all five targets, a uniform clutter component and three nonuniform clutter components.
Assume that the number and the shape of nonuniform clutter components are time-varying and given by
\begin{equation}\label{equ:parameter of varying clutter}
\begin{split}
\lambda_{\tau,k} & =\left( 1+\frac{1}{2} \sin \Big( \frac{k-k^ {\tau}_{\rm start}}{k^ {\tau}_{\rm end}-k^ {\tau}_{\rm start}}\pi \Big) \right) {\lambda}_{\tau,0}, \\
\bm{{D}}_{\tau,k}^{\rm c} & = \Big(\frac{1}{3}+ \frac{k-k^ {\tau}_{\rm start}}{k^ {\tau}_{\rm end}-k^ {\tau}_{\rm start}} \Big) \bm{{D}}_{\tau,0}^{\rm c},
\end{split}
\end{equation}
where the initial number ${\lambda}_{\tau,0}$, initial covariance $\bm{{D}}_{\tau,0}^{\rm c}$, start time $k^ {\tau}_{\rm start}$ and end time $k^ {\tau}_{\rm end}$ of the nonuniform clutter component $\tau$ are represented in Table.~\ref{tab:clutterparam}.
The results reveal that the number of each nonuniform clutter component starts at ${\lambda}_{\tau,0}$ and then varies as a sinusoidal function, and that the shape starts at $\bm{{D}}_{\tau,0}^{\rm c}/3$ and then increases linearly.
The history of detections and the clutter estimation as well as target tracking results are shown in Fig.~\ref{fig:Clutter points_S4}.
This demonstrates that MP-RMTT can obtain target trajectory and nonuniform clutter estimates in cross and close target scenarios.

\begin{figure}[!htbp]
\centering
\includegraphics[scale = 0.5]{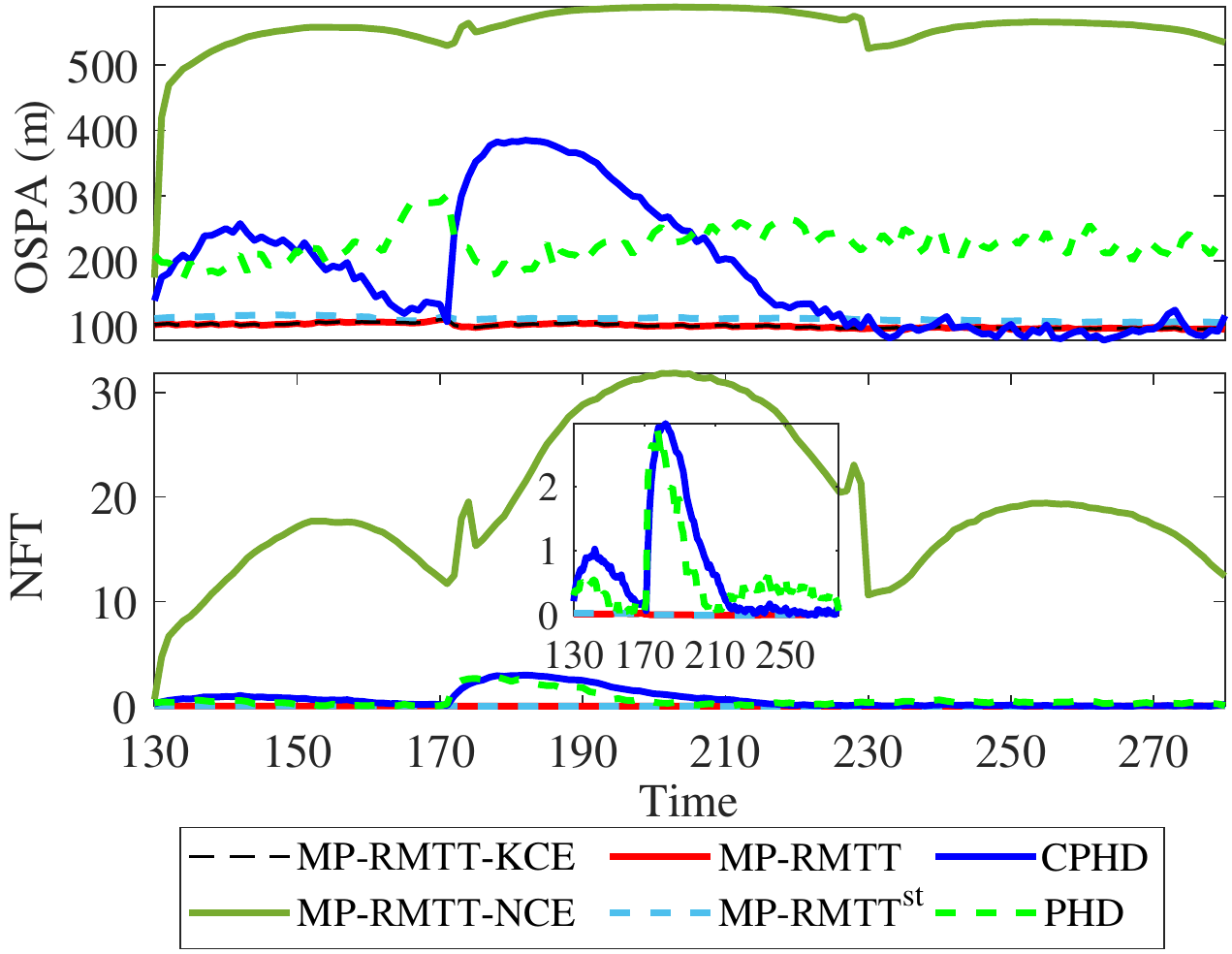}
\caption{Monte Carlo MOSPA (top) and NFT (bottom) for target tracking.}
\label{fig:TarPer_T_S3}
\end{figure}

\begin{figure}[!htbp]
\centering
\includegraphics[scale = 0.5]{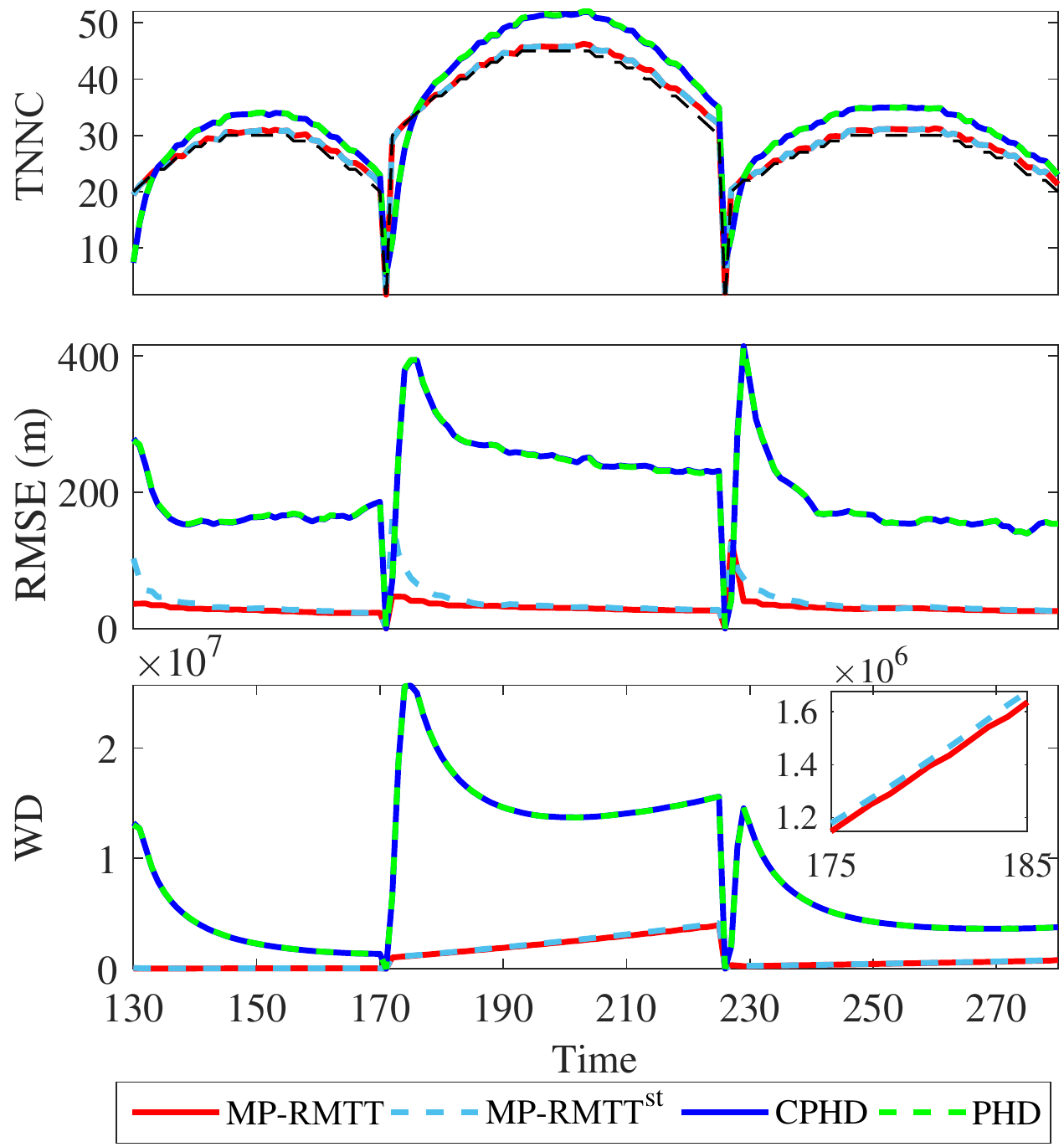}
\caption{Monte Carlo TNNC (top), RMSE (middle) and WD (bottom) for clutter estimation.}
\label{fig:CltPer_T_S3}
\end{figure}

The MOSPA of target position and NFT are given in Fig.~\ref{fig:TarPer_T_S3}.
As the same as in the second scenario, excellent performance is achieved when the clutter distribution is perfectly known, and the performance without clutter estimation is severely degraded.
It is worth noting that as the clutter varies, the NFT of PHD and CPHD increases, even up to two per scan, resulting in a dramatic increase in MOSPA.
The performance of MP-RMTT does not degrade as the clutter varies and is comparable to the optimum.
The TNNC, RMSE and WD metrics for clutter estimation are given in Fig.~\ref{fig:CltPer_T_S3}.
The TNNC metric of PHD and CPHD deteriorates as clutter varies.
It can be seen that MP-RMTT has the best clutter estimation performance.
These results demonstrate that the proposed MP-RMTT algorithm does not degrade as the clutter varies.

\begin{figure}[!htbp]
\centering
\includegraphics[scale = 0.5]{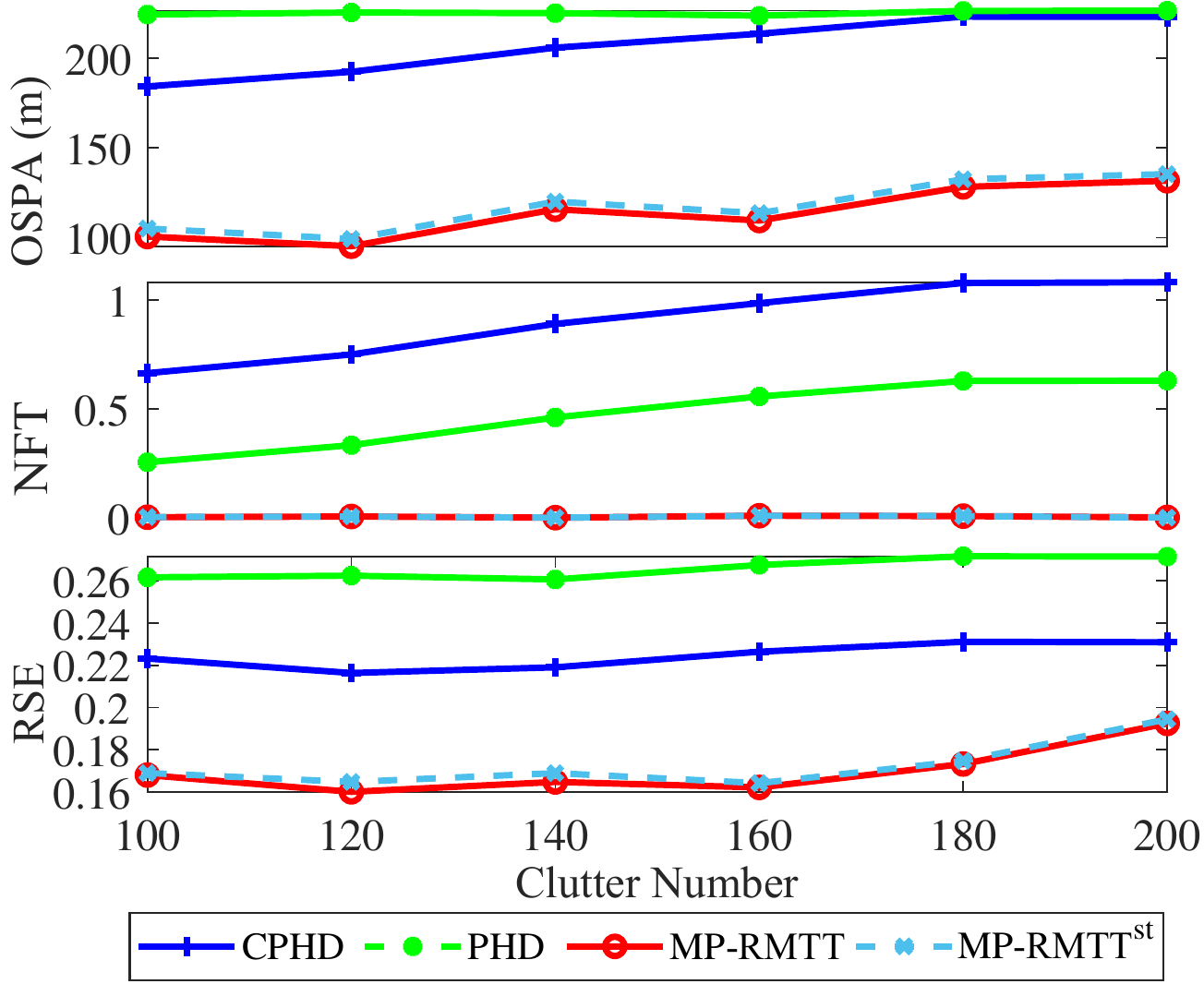}
\caption{Performance comparison of target tracking w.r.t. different number of clutter.}
\label{fig:TarPer_M_S3}
\end{figure}

\begin{figure}[!htbp]
\centering
\includegraphics[scale = 0.5]{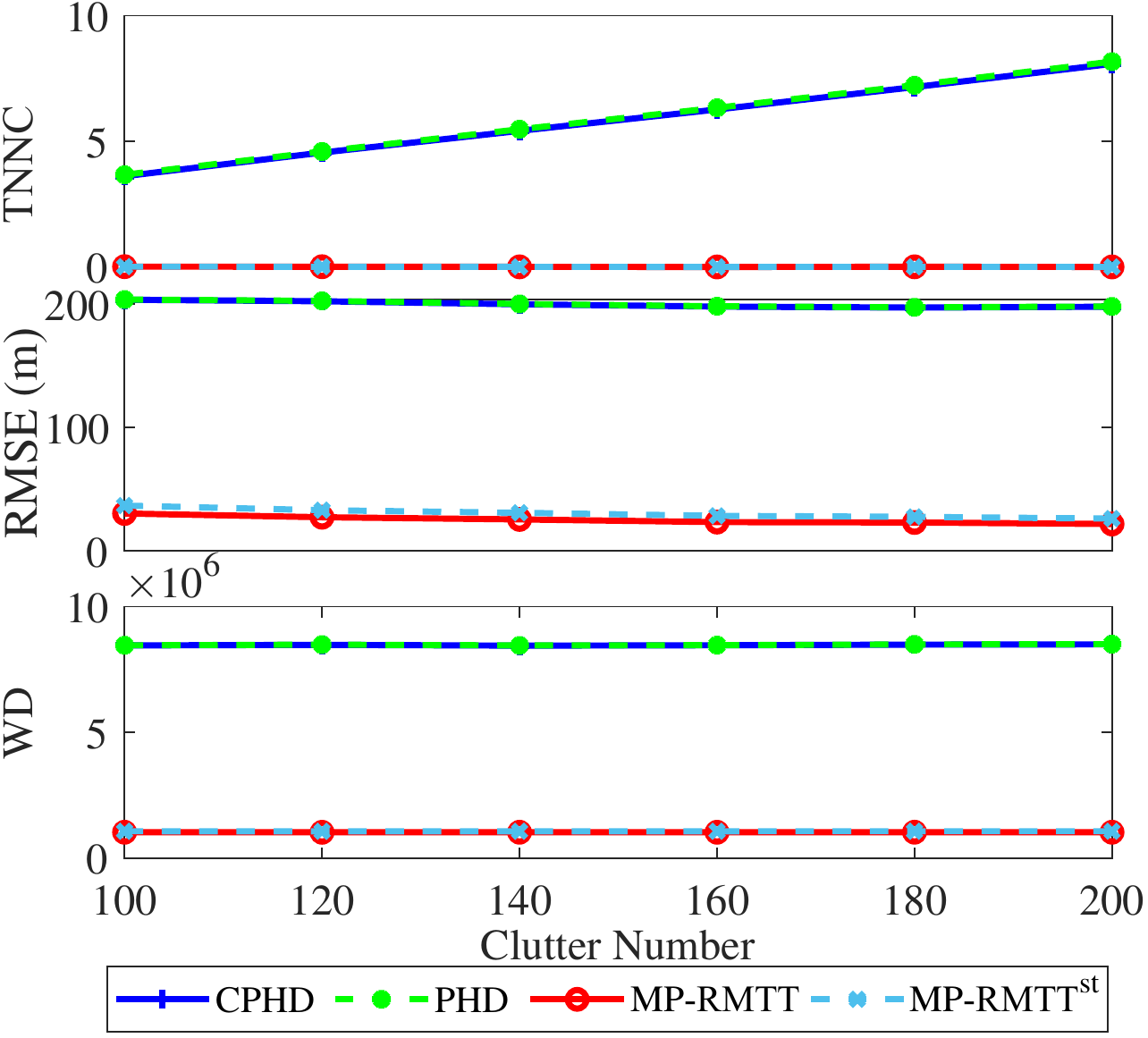}
\caption{Clutter estimation performance comparison w.r.t. different number of clutter.}
\label{fig:CltPer_M_S3}
\end{figure}

Fig.~\ref{fig:TarPer_M_S3} shows the target tracking performance w.r.t. different target SNR.
Fig.~\ref{fig:CltPer_M_S3} illustrates the clutter estimation performance w.r.t. different numbers of clutter.
Likewise in Scenario 2, the results show that MP-RMTT outperforms PHD and CPHD, and improves with the increasing number of iterations, illustrating the robustness of MP-RMTT.

\section{CONCLUSIONS}\label{sec:CONCLUSIONS}
We proposed and demonstrated the application of the MP method to the problem of RMTT by JCETT using measurements with strength information.
The proposed MP-based method exhibits outstanding and robust tracking and clutter estimation performance.
This was achieved through the establishment of a closed-loop iterative framework for target tracking and clutter estimation.
Simulation results in three scenarios with different target and clutter distributions showed that the proposed method is superior to methods without clutter estimation and previously proposed methods.
In particular, the proposed MP algorithm has a significant improvement in clutter estimation performance relative to PHD and CPHD filters.
Promising future research directions are the multi-sensor extension of the proposed method in multi-sensor tracking scenarios as well as using other feature information to discriminate between targets and clutter, such as micro-motion and polarization information.

\section*{Appendix}
\subsection{Derivation of $\mathds{E} \big[ \ln p(m_{j,k};\sigma_{i,k}^{\rm t}) \big]$ and $\mathds{E} \big[ \ln p(m_{j,k};\sigma_{i,k}^{\rm c}) \big]$} \label{sec:APP1}
Because $\mathds{E} \big[ \ln p(m_{j,k};\sigma_{i,k}^{\rm t}) \big]$ and $\mathds{E} \big[ \ln p(m_{j,k};\sigma_{i,k}^{\rm c}) \big]$ are both expectations of Rayleigh distributions, for the sake of brevity, we denote $\sigma_{i,k}^{\rm t}$ and $\sigma_{i,k}^{\rm c}$ here by $\sigma_{i,k}$.
The expectation $\mathds{E} \big[ \ln p(m_{j,k};\sigma_{i,k}) \big]$ is derived as follows.
\begin{equation}\label{equ:Yk2AkC01}
\begin{split}
  & \mathds{E} \big[ \ln p(m_{j,k};\sigma_{i,k}) \big] = \mathds{E} \big[ \ln \mathcal{R}^d(m_{j,k};\sigma_{i,k}) \big] \\
  \overset{\rm{c}}{=} & \mathds{E} \big[ (2n-1) \ln m_{j,k} - n \ln \sigma_{i,k} - n\frac{m_{j,k}^2} {\sigma_{i,k}} \big] \\
  = & (2n-1)\ln m_{j,k} - n \mathds{E} \big[\ln \sigma_{i,k} \big] - n m_{j,k}^2 \mathds{E} \big[\frac{1}{\sigma_{i,k}} \big].
\end{split}
\end{equation}
The closed-form solution of the expectation $\mathds{E} \big[\ln \sigma_{i,k} \big]$ is hard to obtain.
To this end, we use a second-order Taylor expansion of $\ln(\sigma_{i,k})$, given by
\begin{equation}\label{equ:taile}
  \ln(\sigma_{i,k}) \approx \ln(\mathds{E}[\sigma_{i,k}])+ \frac{2\sigma_{i,k}}{\mathds{E}[\sigma_{i,k}]} - \frac{\sigma_{i,k}^2}{2\mathds{E}[\sigma_{i,k}]^2} - \frac{3}{2}.
\end{equation}
Substituting Eq.~\eqref{equ:Yk2AkC01} into Eq.~\eqref{equ:taile}, yields
\begin{equation}\label{equ:Yk2AkC02}
\begin{split}
  \mathds{E} \big[ & \ln p(m_{j,k};\sigma_{i,k}) \big] = (2n-1)\ln m_{j,k} - \\
  & n\ln(\mathds{E}[\sigma_{i,k}]) + \frac{n\mathds{E}[\sigma_{i,k}^2]}{2\mathds{E}[\sigma_{i,k}]^2} -  \mathds{E} \Big[ \frac{n m_{j,k}^2}{\sigma_{i,k}} \Big] - \frac{n}{2},
\end{split}
\end{equation}
where,
\begin{equation}\label{equ:Yk2AkC03}
\begin{split}
  & \mathds{E}[\sigma_{i,k}^{-1}]=\frac{\alpha_{i,k}}{\beta_{i,k}},\quad
\mathds{E}[\sigma_{i,k}]=\frac{\beta_{i,k}}{\alpha_{i,k}-1}, \\
  & \mathds{E}[\sigma_{i,k}^2]=\frac{\beta_{i,k}^2}{(\alpha_{i,k}-1) (\alpha_{i,k}-2)}.
\end{split}
\end{equation}
Substituting Eq.~\eqref{equ:Yk2AkC03} into Eq.~\eqref{equ:Yk2AkC02}, yields
\begin{equation}\label{equ:Yk2AkC04}
\begin{split}
    & \mathds{E} \big[ \ln p(m_{j,k};\sigma_{i,k}) \big] \\
  = & (2n-1)\ln m_{j,k} - n m_{j,k}^2 \frac{\alpha_{i,k}}{\beta_{i,k}} +\frac{n }{2(\alpha_{i,k}-2)}.
\end{split}
\end{equation}

\subsection{Derivation of Data Association}\label{sec:APP3}
We find that it is convenient to define the following messages
\begin{equation}\label{equ:APP31}
\begin{split}
  & \theta_{i,j,k}^{\rm t} = \frac{{m}^{\rm MF}_{{f_{\bm{Y}_{k}} \rightarrow a_{i,j,k}^{\rm t}}}(1)  {m}^{\rm BP}_{{f_{\bm{A}_{k}^{\rm t}} \rightarrow a_{i,j,k}^{\rm t}}}(1)} {{m}^{\rm MF}_{{f_{\bm{Y}_{k}} \rightarrow a_{i,j,k}^{\rm t}}}(0)  {m}^{\rm BP}_{{f_{\bm{A}_{k}^{\rm t}} \rightarrow a_{i,j,k}^{\rm t}}}(0)},\\
  & \theta_{\tau,j,k}^{\rm c} = \frac{{m}^{\rm MF}_{{f_{\bm{Y}_{k}} \rightarrow a_{\tau,j,k}^{\rm c}}}(1)  {m}^{\rm MF}_{{f_{\bm{A}_{k}^{\rm c}} \rightarrow a_{\tau,j,k}^{\rm c}}}(1)} {{m}^{\rm MF}_{{f_{\bm{Y}_{k}} \rightarrow a_{\tau,j,k}^{\rm c}}}(0)  {m}^{\rm MF}_{{f_{\bm{A}_{k}^{\rm c}} \rightarrow a_{\tau,j,k}^{\rm c}}}(0)},\\
\end{split}
\end{equation}
\begin{equation}\label{equ:APP32}
\begin{split}
  & \alpha_{i,j,k}^{\rm t} = \frac{{m}^{\rm BP}_{{f_{\bm{I}_{k}} \rightarrow a_{i,j,k}^{\rm t}}}(1)} {{m}^{\rm BP}_{{f_{\bm{I}_{k}} \rightarrow a_{i,j,k}^{\rm t}}}(0)}, \alpha_{\tau,j,k}^{\rm c} = \frac{{m}^{\rm BP}_{{f_{\bm{I}_{k}} \rightarrow a_{\tau,j,k}^{\rm c}}}(1)} {{m}^{\rm BP}_{{f_{\bm{I}_{k}} \rightarrow a_{\tau,j,k}^{\rm c}}}(0)}, \\
  & \rho_{i,j,k}^{\rm t} = \frac{{n}_{{a_{i,j,k}^{\rm t} \rightarrow f_{\bm{I}_{i,k}}}}(1)} {{n}_{{a_{i,j,k}^{\rm t} \rightarrow f_{\bm{I}_{i,k}}}}(0)}, \rho_{\tau,j,k}^{\rm c} = \frac{{n}_{{a_{\tau,j,k}^{\rm c} \rightarrow f_{\bm{I}_{\tau,k}}}}(1)} {{n}_{{a_{\tau,j,k}^{\rm c} \rightarrow f_{\bm{I}_{\tau,k}}}}(0)},\\
\end{split}
\end{equation}
\begin{equation}\label{equ:APP33}
\begin{split}
  & \eta_{i,j,k}^{\rm t} = \frac{{m}^{\rm BP}_{{f_{\bm{E}_{k}^{\rm t}} \rightarrow a_{i,j,k}^{\rm t}}}(1)} {{m}^{\rm BP}_{{f_{\bm{E}_{k}^{\rm t}} \rightarrow a_{i,j,k}^{\rm t}}}(0)}, \eta_{\tau,j,k}^{\rm c} = \frac{{m}^{\rm BP}_{{f_{\bm{E}_{k}^{\rm c}} \rightarrow a_{\tau,j,k}^{\rm c}}}(1)} {{m}^{\rm BP}_{{f_{\bm{E}_{k}^{\rm c}} \rightarrow a_{\tau,j,k}^{\rm c}}}(0)}, \\
  & \beta_{i,j,k}^{\rm t} = \frac{{n}_{{a_{i,j,k}^{\rm t} \rightarrow f_{\bm{E}_{i,k} ^{\rm t}}}}(1)} {{n}_{{a_{i,j,k}^{\rm t} \rightarrow f_{\bm{E}_{i,k} ^{\rm t}}}}(0)}, \beta_{\tau,j,k}^{\rm c} = \frac{{n}_{{a_{\tau,j,k}^{\rm c} \rightarrow f_{\bm{E}_{\tau,k} ^{\rm c}}}}(1)} {{n}_{{a_{\tau,j,k}^{\rm c} \rightarrow f_{\bm{E}_{\tau,k} ^{\rm c}}}}(0)}.
\end{split}
\end{equation}

Using the definition in Eq.~\eqref{equ:APP32}, $\alpha_{i,j,k}^{\rm t}$ and $\alpha_{i,j,k}^{\rm c}$ can be calculated as follows.
\begin{equation}\label{equ:APP34}
\begin{split}
& \alpha_{i,j,k}^{\rm t} \!=\!\! \frac{\Gamma_{i,j,k}^{\rm t}} {\Upsilon_{i,j,k}^{\rm t} \!+\! \Phi_{i,j,k}^{\rm t}} \!\!=\!\! \frac{1} {\sum_{i'=1 \backslash i}^{N_T} \rho_{i',j,k}^{\rm t} \!+\! \sum_{\tau=0}^{N_C} \rho_{\tau,j,k}^{\rm c}}, \\
& \alpha_{\tau,j,k}^{\rm c} \!=\! \frac{\Gamma_{\tau,j,k}^{\rm c}} {\Upsilon_{\tau,j,k}^{\rm c} \!\!+\!\! \Phi_{\tau,j,k}^{\rm c}} \!\!=\!\! \frac{1} {\sum_{i=1}^{N_T} \rho_{i,j,k}^{\rm t} \!\!+\!\! \sum_{\tau'=0 \backslash \tau}^{N_C} \rho_{\tau',j,k}^{\rm c}},
\end{split}
\end{equation}
where
\begin{equation}\label{equ:APP35}
\begin{split}
& \Gamma_{i,j,k}^{\rm t}=\prod_{i'=1 \backslash i} ^{N_T} \rho_{i',j,k}^{\rm t}(0) \prod_{\tau=0} ^{N_C} \rho_{\tau, j, k} ^{\rm c}(0), \\
& \Upsilon_{i,j,k}^{\rm t} \!=\!\!\! \sum_{i'=1 \backslash i}^{N_T} \rho_{i',j,k}^{\rm t}(1) \!\! \prod_{i''=1 \backslash i,i'} ^{N_T} \!\! \rho_{i'',j,k}^{\rm t}(0) \prod_{\tau=0} ^{N_C} \rho_{\tau,j,k}^{\rm c}(0), \\
& \Phi_{i,j,k}^{\rm t}=\sum_{\tau=0}^{N_C} \rho_{\tau,j,k}^{\rm c}(1) \prod_{i'=1 \backslash i} ^{N_T} \rho_{i',j,k}^{\rm t}(0) \prod_{\tau'=0 \backslash \tau } ^{N_C} \rho_{\tau',j,k}^{\rm c}(0), \\
& \Gamma_{\tau,j,k}^{\rm c}=\prod_{i=1} ^{N_T} \rho_{i, j, k} ^{\rm t}(0) \prod_{\tau'=0 \backslash \tau} ^{N_C} \rho_{\tau',j,k}^{\rm c}(0), \\
& \Upsilon_{\tau,j,k}^{\rm c} \!=\! \sum_{i=1}^{N_T} \rho_{i,j,k}^{\rm t}(1) \prod_{i'=1 \backslash i} ^{N_T} \rho_{i',j,k}^{\rm t}(0) \prod_{\tau'=1 \backslash \tau} ^{N_T} \rho_{\tau',j,k}^{\rm c}(0), \\
& \Phi_{\tau,j,k}^{\rm c}= \!\!\!\! \sum_{\tau'=0 \backslash \tau}^{N_C} \rho_{\tau',j,k}^{\rm c}(1) \prod_{i=1} ^{N_T} \rho_{i,j,k}^{\rm t}(0) \!\! \!\! \prod_{\tau''=0 \backslash \tau,\tau' } ^{N_C} \!\!\!\! \rho_{\tau'',j,k}^{\rm c}(0).
\end{split}
\end{equation}

By Eq.~\eqref{equ:A2AkI2}, Eq.~\eqref{equ:APP32} and Eq.~\eqref{equ:APP33}, we have
\begin{equation}\label{equ:APP36}
\begin{split}
& \beta_{i,j,k}^{\rm t} = \theta_{i,j,k}^{\rm t}\alpha_{i,j,k}^{\rm t}, \ \beta_{\tau,j,k}^{\rm c} = \theta_{\tau,j,k}^{\rm c}\alpha_{\tau,j,k}^{\rm c}, \\
& \rho_{i,j,k}^{\rm t} = \theta_{i,j,k}^{\rm t}\eta_{i,j,k}^{\rm t}, \ \rho_{\tau,j,k}^{\rm c} = \theta_{\tau,j,k}^{\rm c}\eta_{\tau,j,k}^{\rm c}.
\end{split}
\end{equation}
Substituting Eq.~\eqref{equ:APP36} into Eq.~\eqref{equ:APP34}, we obtain
\begin{equation}\label{equ:APP37}
\begin{split}
  \beta_{i,j,k}^{\rm t} =  \frac{\theta_{i,j,k}^{\rm t}}
  {\theta_{0,j,k}^{\rm c} +\!\! \sum\limits_{i' = 1 \backslash i}^{N_T} \theta_{i',j,k}^{\rm t} \eta_{i',j,k}^{\rm t} +\!\! \sum\limits_{\tau = 1}^{N_C} \theta_{\tau,j,k}^{\rm c} \eta_{\tau,j,k}^{\rm c}}, \\
  \beta_{\tau,j,k}^{\rm c} \!\!=\!\!
  \frac{\theta_{\tau,j,k}^{\rm c}}
  {\theta_{0,j,k}^{\rm c} \!\!+\!\! \sum\limits_{i = 1}^{N_T} \theta_{i,j,k}^{\rm t} \eta_{i,j,k}^{\rm t} \!+\!\! \sum\limits_{\tau' = 1 \backslash \tau}^{N_C} \theta_{\tau',j,k}^{\rm c} \eta_{\tau',j,k}^{\rm c}}.
\end{split}
\end{equation}

Using the definition in Eq.~\eqref{equ:APP33}, $\eta_{i,j,k}^{\rm t}$ and $\eta_{i,j,k}^{\rm c}$ can be calculated as follows
\begin{equation}\label{equ:A2AkE222}
  \begin{split}
    \eta_{i,j,k}^{\rm t} = & \frac{\prod_{j'=0 \backslash j} ^{N_{M,k}} \beta_{i,j',k}^{\rm t}(0)} {\sum _{j'=0\backslash j}^{N_{M,k}} \beta_{i,j',k}^{\rm t}(1) \prod_{j''=0 \backslash j,j'} ^{N_{M,k}} \beta_{i,j'',k}^{\rm t}(0) } \\
    = & \frac{1} {\sum _{j'=0\backslash j}^{N_{M,k}} \beta_{i,j',k}^{\rm t}},
  \end{split}
\end{equation}
\begin{equation}\label{equ:A2AkE3}
  \begin{split}
    \eta_{\tau,j,k}^{\rm c} & = \frac{\Psi_{\tau,j,k}^{\rm c}} {\beta_{\tau,0,k}^{\rm c}(1)\prod_{j'=1 \backslash j} ^{N_{M,k}} \beta_{\tau,j,k}^{\rm c}(0)+\Psi_{\tau,j,k}^{\rm c}} \\
    & = \frac{\prod_{j' = 1 \backslash j }^{N_{M,k}} {(1 + \beta_{\tau,j',k}^{\rm c})}}{\beta_{\tau,0,k}^{\rm c} +  \prod_{j' = 1 \backslash j}^{N_{M,k}} {(1 + \beta_{\tau,j',k}^{\rm c})}},
  \end{split}
\end{equation}
\begin{equation}\label{equ:A2AkE4}
\eta_{\tau,0,k}^{\rm c} \!\!=\!\! \frac{\prod_{j'=1} ^{N_{M,k}} \beta_{\tau,j',k}^{\rm c}(0)} {\sum_{{\bm{A} _i^{\rm{c}}} \backslash a_{\tau,0,k}^{\rm c} } \! \prod_{j'=1} ^{N_{M,k}} \!\! \beta_{\tau,j',k}^{\rm c}} \!\!=\!\! \frac{1} {\prod_{j'=1} ^{N_{M,k}} (1\!\!+\!\!\beta_{\tau,j',k}^{\rm c})},
\end{equation}
where
\begin{equation}\label{equ:A2AkE3--where}
 \Psi_{\tau,j,k}^{\rm c} = \beta_{\tau,0,k}^{\rm c}(0) \sum_{{\bm{A} _i^{\rm{c}}} \backslash a_{\tau,j,k}^{\rm c},{a}_{\tau,0,k}^{\rm c} } \prod_{j'=1 \backslash j} ^{N_{M,k}} \beta_{\tau,j',k}^{\rm c}.
\end{equation}

\bibliographystyle{IEEEtran}
\bibliography{IEEEabrv,MTT-AI-MP}

\begin{thebibliography}{10}
\providecommand{\url}[1]{#1}
\csname url@samestyle\endcsname
\providecommand{\newblock}{\relax}
\providecommand{\bibinfo}[2]{#2}
\providecommand{\BIBentrySTDinterwordspacing}{\spaceskip=0pt\relax}
\providecommand{\BIBentryALTinterwordstretchfactor}{4}
\providecommand{\BIBentryALTinterwordspacing}{\spaceskip=\fontdimen2\font plus
\BIBentryALTinterwordstretchfactor\fontdimen3\font minus
  \fontdimen4\font\relax}
\providecommand{\BIBforeignlanguage}[2]{{%
\expandafter\ifx\csname l@#1\endcsname\relax
\typeout{** WARNING: IEEEtran.bst: No hyphenation pattern has been}%
\typeout{** loaded for the language `#1'. Using the pattern for}%
\typeout{** the default language instead.}%
\else
\language=\csname l@#1\endcsname
\fi
#2}}
\providecommand{\BIBdecl}{\relax}
\BIBdecl

\bibitem{Richards2010}
M.~A. Richards, J.~Scheer, W.~A. Holm, and W.~L. Melvin, \emph{Principles of
  modern radar: basic principles}.\hskip 1em plus 0.5em minus 0.4em\relax NC
  Raleigh: SciTech Publising, 2010.

\bibitem{Bar1995Multitarget}
Y.~Bar-Shalom and X.~Li, \emph{Multitarget-multisensor tracking: principles and
  techniques}.\hskip 1em plus 0.5em minus 0.4em\relax YBS publishing, Storrs,
  1995.

\bibitem{Mahler2011}
R.~P. Mahler, B.~T. Vo, and B.~N. Vo, ``{CPHD filtering with unknown clutter
  rate and detection profile},'' \emph{IEEE Transactions on Signal Processing},
  vol.~59, no.~8, pp. 3497--3513, 2011.

\bibitem{Beard2013}
M.~Beard, B.~T. Vo, and B.~N. Vo, ``{Multitarget filtering with unknown clutter
  density using a bootstrap GMCPHD filter},'' \emph{IEEE Signal Processing
  Letters}, vol.~20, no.~4, pp. 323--326, 2013.

\bibitem{Kim2017}
W.~C. Kim and T.~L. Song, ``{Interactive clutter measurement density estimator
  for multitarget data association},'' \emph{IET Radar, Sonar and Navigation},
  vol.~11, no.~1, pp. 125--132, 2017.

\bibitem{Vo2013}
B.~T. Vo, B.~N. Vo, R.~Hoseinnezhad, and R.~P.~S. Mahler, ``{Robust
  multi-Bernoulli filtering},'' \emph{IEEE Journal on Selected Topics in Signal
  Processing}, vol.~7, no.~3, pp. 399--409, 2013.

\bibitem{Gostar2015}
A.~K. Gostar, R.~Hoseinnezhad, and A.~Bab-Hadiashar, ``{Multi-Bernoulli
  sensor-selection for multi-target tracking with unknown clutter and detection
  profiles},'' \emph{Signal Processing}, vol. 119, pp. 28--42, 2015.

\bibitem{Lian2010}
F.~Lian, C.~Han, and W.~Liu, ``{Estimating unknown clutter intensity for PHD
  filter},'' \emph{IEEE Transactions on Aerospace and Electronic Systems},
  vol.~46, no.~4, pp. 2066--2078, 2010.

\bibitem{Chen2012}
X.~Chen, R.~Tharmarasa, M.~Pelletier, and T.~Kirubarajan, ``{Integrated clutter
  estimation and target tracking using Poisson point processes},'' \emph{IEEE
  Transactions on Aerospace and Electronic Systems}, vol.~48, no.~2, pp.
  1210--1235, 2012.

\bibitem{Liu2018}
W.~Liu, Y.~Chen, H.~Cui, and C.~Wen, ``{A nonuniform clutter intensity
  estimation algorithm for random finite set filters},'' \emph{IEEE
  Transactions on Aerospace and Electronic Systems}, vol.~54, no.~6, pp.
  2911--2925, 2018.

\bibitem{Lerro1993}
D.~Lerro and Y.~Bar-Shalom, ``Interacting multiple model tracking with target
  amplitude feature,'' \emph{IEEE Transactions on Aerospace and Electronic
  Systems}, vol.~29, no.~2, pp. 494--509, 1993.

\bibitem{Ehrman2009}
L.~M. Ehrman and P.~R. Mahapatra, ``{Impact of noncoherent pulse integration on
  RCS-assisted tracking},'' \emph{IEEE Transactions on Aerospace and Electronic
  Systems}, vol.~45, no.~4, pp. 1573--1579, 2009.

\bibitem{Clark2010}
D.~Clark, B.~Risti{\'{c}}, B.~N. Vo, and B.~T. Vo, ``{Bayesian multi-object
  filtering with amplitude feature likelihood for unknown object SNR},''
  \emph{IEEE Transactions on Signal Processing}, vol.~58, no.~1, pp. 26--37,
  2010.

\bibitem{Feng2016}
F.~Yang, W.~Zhang, Y.~Liang, Y.~Su, and X.~Yao, ``{Cardinality balanced
  multi-target multi-Bernoulli filter for target tracking with amplitude
  information},'' in \emph{Proceedings of 19th International Conference on
  Information Fusion}.\hskip 1em plus 0.5em minus 0.4em\relax ISIF, 2016, pp.
  958--964.

\bibitem{Mertens2016}
M.~Mertens, M.~Ulmke, and W.~Koch, ``{Ground target tracking with RCS
  estimation based on signal strength measurements},'' \emph{IEEE Transactions
  on Aerospace and Electronic Systems}, vol.~52, no.~1, pp. 205--220, 2016.

\bibitem{Bae2017}
S.~H. Bae, J.~Park, and K.~J. Yoon, ``{Joint estimation of multi-target
  signal-to-noise ratio and dynamic states in cluttered environment},''
  \emph{IET Radar, Sonar and Navigation}, vol.~11, no.~3, pp. 539--549, 2017.

\bibitem{Bae2019}
S.~H. Bae, ``{Survey of amplitude-aided multi-target tracking methods},''
  \emph{IET Radar, Sonar and Navigation}, vol.~13, no.~2, pp. 243--253, 2019.

\bibitem{Sun2019a}
J.~Sun, C.~Liu, Q.~Li, and X.~Chen, ``{Labelled multi-Bernoulli filter with
  amplitude information for tracking marine weak targets},'' \emph{IET Radar,
  Sonar and Navigation}, vol.~13, no.~6, pp. 983--991, 2019.

\bibitem{Yang2018}
B.~Yang, J.~Wang, C.~Yuan, J.~Thiyagalingam, and T.~Kirubarajan,
  ``{Multi-object Bayesian filters with amplitude information in clutter
  background},'' \emph{Signal Processing}, vol. 152, pp. 22--34, 2018.

\bibitem{Ristic2021}
B.~Ristic, L.~Rosenberg, D.~Y. Kim, and R.~Guan, ``{Bernoulli filter for
  tracking maritime targets using point measurements with amplitude},''
  \emph{Signal Processing}, vol. 181, p. 107919, 2021.

\bibitem{Yedidia2005}
J.~S. Yedidia, W.~T. Freeman, and Y.~Weiss, ``{Constructing free-energy
  approximations and generalized belief propagation algorithms},'' \emph{IEEE
  Transactions on Information Theory}, vol.~51, no.~7, pp. 2282--2312, 2005.

\bibitem{Zhang2019}
C.~Zhang, J.~Butepage, H.~Kjellstrom, and S.~Mandt, ``{Advances in variational
  inference},'' \emph{IEEE Transactions on Pattern Analysis and Machine
  Intelligence}, vol.~41, no.~8, pp. 2008--2026, 2019.

\bibitem{Chen2006Data}
L.~Chen, M.~J. Wainwright, M.~Cetin, and A.~S. Willsky, ``Data association
  based on optimization in graphical models with application to sensor
  networks,'' \emph{Mathematical and computer modelling}, vol.~43, no. 9-10,
  pp. 1114--1135, 2006.

\bibitem{Williams2014}
J.~L. Williams and R.~A. Lau, ``{Approximate evaluation of marginal association
  probabilities with belief propagation},'' \emph{IEEE Transactions on
  Aerospace and Electronic Systems}, vol.~50, no.~4, pp. 2942--2959, 2014.

\bibitem{Williams2018}
------, ``{Multiple scan data association by convex variational inference},''
  \emph{IEEE Transactions on Signal Processing}, vol.~66, no.~8, pp.
  2112--2127, 2018.

\bibitem{Sun2016IF}
S.~Sun, H.~Lan, Z.~Wang, Q.~Pan, and H.~Zhang, ``{The application of
  sum-product algorithm for data association},'' in \emph{Proceedings of 19th
  International Conference on Information Fusion}.\hskip 1em plus 0.5em minus
  0.4em\relax ISIF, 2016, pp. 416--423.

\bibitem{Kropfreiter2019}
T.~Kropfreiter, F.~Meyer, and F.~Hlawatsch, ``A fast labeled {multi-Bernoulli}
  filter using belief propagation,'' \emph{IEEE Transactions on Aerospace and
  Electronic Systems}, vol.~56, no.~3, pp. 2478--2488, 2019.

\bibitem{meyer2017scalable}
F.~Meyer, P.~Braca, P.~Willett, and F.~Hlawatsch, ``{A scalable algorithm for
  tracking an unknown number of targets using multiple sensors},'' \emph{IEEE
  Transactions on Signal Processing}, vol.~65, no.~13, pp. 3478--3493, 2017.

\bibitem{Meyer2018}
F.~Meyer, T.~Kropfreiter, J.~L. Williams, R.~Lau, F.~Hlawatsch, P.~Braca, and
  M.~Z. Win, ``{Message passing algorithms for scalable multitarget
  tracking},'' \emph{Proceedings of the IEEE}, vol. 106, no.~2, pp. 121--259,
  2018.

\bibitem{Soldi2019}
G.~Soldi, F.~Meyer, P.~Braca, and F.~Hlawatsch, ``{Self-tuning algorithms for
  multisensor-multitarget tracking using belief propagation},'' \emph{IEEE
  Transactions on Signal Processing}, vol.~67, no.~15, pp. 3922--3937, 2019.

\bibitem{Meyer2020}
F.~Meyer and M.~Z. Win, ``{Scalable data association for extended object
  tracking},'' \emph{IEEE Transactions on Signal and Information Processing
  over Networks}, vol.~6, pp. 491--507, 2020.

\bibitem{meyer2021scalable}
F.~Meyer and J.~L. Williams, ``Scalable detection and tracking of geometric
  extended objects,'' \emph{IEEE Transactions on Signal Processing}, vol.~69,
  pp. 6283--6298, 2021.

\bibitem{Sharma2019}
P.~Sharma, A.~A. Saucan, D.~J. Bucci, and P.~K. Varshney, ``{Decentralized
  Gaussian filters for cooperative self-localization and multi-target
  tracking},'' \emph{IEEE Transactions on Signal Processing}, vol.~67, no.~22,
  pp. 5896--5911, 2019.

\bibitem{Cormack2019}
D.~Cormack, I.~Schlangen, J.~R. Hopgood, and D.~E. Clark, ``{Joint registration
  and fusion of an infra-red camera and scanning radar in a maritime
  context},'' \emph{IEEE Transactions on Aerospace and Electronic Systems},
  vol.~56, no.~2, pp. 1357--1369, 2019.

\bibitem{Gaglione2022}
D.~Gaglione, P.~Braca, G.~Soldi, F.~Meyer, F.~Hlawatsch, and M.~Z. Win,
  ``{Fusion of sensor measurements and target-provided information in
  multitarget tracking},'' \emph{IEEE Transactions on Signal Processing},
  vol.~70, pp. 322--336, 2022.

\bibitem{Riegler2013}
E.~Riegler, G.~E. Kirkelund, C.~N. Manch{\'{o}}n, M.~A. Badiu, and B.~H.
  Fleury, ``{Merging belief propagation and the mean field approximation: A
  free energy approach},'' \emph{IEEE Transactions on Information Theory},
  vol.~59, no.~1, pp. 588--602, 2013.

\bibitem{Lan2019}
H.~Lan, S.~Sun, Z.~Wang, Q.~Pan, and Z.~Zhang, ``{Joint target detection and
  tracking in multipath environment: a variational Bayesian approach},''
  \emph{IEEE Transactions on Aerospace and Electronic Systems}, vol.~56, no.~3,
  pp. 2136--2156, 2019.

\bibitem{Lan-2020-107621}
H.~Lan, J.~Ma, Z.~Wang, Q.~Pan, and X.~Xu, ``{A message passing approach for
  multiple maneuvering target tracking},'' \emph{Signal Processing}, vol. 174,
  p. 107621, 2020.

\bibitem{Lan2020}
H.~Lan, Z.~Wang, X.~Bai, Q.~Pan, and K.~Lu, ``{Measurement-level target
  tracking fusion for over-the-horizon radar network using message passing},''
  \emph{IEEE Transactions on Aerospace and Electronic Systems}, vol.~57, no.~3,
  pp. 1600--1623, 2021.

\bibitem{Skolnik2002}
M.~I. Skolnik, \emph{Introduction to radar systems, third edition}.\hskip 1em
  plus 0.5em minus 0.4em\relax New York: McGraw-Hill, 2002.

\bibitem{2006Pattern}
C.~Bishop, \emph{Pattern recognition and machine learning}.\hskip 1em plus
  0.5em minus 0.4em\relax Stat Sci, 2006.

\bibitem{Huang2018}
Y.~Huang, Y.~Zhang, Z.~Wu, N.~Li, and J.~Chambers, ``{A novel adaptive Kalman
  filter with inaccurate process and measurement noise covariance matrices},''
  \emph{IEEE Transactions on Automatic Control}, vol.~63, no.~2, pp. 594--601,
  2018.

\bibitem{sarkka2013}
S.~S{\"a}rkk{\"a}, \emph{Bayesian filtering and smoothing}.\hskip 1em plus
  0.5em minus 0.4em\relax Cambridge university press, 2013.

\bibitem{Rabiner1989}
L.~R. Rabiner, ``{A tutorial on hidden {Markov} models and selected
  applications in speech recognition},'' \emph{Proceedings of the IEEE},
  vol.~77, no.~2, pp. 257--286, 1989.

\bibitem{Gorji2011}
A.~A. Gorji, R.~Tharmarasa, and T.~Kirubarajan, ``{Performance measures for
  multiple target tracking problems},'' in \emph{Proceedings of the 14th
  International Conference on Information Fusion}.\hskip 1em plus 0.5em minus
  0.4em\relax ISIF, 2011, pp. 1--8.

\bibitem{OSPA200}
D.~Schuhmacher, B.~T. Vo, and B.~N. Vo, ``{A consistent metric for performance
  evaluation of multi-object filters},'' \emph{IEEE Transactions on Signal
  Processing}, vol.~56, no.~8, pp. 3447--3457, 2008.

\end{thebibliography}

\end{document}